\documentclass[twocolumn]{aastex701}

\usepackage{bm}
\usepackage{amsmath}
\shorttitle{\textsc{BIND} Methods}
\shortauthors{M. E. Lee et al.}
\begin{document}

\title{\textsc{BIND} (Baryonic INpainting with Deep learning): A Field-level Emulator for Galaxy Groups and Clusters}

\author[orcid=0000-0002-2318-3087,sname='M.E. Lee']{Max E. Lee}
\affiliation{Department of Astronomy, Columbia University, MC 5246, 538 West 120th Street, New York, NY 10027, USA}
\email[show]{max.e.lee@columbia.edu}  

\author[0000-0002-3185-1540]{Shy Genel}
\affiliation{Center for Computational Astrophysics, Flatiron Institute, 162 Fifth Ave, New York, NY, 10010, USA}
\email{sgenel@flatironinstitute.org}

\author[0000-0003-3633-5403]{Zolt\'an Haiman}
\affiliation{Department of Astronomy, Columbia University, MC 5246, 538 West 120th Street, New York, NY 10027, USA}
\affiliation{Department of Physics, Columbia University, MC 5255, 538 West 120th Street, New York, NY 10027, USA}
\affiliation{Institute of Science and Technology Austria, Am Campus 1, Klosterneuburg 3400 Austria}
\email{Zoltan.Haiman@ista.ac.at}

\author[0000-0003-2630-9228]{Greg L. Bryan}
\affiliation{Department of Astronomy, Columbia University, MC 5246, 538 West 120th Street, New York, NY 10027, USA}
\email{gbryan@columbia.edu}

\author[0000-0001-7964-5933]{Christopher C. Lovell}
\affiliation{Institute of Astronomy, Madingley Road, Cambridge, CB3 0HA, UK}
\affiliation{Kavli Institute for Cosmology Cambridge, Madingley Road, Cambridge, CB3 0HA, UK}
\email{ccl62@cam.ac.uk}

\author[0000-0002-2312-3121]{Boryana Hadzhiyska}
\affiliation{Institute of Astronomy, Madingley Road, Cambridge, CB3 0HA, UK}
\affiliation{Kavli Institute for Cosmology Cambridge, Madingley Road, Cambridge, CB3 0HA, UK}
\email{}

\begin{abstract}
Baryonic feedback is a dominant source of systematic uncertainty for upcoming weak-lensing surveys, but current tools for modeling its effect rely on spherical approximations and density profiles calibrated almost entirely on two-point statistics. We introduce \textsc{BIND} (Baryonic INpainting with Deep learning), a conditional flow-matching model that learns a field-level mapping from dark-matter-only halos to their hydrodynamical counterparts. \textsc{BIND} is trained on halos from the 1024 paired hydrodynamical and dark-matter-only simulations of the CAMELS $50\,h^{-1}\,\mathrm{Mpc}$ SB35 suite and samples dark matter, gas, and stellar mass fields over redshift across the full 35-dimensional $\Lambda$CDM and IllustrisTNG galaxy formation parameter space. \textsc{BIND} recovers dark matter, gas, and stellar masses at the percent level, reproduces azimuthally averaged profiles to $\lesssim10\%$ at all radii, and matches halo shape distributions with high fidelity. The learned parameter dependence captures the rank correlations between the generated fields and the subgrid parameters, and the field-level response to individual parameter variations is recovered in both sign and morphology. Halo mass is never supplied as conditioning, yet the baryon fraction, stellar-to-halo mass relation, inter-component scaling relations, and the joint covariance of their residuals are all reproduced. We finally show that, applied halo-by-halo to a $(50\,h^{-1}\,\mathrm{Mpc})^3$ $N$-body volume with $512^3$ particles, \textsc{BIND} reproduces the projected matter power spectrum suppression to the accuracy ceiling set by pasting in the hydrodynamical halos themselves, in minutes on one GPU. We release the trained \textsc{BIND} models and all generated halos as open-source tools. A companion paper \citep{Lee-2026c} extends \textsc{BIND} to thermodynamic fields and non-Gaussian weak-lensing statistics.
\end{abstract}



\section{Introduction} \label{sec:introduction}

Exploiting the full statistical power of future weak lensing surveys requires accurate modeling of the matter distribution on small, highly non-linear scales. Baryonic processes, including feedback from active galactic nuclei (AGN) and supernovae, radiative cooling, and star formation, have previously been a subdominant effect in cosmological analyses, but for upcoming surveys they will become the dominant source of systematic uncertainty. Therefore, we need modeling approaches that account for these effects \citep[e.g.,][]{Jing-2006,Chisari-2019,Schneider-2019}. In response to this demand, a domain in cosmology and astrophysics has emerged, developing methods for baryonic accounting in cosmological analyses.

The most straightforward and arguably cost-efficient treatment of baryonic effects is to remove small-scale data entirely, a strategy adopted in several weak-lensing analyses, including DES, KiDS, and HSC \citep[e.g.][]{Secco-2022,Li-2023,Dalal-2023}. Such cuts discard the non-linear scales over which baryons dominate, but these same small scales have also been shown to contain a wealth of cosmological information, and their removal leads to a non-negligible loss in cosmological constraining power \citep[e.g., Table V and Section IV B of][]{Fang-2007}.

The alternative extreme is to model baryonic effects directly with cosmological hydrodynamical simulations that self-consistently evolve dark matter and baryons. When appropriately calibrated, this approach serves as an ideal testing ground for baryonic effects. However, these simulations, such as IllustrisTNG, EAGLE, SIMBA, BAHAMAS, FLAMINGO or COLIBRE \citep[e.g.][]{Vogelsberger-2014,Schaye-2015,Pillepich-2018,Dave-2019,Schaye-2023, Colibre}, have limited mass and spatial resolution, and the galaxy-formation physics below the resolution scale is implemented through uncertain subgrid prescriptions that differ from code to code \citep{vanDaalen-2011,Vogelsberger-2020,VillaescusaNavarro-2021}. In addition, their high computational cost precludes their use in forward-modeling pipelines that require simultaneous sampling of cosmological and galaxy-formation parameters.

To reduce the computational cost of hydrodynamical simulations while avoiding small-scale data cuts, semi-analytic baryonic correction models (BCMs) have been proposed. BCMs modify dark-matter-only (DMO) simulations, which are far cheaper than hydrodynamical simulations, by applying physically motivated empirical prescriptions to halo density profiles \citep{Schneider-2015,Schneider-2019,Arico-2020}. In these models, a baryonic halo profile is decomposed into components such as cold bound gas, ejected gas, and relaxed dark matter, each with an analytic form. DMO halos from N-body simulations can then have their particles radially shifted so that the resulting halo profile matches the sum of the baryonic decomposition. Ideally the parameters from these models could be fit to X-ray and CMB observations, but as of yet, they have primarily been instead calibrated to hydrodynamical simulations \citep[however, see][for approaches fitting to data]{Kovac-2025}.

BCMs have proven highly effective for cosmological analyses. The A20-BCM introduced in \citet{Arico-2020} reproduces the suppression of the 3D matter power spectrum due to baryons at the percent level across several state-of-the-art hydrodynamical simulations. Building on this, the BACCO project constructed a neural-network-based emulator over a multi-dimensional baryonic and cosmological parameter space, enabling fast evaluation of BCM predictions for $P(k)$ \citep{Arico-2021}. 

Similar to BCM models, halo-model approaches such as HMcode-2020 introduce a small number of baryonic feedback parameters into an augmented halo model to match the matter power spectrum to percent-level accuracy \citep{Mead-2021}, but they do not directly specify the underlying field-level redistribution of mass within and around halos. As a result, their impact on higher-order and non-Gaussian statistics must be calibrated a posteriori and may not faithfully represent the true baryon-induced spatial structure \citep{Lee-2026a}.

The more recent component-wise baryonification (BFC) framework of \citet{Schneider-2025} extended previous BCMs by assigning each DMO particle from an $N$-body simulation a dark matter, gas, and stellar component, which are displaced individually, as well as thermodynamic properties using hydrostatic and ideal gas equations. The BFC model was validated against the FLAMINGO and IllustrisTNG simulation suites, achieving $\lesssim 2\%$ agreement in the matter power spectrum suppression up to $k = 5~h\,\mathrm{Mpc}^{-1}$ across a range of feedback prescriptions. In a companion paper, \citet{Kovac-2025} applied the BFC framework to jointly fit kinematic Sunyaev-Zel'dovich (kSZ) observations from the Atacama Cosmology Telescope (ACT) and X-ray gas fractions from eROSITA, finding that current multiwavelength data prefer a stronger feedback model than assumed in most hydrodynamical simulations, with a resulting power spectrum suppression of $20$--$25\%$ at $k = 5~h\,\mathrm{Mpc}^{-1}$. This is consistent with recent findings by \citet{Hadzhiyska-2024, Bigwood-2024} and \citet{Siegel-2025}, who also find that feedback models in state-of-the-art simulations appear to underpredict the level of baryon suppression inferred from observations.

A complementary strategy is to account for baryons in cosmic shear analyses directly from data rather than through a model. Cross-correlating shear with tracers of the ionized gas, such as the kSZ effect \citep{Bigwood-2024} or fast radio burst dispersion measures \citep{Leung-2025, Wayland-2026}, constrains, or in principle nulls, the baryonic contribution to the lensing signal, and joint kSZ and CMB-lensing measurements around galaxies and groups now recover the gas fraction as a function of halo mass directly \citep{Hadzhiyska-2025}. These approaches are largely independent of the subgrid physics of any one simulation, and they provide the empirical targets that any baryonification model, including the one presented here, must eventually be calibrated against.

BCMs, halo-model approaches, and BFCs have been calibrated and tested almost exclusively on two-point statistics. The first steps beyond this regime, such as joint fits to the power spectrum and bispectrum, have shown that while BCMs can achieve $\lesssim 3\%$ agreement with the bispectrum, discrepancies persist on small scales and for certain triangle configurations \citep{Arico-2021a}. Moreover, when tested on non-Gaussian statistics such as peak counts, BCMs can fail more dramatically. \citet{Lee-2023} found that while the A20-BCM reproduces weak-lensing peak counts at the percent level for peaks with signal-to-noise $S/N < 4$, it systematically underpredicts the number of the highest peaks, particularly for deep, wide-area surveys such as the Vera C. Rubin Observatory LSST \citep{Ivezic-2019} and \textit{Euclid} \citep{Laureijs-2011}.

More recently, several field-level and map-level approaches have been developed to go beyond analytic halo profiles. \citet{Sharma-2024} proposed a model for baryonic effects where DMO simulation modes are multiplied by an isotropic Fourier-space transfer function, $\sqrt{P_{\rm hydro}(k)/P_{\rm DMO}(k)}$. This approach is motivated by the observation that the cross-correlation coefficient between matched hydrodynamical and $N$-body simulations remains close to unity down to small scales and is accurate for two-point statistics across thousands of simulations \citep{Sharma-2024}. However, because the transfer function is isotropic in $k$-space, this approach cannot capture anisotropic, halo-centric mass redistribution. 

Several other approaches to baryonic modeling have been explored that leverage machine learning. At the halo level, \citet{Chadayammuri-2023} used a U-Net to map DMO cluster fields to gas density, temperature, and X-ray observables, reproducing observed mass-observable scaling relations for massive clusters. Similar halo-based approaches have used machine learning to infer baryonic properties such as gas mass and temperature from dark-matter halo properties in simulations such as EAGLE \citep[e.g.][and the references therein]{Moews-2021, Lovell-2022}. Other generative models such as diffusion models \citep{ho2020ddpm, song2021sde} have also been explored for cosmological fields, for example, Cosmo-FOLD, introduced by \citet{Mishra-2026}, uses a latent diffusion model to generate 3D dark-matter and gas-density fields conditioned on lower-resolution or partial information, achieving $\lesssim 10\%$ accuracy in power spectra up to $k \lesssim 5~h\,\mathrm{Mpc}^{-1}$.

Generative models are particularly attractive because they sample a learned posterior of baryonic effects rather than return a single deterministic prediction. This means they can represent the intrinsic physical scatter, and this has now been demonstrated across a range of astrophysical inverse problems. Score-based priors recover posterior samples of lensed source galaxies \citep{Adam-2022} and of the weak lensing convergence field from noisy shear \citep{Remy-2023}, they recover the initial conditions of the Universe from the evolved density field \citep{Legin-2023}, diffusion models reconstruct dark matter fields from galaxies in CAMELS \citep{VillaescusaNavarro-2021, Ni-2023, Genel-2026} -- a suite of thousands of cosmological hydrodynamical simulations spanning various galaxy formation parameter spaces -- while marginalizing over cosmology and feedback \citep{Ono-2024}, they reconstruct cluster gas and dark matter maps from mock SZ and X-ray images \citep{Hsu-2024}, and they have been used directly for cosmological parameter inference \citep{Mudur-2024}. 

Closest in spirit to the present work, \citet{CuestaLazaro-2024} generate halo point clouds at the field level with a diffusion model conditioned on cosmology, and \citet{Pandey-2025} paint galaxies onto dark-matter-only simulations with a transformer-based model trained on CAMELS, reproducing galaxy clustering and one-point statistics at a fraction of the cost of a hydrodynamical run. These neural network and generative methods are an important step toward field-level emulation of baryonic effects, but they either operate on full volumes without explicit halo-level conditioning, or target the galaxy population rather than the continuous fields that set the lensing signal, making it nontrivial to connect them to the halo-level observables associated with these baryonic effects.

Given the rapid progress and successes of the methods described above, several opportunities remain to expand the current state of baryonic modeling. 

First, BCMs and BFCs, while halo-centric, rely on fixed analytic profile prescriptions and are not generative, so they cannot sample from a posterior over baryonic field configurations at fixed dark matter and model parameters. Transfer-function approaches are field-level, isotropic in $k$-space, and similarly non-generative. Map-level ML approaches operate on full volumes rather than individual halos. The advantage of halo-centric approaches is that the source of baryonic effects lies in processes within halos, and by applying full-volume corrections, one loses the halo-centric resolution needed to capture detailed mass redistribution within and around groups and clusters. Further, because most particles in simulations reside outside halos, correcting full volumes is computationally expensive and does not necessarily affect the scales we truly care about.

Second, higher-order and non-Gaussian weak-lensing statistics, such as peak and minimum counts, Minkowski functionals, and field-level summaries, have been shown to add substantial cosmological information beyond the power spectrum, but existing baryonic correction tools have not been systematically validated for these observables \citep{Lee-2023,Lee-2026a}. In fact, \citet{Lee-2026a} shows explicitly that different lensing statistics are sensitive to different halo masses and radial ranges, implying that models fit to one lensing statistic may fail to capture the profile changes required by another. The best way to ensure all statistics are captured well is with a field-level approach that accurately captures the anisotropic redistribution and full field inside and around halos.

Third, the effect of galaxy-formation model parameters on matter redistribution and its propagation to weak lensing statistics is poorly understood. The 35-dimensional IllustrisTNG galaxy formation parameter space (30 astrophysical + 5 $\Lambda$CDM cosmological parameters) contains complex, nonlinear degeneracies between both the astrophysical parameters themselves, and between the subgrid and cosmological parameters together (for example, between supernova and AGN feedback), which analytic prescriptions are potentially ill-equipped to capture \citep{Lee-2024}.

No existing method simultaneously operates at the field level, conditioned on galaxy formation model parameters, in a halo-centric manner while also being generative, enabling proper posterior sampling over baryonic field realizations at fixed inputs. Table~\ref{tab:capabilities} summarizes where existing approaches stand on these axes.

\begin{deluxetable*}{lcccc}
\tabletypesize{\footnotesize}
\tablecaption{Capabilities of currently implemented baryonic-modeling approaches across the gaps
identified in \S~\ref{sec:introduction}.\label{tab:capabilities}}
\tablewidth{0pt}
\tablehead{\colhead{Approach} & \colhead{Halo-centric} & \colhead{Field-level} &
           \colhead{Generative} & \colhead{Param.-cond.} }
\startdata
BCM / halo model$^{a}$ & yes & no (profiles) & no & BCM params \\
BFC$^{b}$ & yes & particle-level & no & BFC params  \\
Isotropic transfer function$^{c}$ & no & yes & no & suite params \\
Volume-level generative$^{d}$ & no & yes & yes & no  \\
Halo image-to-image$^{e}$ & yes & yes & no & fixed physics \\
Galaxy painting$^{f}$ & no (galaxies) & no (catalog) & yes & 6 (cosmo + astro)  \\
\textsc{BIND} (this work) & yes & yes (2D patches) & yes & 35 (cosmo + TNG) 
\enddata
\tablecomments{$^{a}$\citet{Schneider-2015, Arico-2020, Mead-2021};
$^{b}$\citet{Schneider-2025}; $^{c}$\citet{Sharma-2024};
$^{d}$\citet{Mishra-2026}; $^{e}$\citet{Chadayammuri-2023};
$^{f}$\citet{Pandey-2025}.}
\end{deluxetable*}

In this work, we introduce \textsc{BIND} (Baryonic INpainting with Deep learning), a conditional flow matching model that learns the mapping from dark-matter-only halos to their hydrodynamical counterparts at the field level. \textsc{BIND} is trained on $151,685$ halos from matched hydrodynamical and dark-matter-only simulations across $8$ redshifts in the CAMELS IllustrisTNG suite of $50\,h^{-1}\,\mathrm{Mpc}$ simulations \citep{Genel-2026} and is conditioned on redshift and the full set of 35 cosmological and astrophysical parameters from $\Lambda$CDM and the IllustrisTNG galaxy formation model. 

On a halo-by-halo basis, \textsc{BIND} takes as input a projected two-dimensional DMO surface density field centered on a halo from an N-body simulation and outputs simultaneous predictions for the dark matter, gas, and stellar mass distributions expected in a hydrodynamical simulation with a given set of astrophysical and cosmological parameters. Because the model is generative, it naturally provides samples from the posterior distribution of baryonic fields given a DMO realization and a parameter vector, representing stochastic draws for the same halo. In a companion paper \citep{Lee-2026c}, \textsc{BIND} is extended to output field-level thermodynamic quantities for gauging feedback at the halo level, such as temperature, pressure, and entropy maps.

By construction, \textsc{BIND} is halo-centric, field-level, and parameter-conditioned. This lets it capture anisotropic mass redistribution in and around halos, learn how different mass components vary across the galaxy-formation parameter space, and preserve higher-order information in the output fields, enabling direct application to downstream statistics. Because the only spatial input to the model is the projected dark-matter field, the generated fields do not depend on any halo mass or radius definition; a mass threshold enters only in choosing which halos to process. This sidesteps one of the persistent ambiguities of halo-based baryonification models, whose profiles are anchored to a particular mass and radius convention. By applying \textsc{BIND} patch-by-patch to the halos of an arbitrary $N$-body simulation, we show that one can baryonify large cosmological volumes across the IllustrisTNG parameter space in minutes on a single GPU, orders of magnitude faster than re-running a hydrodynamical simulation. This makes \textsc{BIND} suitable for field-level analysis pipelines.

In this paper, we introduce \textsc{BIND} and provide a comprehensive suite of validations exploring its ability to generate accurate density fields with correct parameter dependencies. We are primarily interested in answering:
\begin{enumerate}
    \item Can a conditional flow matching model learn the mapping from dark-matter-only to hydrodynamical fields at the halo level, conditioned on the joint $\Lambda$CDM $+$ IllustrisTNG galaxy-formation parameter space?
    \item Does \textsc{BIND} learn the appropriate astrophysical parameter dependencies for the generated hydrodynamical fields?
    \item Is \textsc{BIND} able to learn the correlations between field-level observables and how they change as a function of astrophysical parameters?
    \item Can \textsc{BIND} be deployed as a practical baryonification engine for two-point statistics such as the matter power spectrum? In a companion paper, we explore the efficacy of \textsc{BIND} on non-Gaussian statistics. 
\end{enumerate}

The paper is organized as follows. In \S~\ref{sec:simulations} we introduce the training and testing suites ingested by \textsc{BIND}. We then discuss the flow-matching model and the data-processing pipeline in \S~\ref{sec:model}. We tackle the first and second questions above and validate the model both with fixed parameters and across the parameter space in \S~\ref{sec:validation} and \S~\ref{sec:param_response}, respectively. We then explore the third question and joint distributions of generated fields in \S~\ref{sec:joint_structure} and consider the final question above with the application of \textsc{BIND} to a full N-body simulation in \S~\ref{sec:bind_in_the_wild}. We close with a list of caveats in \S~\ref{sec:caveats}, and conclusions in \S~\ref{sec:conclusions}.

\section{Simulations}
\label{sec:simulations}

All training and testing data in this work come from the Cosmology and Astrophysics with MachinE-Learning Simulations \citep[CAMELS;][]{VillaescusaNavarro-2021, Villaescusa-Navarro-2023} project. CAMELS is a suite of hydrodynamical and $N$-body simulations designed to train and test machine-learning models on cosmological data. Beyond machine learning, the suite has been used extensively to characterize how subgrid feedback reshapes the baryon distribution, including its effect on the warm-hot circumgalactic medium \citep{Medlock-2024b}, on X-ray observables of the circumgalactic gas \citep{Lau-2025}, on fast radio burst dispersion measures \citep{Medlock-2024, Medlock-2025}, and on the dark matter distribution itself \citep{Gebhardt-2026}. In the updated $50\,h^{-1}\,\mathrm{Mpc}$ suite of \citet{Genel-2026}, each simulation follows \(512^3\) dark-matter particles and, in the hydrodynamical runs, an equal number of gas resolution elements in a periodic comoving box of side length $L = 50\,h^{-1}\,{\rm Mpc}$. The hydrodynamical runs employ the IllustrisTNG galaxy-formation model \citep{weinberger2017supermassive,pillepich2018simulating}, implemented in the \textsc{arepo} moving-mesh code \citep{springel2010arepo}. For a description of the flagship IllustrisTNG runs and their public data products, see \citet{Nelson-2019}.

\subsection{The SB35 Sobol Sequence Set}
\label{subsec:sb35}

We train \textsc{BIND} on the SB35 suite \citep{Genel-2026}, consisting of 1024 paired IllustrisTNG hydrodynamical and $N$-body simulations spanning a 35-dimensional parameter space of cosmological and galaxy formation parameters sampled with a Sobol sequence \citep{Sobol-1967}. The parameter space includes five cosmological parameters and 30 galaxy-formation subgrid parameters governing stellar feedback, AGN feedback, star formation, and interstellar medium processes in the IllustrisTNG model (for a full breakdown of the parameters, see \citealt{Genel-2026}, particularly Table~1). The Sobol sequence ensures space-filling coverage of this high-dimensional parameter space, with each simulation using a unique combination of parameter values and random initial phases. From every simulation we use eight snapshots spanning $z = 0$--$4$; however, the evaluations of Sections~\ref{sec:validation} and \ref{sec:bind_in_the_wild} are performed at $z = 0$, and \S~\ref{subsec:z_response} validates the model across the full redshift range.

Each hydrodynamical simulation and its gravity-only $N$-body twin is evolved from identical initial conditions. Therefore, by holding the initial conditions fixed and aside from intrinsic randomness due to the butterfly effect \citep{Genel-2019}, differences between the $N$-body and hydrodynamical outputs arise solely from baryonic physics and provide a clean training signal for the mapping of pixelated mass maps $\bm{M}_{\rm DMO} \rightarrow \bm{M}_{\rm hydro}$.

\subsection{Evaluation Suites}
\label{subsec:eval_suites}

We use three independent simulation sets for evaluation.

The \textbf{Sobol sequence (SB35) test set} consists of the 102 SB35 simulations withheld from training. Each simulation contains a different set of cosmological and astrophysical parameters and initial conditions.

The \textbf{One-Parameter (1P) set} contains simulations in which a single parameter is varied at a time while all others are held at their fiducial values, and all runs share the same initial conditions. 

The \textbf{Cosmic Variance (CV) set} consists of 27 simulations run at the fiducial IllustrisTNG parameter values but with different random initial conditions.

%=============================================================
\section{Model and Training}
\label{sec:model}
%=============================================================

Our goal is to learn the conditional distribution,
\begin{equation}
    p(\bm{M}_{\rm hydro} \mid \bm{M}_{\rm DMO},\,
    \boldsymbol{\theta}_{\rm TNG},\, z).
    \label{eq:target_distribution}
\end{equation}
Each channel of $\bm{M}_{\rm hydro}$ is a halo-centered projected mass map on an $H \times W$ pixel grid, for $c \in \{\rm dm,\, gas,\, stars\}$, where the value in a given pixel is the total mass in that pixel's column through the box,
\begin{equation}
    M_{c,\,ij} = \sum_{k} m_{c, ijk},
    \label{eq:mass_map}
\end{equation}
where the sum runs over all particles of species $c$ whose projected position falls in pixel $(i,j)$, along the full $50\,h^{-1}\,\mathrm{Mpc}$ line of sight. Throughout, bold symbols denote pixelated two-dimensional maps, so that $\bm{M}_{\rm hydro} \in \mathbb{R}^{3 \times H \times W}$ with one channel per species. $\bm{M}_{\rm DMO} \in \mathbb{R}^{1 \times H \times W}$ is the conditioning input computed in the same way and is the dark-matter-only patch centered on a halo. Note that we have chosen the fields as masses here, though this choice is somewhat arbitrary. In our companion paper \citep{Lee-2026c}, we include gas pressure, temperature, and entropy fields such that $\bm{M}_{\rm hydro} \in \mathbb{R}^{6 \times H \times W}$, but for clarity in this work, we limit the channel outputs to just the three mass fields. 

The parameter vector $\boldsymbol{\theta}_{\rm TNG} \in \mathbb{R}^{35}$ encodes the full set of cosmological and astrophysical parameters of the simulation (Section~\ref{subsec:sb35}); however, in general, this vector can be any conditioning parameter vector that is associated with the halo. One future plan discussed in \S~\ref{sec:caveats} is to train \textsc{BIND} using physical inputs measured from each simulation, rather than the TNG parameter vector. For this first iteration of \textsc{BIND}, we stick to the parameter set handed down by the simulation. The scalar $z$ denotes the snapshot redshift, supplied to the network as the scale factor $a = 1/(1+z)$. \textsc{BIND} is trained on halos from multiple snapshots, and its redshift conditioning is validated in Section~\ref{subsec:z_response}.

The \textsc{BIND} framework is built on a conditional flow matching network that learns the mapping from dark-matter-only halo fields and IllustrisTNG model parameters to the baryonic mass fields of the paired hydrodynamical simulation. In the following subsections, we detail the data-processing pipeline, review the flow-matching formalism, describe the network architecture, and specify the training procedure.

%-------------------------------------------------------------
\subsection{Data Processing}
\label{subsec:data_processing}
%-------------------------------------------------------------

Converting raw simulation particle data into training samples for \textsc{BIND} involves four main steps: halo identification and centering, random rotation, line-of-sight projection, and field normalization.

We treat the $N$-body $50\,h^{-1}\,\mathrm{Mpc}$ CAMELS suite of simulations as the base set from which halos with $M_{200c} \geq 10^{13}\,M_{\odot}\,h^{-1}$ are identified\footnote{$M_{200c}$ is the mass enclosed within the radius $R_{200c}$, inside which the mean density is $200\rho_c$, with $\rho_c$ the critical density of the universe.} at 8 redshifts between $z=0$ and $z=4$. We find halos with the friends-of-friends algorithm \citep{Davis-1985}, and we take their centers and bound substructure from the \textsc{subfind} catalogs \citep{Springel-2001}. For each halo, we exploit the periodic boundary conditions of the simulation box and re-center the volume on the particle with the deepest gravitational potential (\textsc{GroupPos} in the FoF catalog), placing the halo at $(x,y,z) = (25,25,25)\,h^{-1}\,\mathrm{Mpc}$.

We place a stationary observer at $(x,y,z) = (25,25,0)\,h^{-1}\,\mathrm{Mpc}$ and apply a random rotation about the halo center. The periodic boundary conditions ensure that, with appropriate tiling of the simulation box, a $50\,h^{-1}\,\mathrm{Mpc}$ cube remains volume-filling after the rotation. We store the three Euler angles defining each rotation and apply them to the paired hydrodynamical simulation to ensure exact spatial correspondence between the $N$-body and baryonic projections. We perform this procedure 10 times per halo, drawing independent rotation matrices each time to augment the training set.

We project each rotated volume grid along the $z$-axis using cloud-in-cell (CIC) interpolation \citep{hockney}. We retain the central $128\times128$ pixel patch as the primary $N$-body conditioning field $\bm{M}_{\rm DMO}$. For the hydrodynamical simulations, we use the same procedure to project the dark-matter, gas, and stellar density fields. All fields are then defined on a regular $H \times W = 128 \times 128$ grid with $0.049\,h^{-1}\,\mathrm{Mpc}$ resolution per pixel extending $6.25\,h^{-1}\,{\rm Mpc}$.

We highlight a few of the design choices described above. First, we do this in 2D, which lets us explore the architecture and test our procedures at reduced computational cost. 3D fields increase the memory and computational requirements by orders of magnitude, so this first iteration of \textsc{BIND} remains 2D. This choice is a well-motivated first step, though, as baryonification applied directly at the map level reproduces weak lensing two-point and higher-order statistics to within a few percent \citep{Anbajagane-2024, Zhou-2025}. Extending the procedure to 3D fields is a goal saved for future development of \textsc{BIND}. The other important caveat is the projection depth. Each training sample is $6.25\times6.25\,h^{-1}\,\mathrm{Mpc}$ and $128^2$ pixels, computed as the projected mass over the full $50\,h^{-1}\,\mathrm{Mpc}$ line of sight. While the majority of the mass within a halo will be from material associated with that halo, there will, of course, be foreground and background matter not associated with the halo included in this projection. The projection may also include multiple halos along the line of sight, which could bias our conditioning vector. Also note that we center our training data only on halo centers; subhalos are not used as explicit training data, but because each massive halo has subhalos in the field, they are implicitly included in the training. We made this projection choice to facilitate pasting the baryonified halos back into $N$-body simulations, which we discuss in \S~\ref{sec:bind_in_the_wild}. However, we have also explored the effect of this projection and trained an equivalent model using projections of the $6.25\,h^{-1}\,\mathrm{Mpc}$ cube centered on each halo (see Appendix~\ref{cube_comp}). 

Because projected surface density fields span several orders of magnitude in intensity and contain many near-zero pixels, all map channels, $\bm{M}_c$, are compressed with
\begin{equation}
    \bm{M}_c' = \log_{10}(1 + \bm{M}_c),
    \label{eq:logtransform}
\end{equation}
which maps zero-density pixels to zero while suppressing the extreme dynamic range. Each transformed channel $\bm{M}_c'$ is then standardized to zero mean and unit variance,
\begin{equation}
    \tilde{\bm{M}}_c
    = \frac{\bm{M}'_c - \mu_c}{\sigma_c + \epsilon},
    \label{eq:zscore}
\end{equation}
where $\mu_c$ and $\sigma_c$ are estimated from a random subsample of 10,000 training maps, and $\epsilon = 10^{-8}$ prevents division by zero. This normalization is applied identically to the target baryonic fields and $\bm{M}_{\rm DMO}$, but with their own computed $\mu_c$ and $\sigma_c$ per field.

The parameter vector $\boldsymbol{\theta}_{\rm TNG}$ is normalized component-wise and in accordance with the parameter sampling to generate the Sobol sequence in \citet{Genel-2026}. This requires many parameters (see Table 1 of \citealt{Genel-2026} for the list of log-space parameters) to first be log-transformed before linear rescaling to $[0,1]$:
\begin{equation}
    \tilde{\theta}_j
    =
    \frac{\theta_j^{(\log)} - \theta_{j,\min}^{(\log)}}
         {\theta_{j,\max}^{(\log)} - \theta_{j,\min}^{(\log)} + \epsilon},
    \label{eq:paramnorm}
\end{equation}
where
\begin{equation}
    \theta_j^{(\log)}
    =
    \begin{cases}
        \log_{10}(\theta_j), & \text{if parameter } j \text{ is log-scaled},\\
        \theta_j,            & \text{otherwise.}
    \end{cases}
\end{equation}
The bounds $\theta_{j,\min}$ and $\theta_{j,\max}$ are taken from the SB35 parameter tables \citep{Genel-2026}, so the normalization range is well-defined for any simulation in the suite, not only those in the training subset.

We split the 1024 SB35 simulations into $90\%$ for training (922 simulations) and $10\%$ for testing (102 simulations). Within the training set, we reserve $20\%$ (184 simulations) for validation, and use the remaining $80\%$ (738 simulations) for training. After applying 10 random rotations per halo, the training set contains approximately 360,000 projected halo maps spanning the full 35-dimensional parameter space.

%-------------------------------------------------------------
\subsection{Flow Matching}
\label{subsec:fm}
%-------------------------------------------------------------
\begin{deluxetable}{lcc}
\tabletypesize{\footnotesize}
\tablecaption{Measured computational cost for training and generating with \textsc{BIND}.\label{tab:timing}}
\tablewidth{0pt}
\tablehead{\colhead{Task} & \colhead{Hardware} & \colhead{Wall time}}
\startdata
Model Training & $8\times$H100 & $19$ h (${\sim}150$ GPU-h) \\
Load model + normalization & A100 (MIG $2/7$) & $29$ s \\
Generate one halo & A100 (MIG $2/7$) & $0.9$ s \\
Baryonify a $50\,h^{-1}$Mpc box & A100 (MIG $2/7$) & $\sim40$ s \\
Halo-pasting step alone (CPU) & 1 core & $0.05$ s \\
100-draw posterior, one halo & A100 (MIG $2/7$) & $88$ s \\
\enddata
\end{deluxetable}

The overall goal in our conditional flow matching is to learn a velocity field $\bm{v}_{\phi}$ that transports samples from an isotropic Gaussian prior, $\bm{M}_0 \sim \mathcal{N}(\bm{0},\bm{I})$, to the target baryonic field $\bm{M}_{\rm hydro}$ conditioned on the paired dark-matter-only halo $\bm{M}_{\rm DMO}$, a parameter vector $\boldsymbol{\theta}_{\rm TNG}$ and the scale factor $a$. In this section, we now describe how we build the velocity model.

The trajectory of the field from its Gaussian initial state to the final field follows a path such that intermediate states on the time interval, $t \sim \mathcal{U}(0,1)$, are simply a linear interpolation of the initial and final fields,
\begin{equation}
\begin{split}
    \bm{M}_t &= (1-t)\,\bm{M}_0 + t\,\bm{M}_{\rm hydro}.
\end{split}
    \label{eq:fm_path}
\end{equation}
This defines a straight-line probability path from noise at $t=0$ to data at $t=1$ \citep{lipman2023flowmatching}, an objective developed concurrently as stochastic interpolants \citep{albergo2023interpolants} and as rectified flow \citep{liu2022rectified}. We use this formulation rather than the diffusion one \citep{ho2020ddpm, song2021sde} because the interpolation in Eq.~\ref{eq:fm_path} fixes the probability path in advance and makes the regression target in Eq.~\ref{eq:fm_target} constant along each training trajectory. However, we have also explored similar architectures using diffusion models and found that, for this problem, which involves highly sparse stellar fields, the flow-matching model yields the highest-fidelity reconstructions. 

The target velocity along this path is the difference between the target and noise fields with respect to the time, which by construction is $1$,
\begin{equation}
    \bm{v}^{\ast} = \bm{M}_{\rm hydro} - \bm{M}_0.
    \label{eq:fm_target}
\end{equation}
We parameterize the velocity field with a neural network $\bm{v}_{\phi}(\bm{M}_t, t, \bm{M}_{\rm DMO}, \boldsymbol{\theta}_{\rm TNG}, a)$, with neural network parameters $\phi$, and we write $\bm{v}_{\phi}$ for brevity, suppressing the scale-factor argument in the equations below. The network is trained to minimize the conditional flow matching objective
\begin{equation}
    \mathcal{L}_{\rm FM}(\phi)
    =
    \mathbb{E}_{t,\,\bm{M}_0,\,\bm{M}_{\rm hydro}}
    \left[\,
    \left\|
        \bm{v}_{\phi}
        -
        \left(\bm{M}_{\rm hydro} - \bm{M}_0\right)
    \right\|_2^2
    \right],
    \label{eq:fm_loss}
\end{equation}
where the conditioning inputs $(\bm{M}_{\rm DMO}, \boldsymbol{\theta}_{\rm TNG}, a)$ are held fixed for each training sample.

At inference, we generate baryonic fields by integrating the learned probability-flow ODE from $t=0$ to $t=1$,
\begin{equation}
    \frac{d\bm{M}}{dt}
    =
    \bm{v}_{\phi}\left(
        \bm{M},\,t,\,
        \bm{M}_{\rm DMO},\,\boldsymbol{\theta}_{\rm TNG}, a
    \right),
    \label{eq:ode}
\end{equation}
starting from $\bm{M}(0) \sim \mathcal{N}(\bm{0},\bm{I})$. We integrate Eq.~(\ref{eq:ode}) with a first-order Euler scheme using $N_{\rm steps}=20$ uniform time steps $\Delta t = 1/N_{\rm steps}$,
\begin{equation}
    \bm{M}_{k+1}
    =
    \bm{M}_{k}
    +
    \Delta t\;
    \bm{v}_{\phi}\left(
        \bm{M}_k,\,t_k,\,
        \bm{M}_{\rm DMO},\,\boldsymbol{\theta}_{\rm TNG}, a
    \right).
    \label{eq:euler}
\end{equation}
Because the probability path in Eq.~(\ref{eq:fm_path}) is a straight line, the learned velocity field is nearly constant along each trajectory, and the Euler integrator remains accurate even with this modest number of function evaluations \citep{lipman2023flowmatching}. In Table~\ref{tab:timing} we show that generating a single $128\times128$ halo patch with 20 steps takes approximately one second.

\subsection{Architecture}
\label{subsec:architecture}
%-------------------------------------------------------------
We parameterize the velocity field $\bm{v}_\phi$ with a conditional U-Net \citep{ronneberger2015unet}. Appendix~\ref{appendix:architecture} gives the full specification of widths, depths, and attention placement; here we briefly describe the model architecture. 

\textsc{BIND} is conditioned on two different kinds of information, the first of which is spatial. The dark-matter-only patch $\bm{M}_{\rm DMO}$ is a map where each of its pixels carries conditional information about where the resulting baryon fields should end up. We preserve the alignment between pixels in this spatial conditioning and the hydro conditioning by simply concatenating it channel-wise with the noisy field at each integration time $t$,
\begin{equation}
    \bm{X}^{\rm in}_t
    = \left[\bm{M}_t \;\middle|\; \bm{M}_{\rm DMO}\right]
    \in \mathbb{R}^{(3+1) \times H \times W}.
    \label{eq:model_input}
\end{equation}
Because both noise and the dark-matter-only conditioning are stacked pixel by pixel, the convolutions see the local dark matter morphology at every encoding step, and the generated gas, stellar, and dark matter channels stay tied to the structure of the halo they are being painted onto.

The second kind of conditioning has no spatial structure and is simply a vector of numbers including the flow time $t$, the parameter vector $\boldsymbol{\theta}_{\rm TNG}$, and the scale factor $a$. These numbers are global scalars that apply to the whole patch. To incorporate them, rather than concatenating them, which risks the values being overwhelmed by spatial conditioning, we inject them via feature-wise linear modulation \citep[FiLM;][]{perez2018film}. Each of $t$, $a$, and $\boldsymbol{\theta}_{\rm TNG}$ is embedded separately to the same dimension, so that each carries the same conditioning weight, and the embeddings are summed into a single conditioning vector $\bm{e}$ (Appendix~\ref{appendix:architecture}). The flow time and the redshift are encoded with the standard sinusoidal features before their embeddings, as both are single scalars whose effect on the output is smooth but strongly nonlinear, while the parameter vector is passed through its own multilayer perceptron.
\begin{equation}
    \bm{\theta_{TNG}} \;\longmapsto\;
    \bigl(1 + \boldsymbol{\gamma}(\bm{e})\bigr) \odot
    {\rm Norm}(\bm{\theta_{\rm TNG}}) + \boldsymbol{\beta}(\bm{e}),
    \label{eq:film}
\end{equation}
where $\boldsymbol{\gamma}$ and $\boldsymbol{\beta}$ are learned linear projections of $\bm{e}$ and ${\rm Norm}$ is group normalization \citep{wu2018groupnorm}. This conditioning vector then acts at every depth in both the encoder and the decoder, modulating the features themselves. 

The network architecture is a U-Net that is not purely convolutional. Rather than relying on network depth and additional convolutions to carry spatial information across the map, we insert multi-head self-attention layers \citep{vaswani2017attention} at the two coarsest resolutions of the U-Net, $32^2$ and $16^2$, where every position in the feature map can interact directly with every other position. We introduced this feature to enhance small-scale fidelity within halos of the resulting generations and found that, without it, our results degrade. We therefore traded the generalizability of a purely convolutional network for accuracy within the generated halos we care about. 

%-------------------------------------------------------------
\subsection{Training}
\label{subsec:training}
%-------------------------------------------------------------

We train with the AdamW optimizer \citep{loshchilov2019decoupled} for up to 200 epochs, using a batch size of 64 maps per GPU across 8 H100 GPUs in mixed \texttt{bfloat16} precision (which enhances both speed and memory efficiency), implemented in PyTorch \citep{paszke2019pytorch} with PyTorch Lightning's Distributed Data Parallel strategy. The learning rate warms up linearly and then anneals to zero on a cosine schedule, and we keep an exponential moving average of the network weights, which we use at inference. Checkpoints are saved every 10 epochs. The schedule, the averaging, and the complete set of hyperparameters are given in Appendix~\ref{appendix:training}. 

In Table~\ref{tab:timing}, we show that training a single \textsc{BIND} model in the above fashion takes ${\sim}150$ GPU-hours, and once trained, it can produce baryonic halos in under a second. For baryonifying a $50\,h^{-1}\,\mathrm{Mpc}$ box with $\sim45$ halos, the cost scales roughly linearly with generation time and takes about $40$ seconds. 

\begin{deluxetable*}{llll}
\tabletypesize{\footnotesize}
\tablecaption{Summary of the validation suite.\label{tab:validation_summary}}
\tablewidth{0pt}
\tablehead{\colhead{Test} & \colhead{Sets} & \colhead{Figure(s)} &
           \colhead{Headline result}}
\startdata
Integrated masses & CV, 1P, SB35 & Figs.~\ref{fig:each_halo_mass_comp}--\ref{fig:mass_comp} & $<1\%$ DM/total, $\sim1\%$ gas, $\sim10\%$ stars \\
Radial profiles & CV, 1P, SB35 & Fig.~\ref{fig:total_density} & $\lesssim10\%$ at all radii to $3.125\,h^{-1}\,$Mpc \\
Projected shapes & CV, 1P, SB35 & Fig.~\ref{fig:axes_ratios} & $q$ medians and widths matched \\
Mass--parameter correlations & SB35 & Fig.~\ref{fig:mass_corr} & Spearman matched, 15 strongest params \\
Profile--parameter correlations & SB35 & Fig.~\ref{fig:dens_corr} & matched at all radii \\
1P field-level response & 1P & Fig.~\ref{fig:butterfly} & sign and morphology reproduced \\
Baryon fraction & CV, 1P, SB35 & Fig.~\ref{fig:baryon_fraction} & median and scatter, no mass conditioning \\
Scaling relations, residual corner & CV & Figs.~\ref{fig:relations}--\ref{fig:corner} & slopes, normalizations, joint scatter \\
Power spectrum suppression & CV, SB35 & Fig.~\ref{fig:suppression} & tracks the hydro-replace ceiling
\enddata
\end{deluxetable*}

\section{Fidelity at Fixed Parameters}\label{sec:validation}
This section addresses the first question posed in the introduction. \textit{Can \textsc{BIND} reproduce baryonic fields of individual halos at fixed parameters?} We start with qualitative field-level comparisons and move to more quantitative comparisons of integrated masses, radial profiles, and projected shapes. The response of those fields across the parameter space is the subject of Section~\ref{sec:param_response}. Table~\ref{tab:validation_summary} indexes the full validation suite of this paper.

\subsection{Qualitative Field-Level Comparisons}
\label{subsec:qualitative}
%-------------------------------------------------------------

\begin{figure*}[!t]
    \centering
    \includegraphics[width=\linewidth]{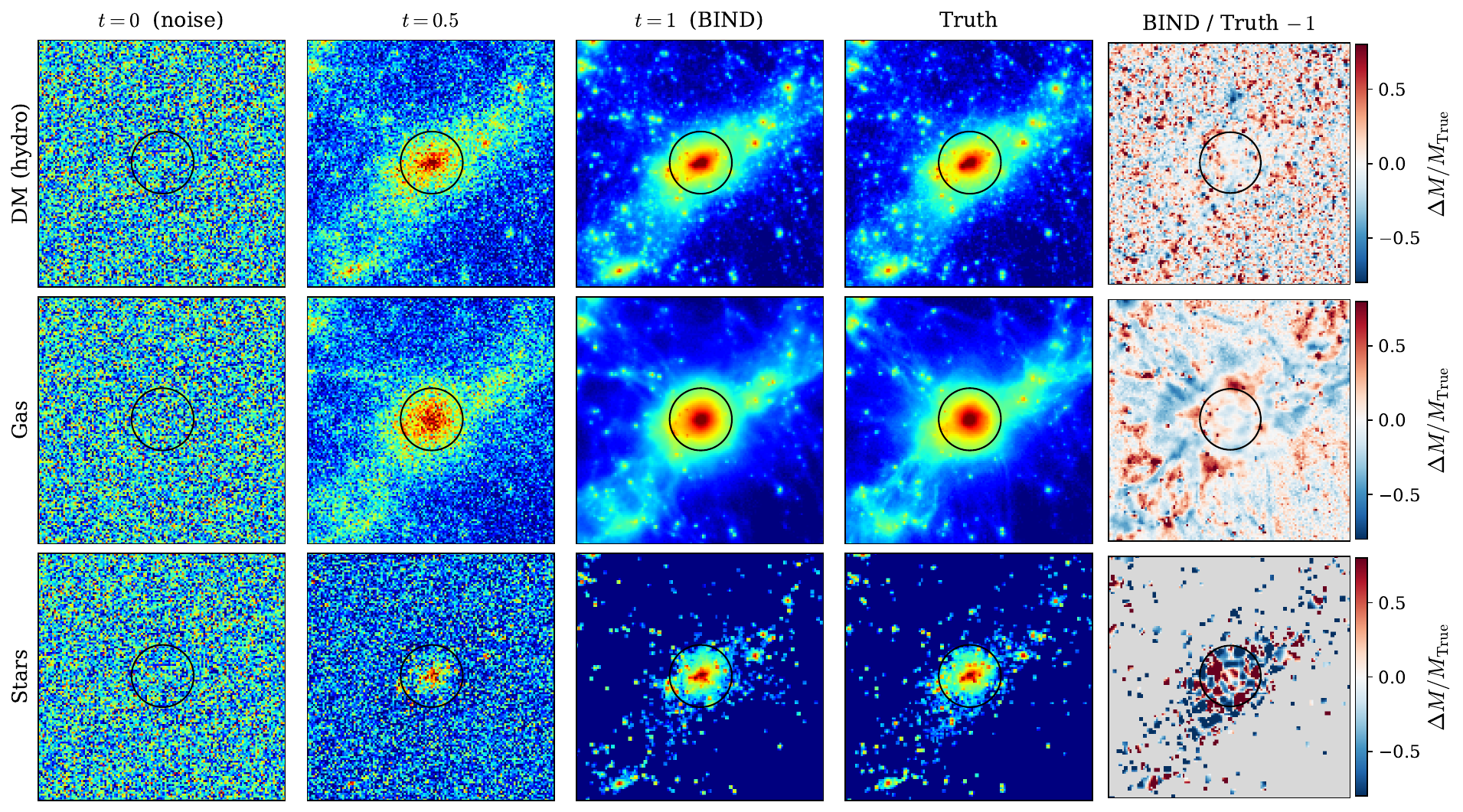}
    \caption{Flow matching trajectory and field-level comparison for a representative cluster-mass halo ($M_{200c} \sim 10^{14}\,M_{\odot}\,h^{-1}$) drawn from the CV test set, conditioned on the fiducial IllustrisTNG parameter vector $\boldsymbol{\theta}_{\rm fid}$. \emph{Columns 1--3:} Snapshots of projected surface density distributions of the generative trajectory at $t=0$ (Gaussian noise), $t=0.5$ (intermediate state), and $t=1$ (final generated field). \emph{Column 4:} Ground-truth hydrodynamical fields from the paired IllustrisTNG simulation. \emph{Column 5:} Signed residual $(M_{\rm BIND} - M_{\rm hydro})/M_{\rm hydro}$. \emph{Rows:} Projected dark-matter, gas, and stellar masses, respectively. The residual panel for the stellar-mass comparison includes 1-pixel-level Gaussian smoothing. The generated fields reproduce the large-scale morphology of all three components. Residuals in the dark-matter channel are noise-like, while the gas channel retains coherent structure in the residuals, and the stellar channel shows significant pixel-level scatter owing to its sparsity.}
    \label{fig:flow_matching}
\end{figure*}

\begin{figure*}[!t]
    \centering
    \includegraphics[width=\linewidth]{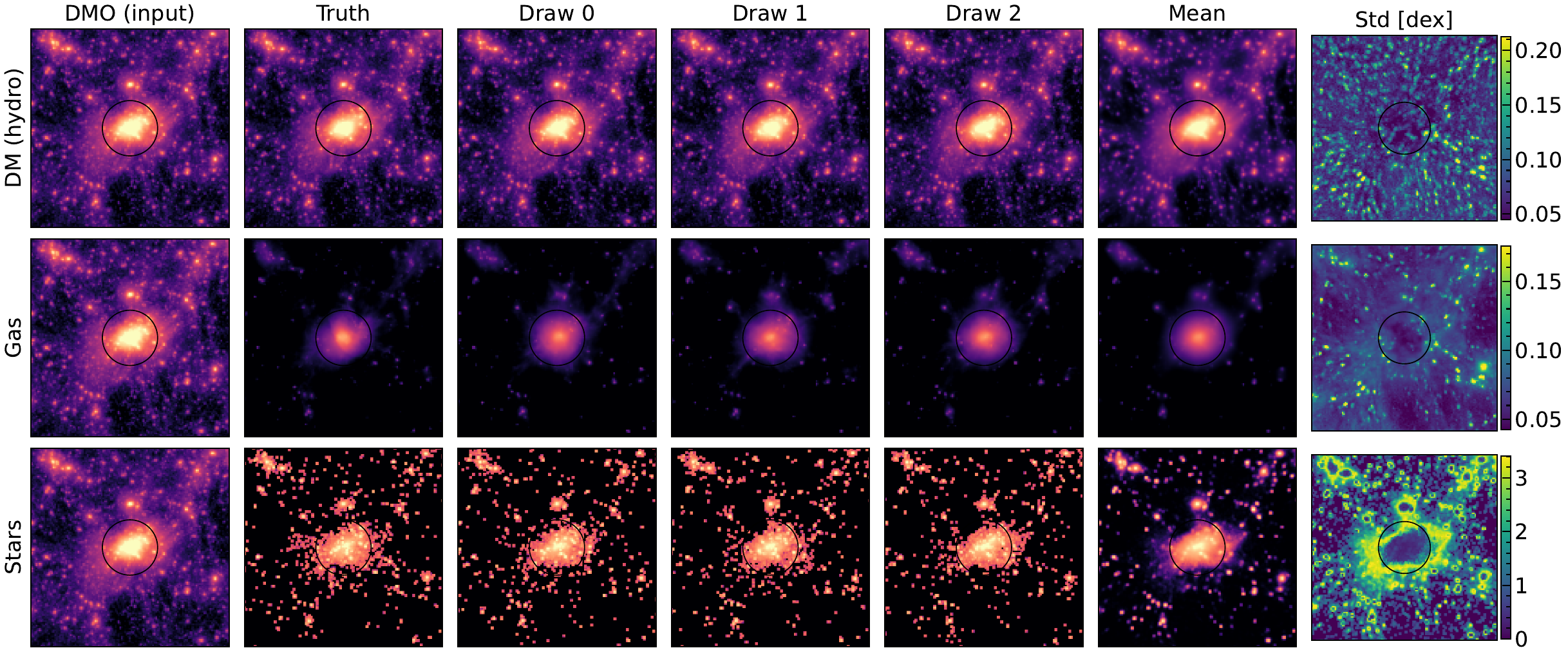}
    \caption{We show various generative draws for a halo. The \emph{first column} shows the DMO conditioning for the halo, the \emph{second column} shows the truth, and the \emph{third-fifth columns} show individual draws with \textsc{BIND}, where the only change was the initial Gaussian field used in sampling. The \emph{sixth column} shows the mean over 100 draws, and the \emph{seventh column} shows the standard deviation over the 100 draws, in units of dex. The rows show the DM (hydro), gas, and stellar channels, respectively.}
    \label{fig:halo_samples}
\end{figure*}

\begin{figure}
    \centering
    \includegraphics[width=\linewidth]{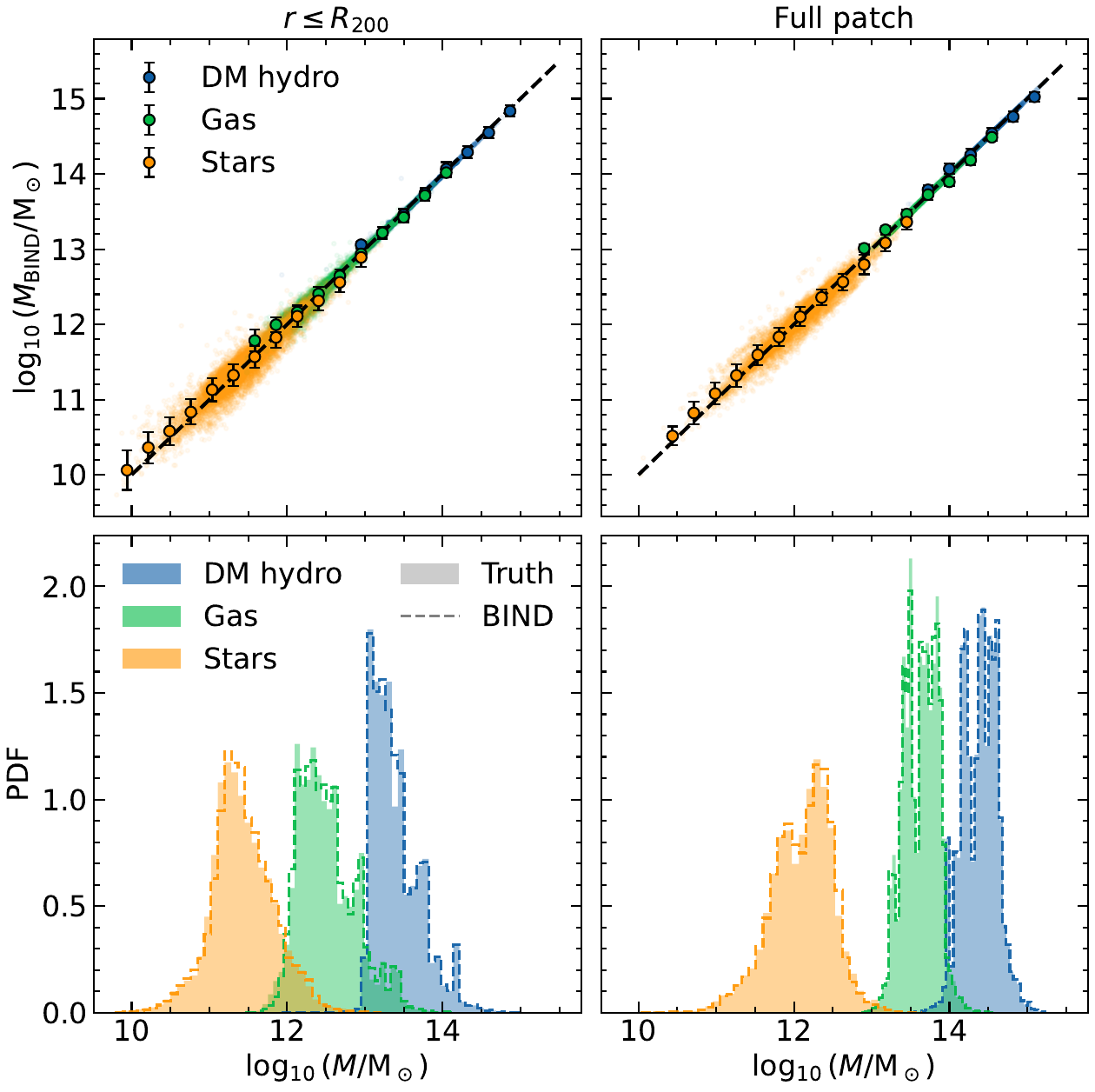}
    \caption{Per-halo integrated-mass comparison between the \textsc{BIND}ed and true hydrodynamical halos for the stellar, gas, and dark matter channels, integrated within each halo's $R_{200c}$ (left column) and over the full $6.25\times6.25\,h^{-1}\,\mathrm{Mpc}$ patch (right column). Points show individual halos from the SB35 test, CV, and 1P suites. Black outlined points show binned medians with $1\sigma$ error bars. Lower panels compare the mass distributions of the true (solid) and generated (outlined) populations. The dashed line corresponds to perfect prediction.
    }
    \label{fig:each_halo_mass_comp}
\end{figure}

\begin{figure}
    \centering
    \includegraphics[width=\linewidth]{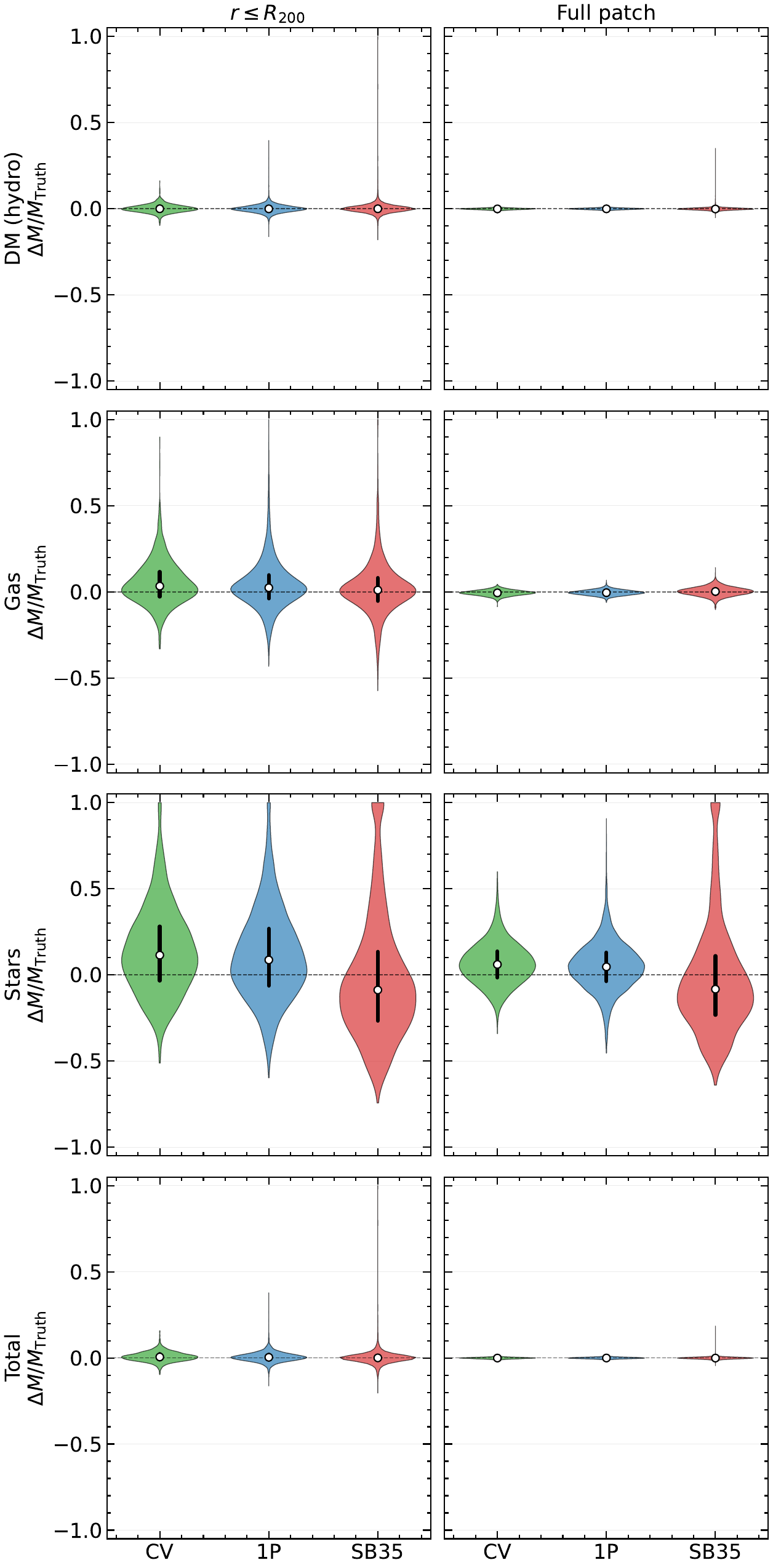}
    \caption{Relative error in integrated mass, $M_{\rm BIND}/M_{\rm hydro}-1$, for each matter component (\emph{rows}: dark matter, gas, stars, and total) and two apertures (\emph{columns}: within $R_{200c}$ and within the full $128\times128$ patch). We show results separately for the CV (green), 1P (blue), and SB35 test suites (red). Lines are centered on the median difference, and the extent marks the 25th and 75th percentiles. The violin shows the full distribution, which is mostly Gaussian for each residual, though the stellar channel deviates and includes a population of outliers. \textsc{BIND} recovers integrated masses across all components and test sets, with median errors below $1\%$ for dark matter and total mass, ${\sim}1\%$ for gas, and ${\sim}10\%$ for stars. We show the bias and scatter values in Table~\ref{tab:mass_error}.}
    \label{fig:mass_comp}
\end{figure}

We begin with a qualitative examination of \textsc{BIND}'s outputs for a representative cluster-mass halo ($M_{200c} \sim 10^{14}\,M_{\odot}\, h^{-1}$) drawn from the CV test set. The $N$-body dark-matter field, centered on the halo as described in Section~\ref{subsec:data_processing}, is provided as the conditioning input together with the fiducial IllustrisTNG parameter vector $\boldsymbol{\theta}_{\rm fid}$.

Fig.~\ref{fig:flow_matching} illustrates both the generative trajectory and the final field-level comparison between \textsc{BIND} and the true hydrodynamical realization of the halo. Reading left to right, the first three columns show snapshots of the flow at $t=0$ (isotropic Gaussian noise), $t=0.5$ (an intermediate state), and $t=1$ (the fully denoised output). The fourth and fifth columns show the ground-truth hydrodynamical fields and the residuals between the \textsc{BIND}ed and true fields, respectively. By $t=0.5$, the network has already established the large-scale mass distribution, and the remaining integration steps primarily sharpen the morphology and populate small-scale features. 

All three baryonic components are reproduced with good qualitative agreement. The errors associated with the channels appear to increase with respect to the distance a field is from its conditioning DMO input. For example, the generated dark-matter field closely traces the $N$-body input, and the residuals are approximately noise-like, with no coherent large-scale structure. The gas channel residuals retain more coherent spatial structure, reflecting the sensitivity of the intracluster gas to pressure forces and AGN and stellar feedback processes that are not fully encoded in the dark-matter morphology alone and for which the parameters were not able to fully communicate. The stellar channel, being highly sparse and concentrated in a small number of bright central pixels, shows the largest pixel-level scatter between the generated and true fields. For this row, we apply a small $1$-pixel Gaussian smoothing to highlight differences in the residual, but even then the differences in the stellar channel are saturated and hard to discern. This is expected, as the precise location and amplitude of individual stellar clumps are stochastic for a given dark matter conditioning field and parameter set, and a generative model is not expected to reproduce a particular realization exactly, but rather the correct statistical ensemble. Beyond the stochasticity of the stellar channel is its sparsity. Individual pixel shifts relative to the truth cause large differences, whereas the DMO channel, which has a much smoother density distribution around density peaks, is less sensitive to this effect. 

To highlight this point, we show in Fig.~\ref{fig:halo_samples} a single halo generated several times with \textsc{BIND} using different initial Gaussian noise realizations. The mean over $100$ realizations is shown in the sixth column and highlights that, across draws, the field is much smoother than the truth, which is itself just a single draw from a distribution. The final column of Fig.~\ref{fig:halo_samples} shows the standard deviation of masses over 100 generations for a single halo. For the gas and dark matter channel, this amounts to noise from different placements of satellite galaxies within the patch. However, for the stellar channel, the spread is much larger, particularly around the edges of halos and subhalos. The typical shapes, density profiles, and masses for each individual draw roughly match the truth (see \S~\ref{sec:validation}); however, the stellar channel is itself stochastic and has a large spread. 

These qualitative observations motivate the quantitative analyses that follow. The next sections systematically characterize \textsc{BIND}'s accuracy through increasingly difficult tests, starting with cumulative mass comparisons (Section~\ref{subsec:integrated_mass}), moving to radially resolved surface density profiles (Section~\ref{subsec:radial_profiles}), then looking at projected shapes of halos in each channel (Section~\ref{subsec:shapes}), before turning to the learned dependence on cosmological and astrophysical parameters (Section~\ref{sec:param_response}).

%-------------------------------------------------------------
\subsection{Integrated Mass}
\label{subsec:integrated_mass}
%-------------------------------------------------------------

We begin with the total integrated mass, which measures whether \textsc{BIND} allocated the correct matter proportions to each channel. For every halo with $M_{200c} \geq 10^{13}\,M_{\odot}\,h^{-1}$ in the full test set (CV, 1P and Sobol test set), we sum the projected pixel values within $R_{200c}$ of the halo center and within the full $128\times128$ patch for each matter component. We then compute the relative error,
\begin{equation}
    \dfrac{\Delta M}{M_{\rm hydro}} = \frac{M_{\rm BIND}}{M_{\rm hydro}} - 1,
    \label{eq:mass_err}
\end{equation}
where $M_{\rm BIND}$ is the generated mass and $M_{\rm hydro}$ is the true mass from the hydrodynamical simulation. 

In Fig.~\ref{fig:each_halo_mass_comp}, we show the match between the stellar, gas, and dark matter channels integrated in both the radial (left panels) and full patch (right panels) apertures. All channels for each of the halos from the SB35 test set, CV set, and 1P set are shown as points in the upper panels, with binned medians and $1\sigma$ error bars shown as the black-outlined points. The lower panels show each channel's mass distribution for the truth (solid) vs. the generated (dashed line). The agreement at this total-mass level for both radial and full patches is spectacular and shows that \textsc{BIND} can capture each channel's mass distribution separately. We also see that the scatter in the predictions vs. the true values is smaller for the channels that have a higher correlation with the dark matter conditioning field. The DM-mass predictions are the most precise with respect to the true halos, while the stellar predictions have the largest scatter. Note that this plot does not show sensitivities to different parameters; instead, it combines all halos from all test simulations. In doing so, the presentation inherently shows the distribution of errors across parameter space under all parameter conditioning.

To expand this into a more parameter-sensitive comparison, Fig.~\ref{fig:mass_comp} shows this relative difference for each matter component, computed independently for each test set, and Table~\ref{tab:mass_error} shows the percent of bias and scatter for each of the channels and suites. Across all three sets, the predicted integrated masses remain unbiased and have small scatter. The median relative error for the hydro dark-matter and total-mass channels is sub-percent, while gas and stellar masses are recovered at the ${\sim}1\%$ and ${\sim}10\%$ level, respectively. Enlarging the aperture from $R_{200c}$ to the full patch (right column) produces little change in the bias but noticeably reduces the scatter, consistent with pixel-level noise averaging down over larger areas. As we will show in \S~\ref{subsec:radial_profiles}, the majority of the stellar mass is determined by the central pixels' predictions, which host the brightest central galaxies of our halos. So it stands to reason that our $10\%$ stellar-mass bias within $R_{200c}$ persists even as we increase the aperture to the full patch. The gas channel, however, shows that for each suite, the integrated masses are sub-percent accurate.

\begin{table}[t]
\centering
\footnotesize
\setlength{\tabcolsep}{4pt}
\caption{Bias and scatter of the integrated-mass residual $\Delta M/M_{\rm Truth} \equiv (M_{\rm BIND}-M_{\rm Truth})/M_{\rm Truth}$ summarized from Fig.~\ref{fig:mass_comp}. Bias is the per-halo median, and scatter is reported as half the distance between the $16^{th}$ and $84^{th}$ percentiles. Residuals are clipped to $[-1,1]$ before summarizing, matching the violin display. \label{tab:mass_error}}
\begin{tabular}{l|r@{\,/\,}lr@{\,/\,}lr@{\,/\,}l}
\toprule
 & \multicolumn{2}{c}{CV} & \multicolumn{2}{c}{1P} & \multicolumn{2}{c}{SB35} \\
Channel & \multicolumn{2}{c}{bias / scat.\ [\%]} & \multicolumn{2}{c}{bias / scat.\ [\%]} & \multicolumn{2}{c}{bias / scat.\ [\%]} \\
\hline
\hline
\multicolumn{7}{l}{\emph{$r\leq R_{200c}$}} \\
\hline
DM (hydro) & $+0.07$ & $1.94$ & $-0.02$ & $1.88$ & $+0.02$ & $1.76$ \\
Gas & $+3.42$ & $10.60$ & $+2.24$ & $10.95$ & $+1.15$ & $10.62$ \\
Stars & $+10.51$ & $23.39$ & $+8.38$ & $24.66$ & $-8.95$ & $32.42$ \\
Total & $+0.48$ & $2.45$ & $+0.30$ & $2.33$ & $-0.00$ & $2.24$ \\
\hline
\multicolumn{7}{l}{\emph{Full patch}} \\
\hline
DM (hydro) & $-0.04$ & $0.45$ & $-0.03$ & $0.46$ & $-0.09$ & $0.54$ \\
Gas & $-0.66$ & $1.52$ & $-0.52$ & $1.45$ & $+0.16$ & $2.42$ \\
Stars & $+6.20$ & $11.10$ & $+4.17$ & $12.44$ & $-8.26$ & $27.26$ \\
Total & $-0.09$ & $0.45$ & $-0.08$ & $0.45$ & $-0.12$ & $0.53$ \\
\end{tabular}
\end{table}

This serves as the most fundamental test of \textsc{BIND}. For virtually all halos across a parameter space entirely unseen during training, the model has learned to correctly partition the mass of the $N$-body conditioning halo among the dark matter, gas, and stellar components across a variety of cosmologies and astrophysical parameter variations. This implies that the network has learned the baryon fraction and stellar-to-halo mass relation in some implicit sense, which we investigate further in Section~\ref{sec:joint_structure}.

%-------------------------------------------------------------
\subsection{Radial Profiles}
\label{subsec:radial_profiles}
%-------------------------------------------------------------

\begin{figure*}[!t]
    \centering
    \includegraphics[width=\linewidth]{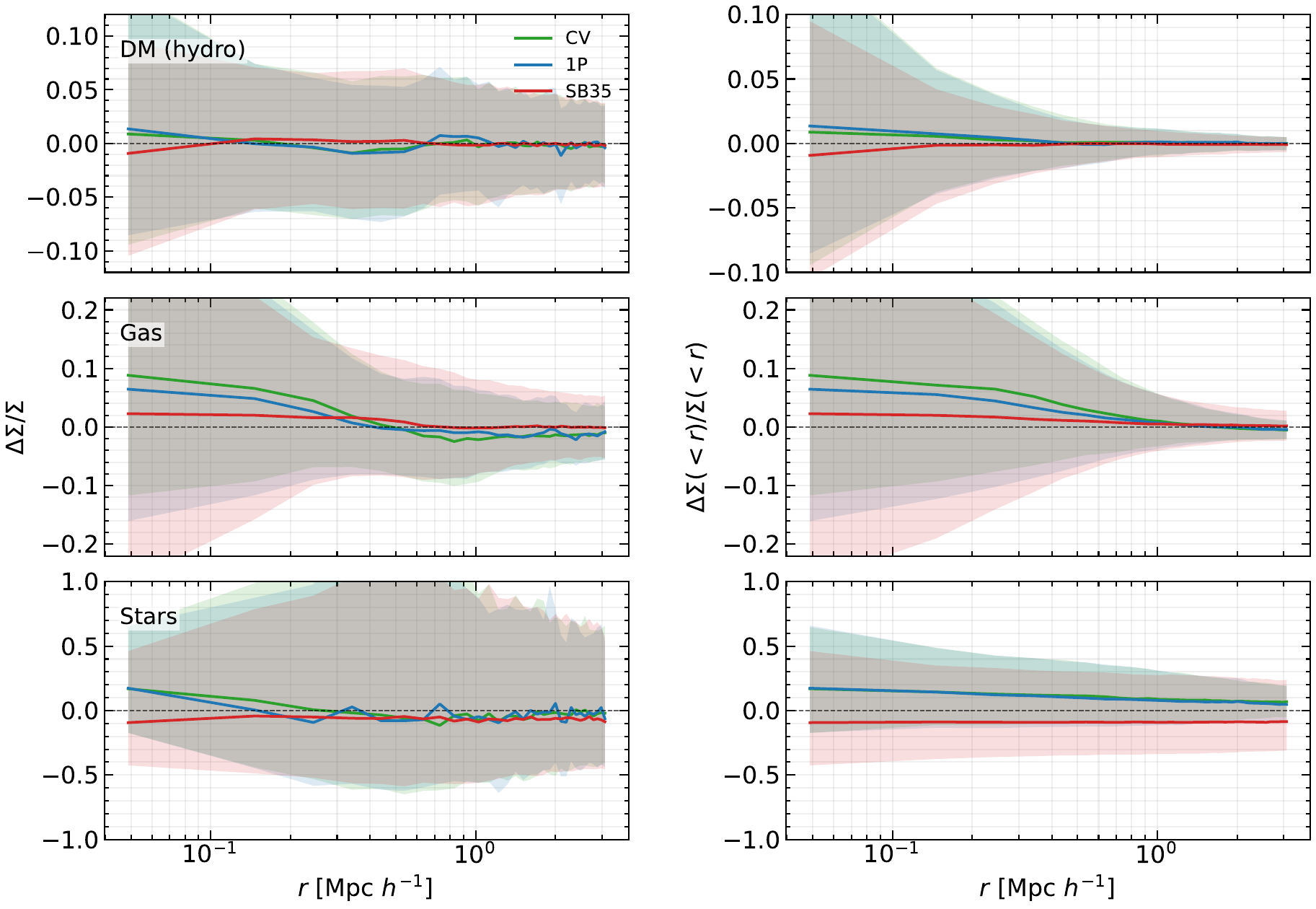}
    \caption{\emph{Left column:} mean fractional error in the azimuthally averaged projected surface density profile, $\langle\Sigma_{\rm BIND}/ \Sigma_{\rm hydro} - 1\rangle$, and \emph{right column:} cumulative surface density profile, $\langle\Sigma_{\rm BIND}(<r)/ \Sigma_{\rm hydro}(<r) - 1\rangle$ as a function of projected radius for each matter component and test set (CV: green; 1P: blue; SB35 test: red). Profiles are computed in 19 logarithmically spaced annuli out to $r = 3.125\,h^{-1} \,{\rm Mpc}$. Shaded bands indicate the halo-to-halo scatter (16th--84th percentile) instead of the error on the median to reflect the intrinsic halo-to-halo scatter, and the $y$-axis is scaled differently for each quantity in order to show its full range. \textsc{BIND} achieves $\lesssim10\%$ accuracy at all radii across all three test sets, with the largest deviations confined to the innermost bins where AGN feedback and steep stellar cusps are hardest to reproduce.}
    \label{fig:total_density}
\end{figure*}

Rather than integrating within a fixed aperture, we now examine \textsc{BIND}'s accuracy as a function of projected radius. For each halo, we compute the azimuthally averaged projected surface density profile and cumulative surface density profile in 19 logarithmically spaced annuli extending to the edge of the patch ($r = 3.125\,h^{-1}\,{\rm Mpc}$), and, similar to the integrated mass, compute the mean fractional error
\begin{equation}
    \langle \Delta\Sigma(r) \rangle
    = \left\langle
        \frac{\Sigma_{\rm BIND}(r)}{\Sigma_{\rm hydro}(r)} - 1
      \right\rangle,
    \label{eq:profile_err}
\end{equation}
averaged over all halos in each test set.

Fig.~\ref{fig:total_density} shows this fractional error for both the marginal (left column) and cumulative (right column) profiles for each matter component and test set. \textsc{BIND} achieves $\lesssim10\%$ accuracy across all radii and all three evaluation sets. The largest deviations are confined to the innermost radial bins, which span $r < 97.7\,h^{-1}\,$kpc and contain only two pixels at our $48.8\,h^{-1}\,$kpc resolution, where the stellar surface density profile is steepest, and the model finds it most difficult to learn. Comparing the left and right columns, the gas distributions, while overpredicted in the cores at the $10\%$ level for the CV and 1P sets, converge on the correct cumulative masses as the profile approaches the virial radius. Conversely, the stars, which are overpredicted in the central pixels but seemingly well predicted outside of this region, are the sole cause of the stellar mass bias that we observed in Fig.~\ref{fig:mass_comp}.

%-------------------------------------------------------------
\subsection{Projected Halo Shapes}
\label{subsec:shapes}
%-------------------------------------------------------------
\begin{figure*}[!t]
    \centering
    \includegraphics[width=\linewidth]{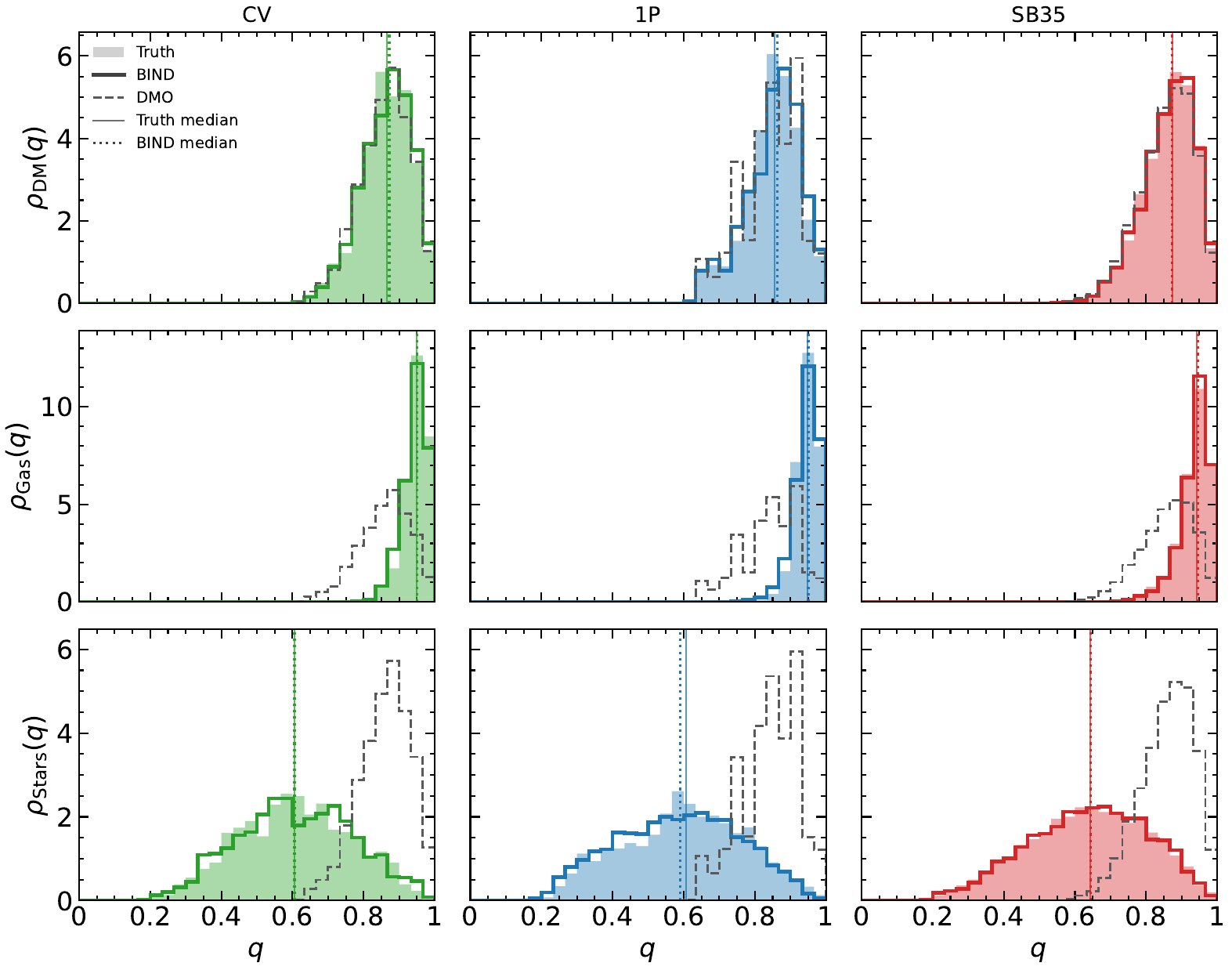}
        \caption{Distributions of the projected axis ratio $q = \sqrt{\lambda_{\min}/\lambda_{\max}}$ measured within $R_{200c}$. Rows show the dark-matter, gas, and stellar channels, while columns show the CV (green), 1P (blue), and SB35 test (red) suites. The filled histograms are measured from the true CAMELS simulations, and the outlines are measured from \textsc{BIND}ed halos. We also show the DMO conditioning field in gray dashed, with its single $q$ distribution repeated in every row as the no-learning baseline. Thin solid and dotted vertical lines mark the truth and \textsc{BIND} medians. In the dark-matter row, the DMO baseline already lies on top of the truth, so the agreement largely comes from the conditioning field rather than learning. In the gas rows, the DMO field is more elliptical; in the stellar fields, it is more round. \textsc{BIND} matches the true medians to $\leq 0.007$ (dark matter), $\leq 0.002$ (gas), and $\leq 0.017$ (stars).}

    \label{fig:axes_ratios}
\end{figure*}
\begin{figure*}[!t]
    \centering
    \includegraphics[width=\linewidth]{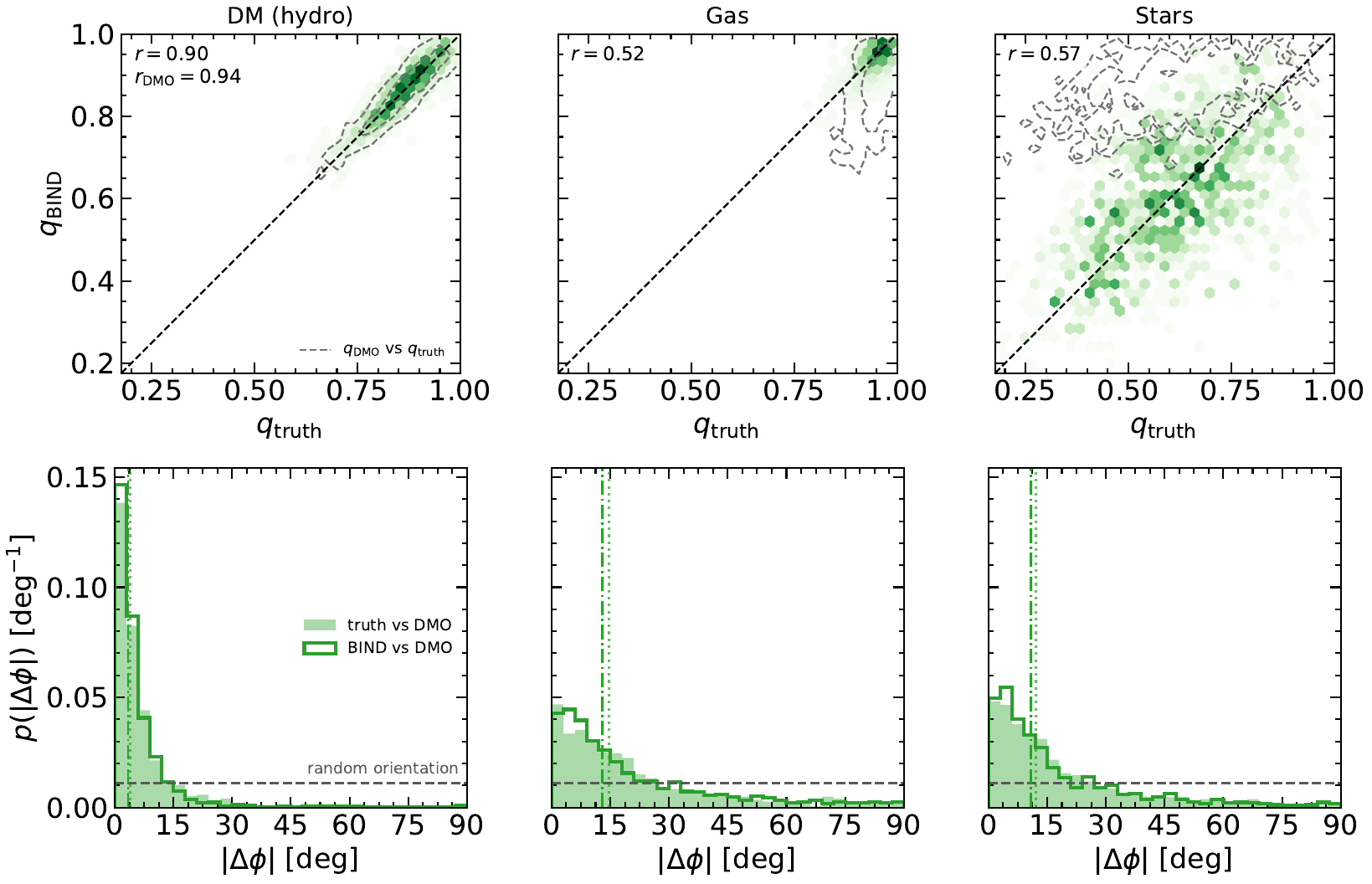}
        \caption{Per-halo shape fidelity measured in the CV suite. \emph{Top row:} generated versus true axis ratio per channel, shown as a 2D density with the 1:1 line, and a black dashed contour showing the DMO baseline $q_{\rm DMO}$ versus $q_{\rm truth}$. The Pearson correlation is shown for the comparison. Note that the DMO baseline hugs the 1:1 line for dark matter but sits below it for gas and far above it for stars, implying that the conditioning field is more elliptical than the gas and more round than the stars. \emph{Bottom row:} distribution of the major-axis misalignment angle $|\Delta\phi|$ between the generated and DMO axes (solid line) compared to that between the true and DMO axes (filled histogram). The medians are marked by vertical lines (truth dotted, \textsc{BIND} dash-dotted), and the flat random-orientation expectation (median $45^\circ$) is shown for reference.}

    \label{fig:shape_perhalo}
\end{figure*}

The previous tests explore integrated pixel quantities that are axisymmetric or cumulative. In this section, we explore whether \textsc{BIND} also reproduces the \textit{shapes} of the baryonic fields. This is a stronger test of \textsc{BIND} as it requires the model to learn to partition mass across the field asymmetrically. It is also interesting to explore shape measures here, which are directly relevant to weak-lensing intrinsic alignment studies \citep[e.g.,][]{troxel2015intrinsic, joachimi2015intrinsic, Lee-2026}.

For each halo in the test and generated catalogs, we measure the projected 2D shape by computing the intensity-weighted inertia tensor within $R_{200c}$ following the common approach in galaxy ellipticity measurements and intrinsic alignment studies,
\begin{equation}
    Q_{ij} = \frac{\displaystyle\sum_{\rm pix}
                   (r_i - \bar{r}_i)(r_j - \bar{r}_j)\,\sigma}
                  {\displaystyle\sum_{\rm pix} \sigma},
    \label{eq:inertia}
\end{equation}
where $\sigma$ is the projected surface density (mass per pixel) and the sum runs over all pixels within $R_{200c}$. The eigenvalues $\lambda_{\max} \geq \lambda_{\min}$ of $Q_{ij}$ define the projected axis ratio and ellipticity,
\begin{equation}
    q = \sqrt{\frac{\lambda_{\min}}{\lambda_{\max}}}\\
    \epsilon = \dfrac{1-q}{1+q},
    \label{eq:shapes}
\end{equation}
where $q=1$ corresponds to a perfectly circular projected profile and $q \to 0$ to a maximally elongated one.

Fig.~\ref{fig:axes_ratios} shows the resulting $q$ distributions for each channel and test set, with the true histograms in solid and the \textsc{BIND} predictions as unfilled histogram lines. The generated histograms match the truth in median (to $\leq 0.007$ for dark matter, $\leq 0.002$ for gas, and $\leq 0.017$ for stars) and in width across all three channels and evaluation sets. \textsc{BIND} also captures the clear ordering among the channels, which hydrodynamical simulations more generally show \citep{Velliscig-2015}. The dark-matter maps have median $q \simeq 0.87$, reflecting the mildly elliptical projected shapes of group-to-cluster-mass halos. The gas channel is more strongly concentrated toward circular configurations, consistent with the intracluster medium's tendency to thermalize and fill the halo potential isotropically, and the stellar channel shows the broadest distribution with the largest ellipticities, driven by the compact, asymmetric morphologies of brightest cluster galaxies. Notably, the SB35 test set, which spans the full 35-dimensional parameter space, shows no appreciable shift in median or spread relative to the CV halos at the fiducial parameters, suggesting that halo shape is determined primarily by the dark-matter structure and is largely insensitive to the subgrid physics parameters over the ranges explored here. 

Instead of looking at the distribution of shapes over the entire set of simulations, Fig.~\ref{fig:shape_perhalo} correlates the shapes halo-by-halo in the top row. The dark-matter axis ratio is recovered at $r = 0.90$ per halo, but we also find that the DMO conditioning field alone achieves $r = 0.94$. We conclude that the dark-matter shape agreement is inherited from the input, so the network has learned to leave the DM field mostly alone. Comparing the gas and star channels, though, we see that the DMO baseline is structurally wrong. We show the halo-by-halo correlation of axis ratio for the DMO halo vs. the gas and stellar halos in black dashed contours, which show clear biases for the gas and stellar channels, compared to \textsc{BIND}, which obtains a relatively tight correlation for the predicted stellar and gas fields. Computing the median ellipticity for the DMO fields and the gas and stellar fields following Eq.~\ref{eq:shapes}, we find that the dark matter is ${\sim}2.8$ times more elliptical than the gas ($\epsilon_{\rm DMO}/\epsilon_{\rm gas}$) and ${\sim}3$ times rounder than the stars ($\epsilon_{\rm star}/\epsilon_{\rm DMO}$). In these channels, \textsc{BIND} recovers the per-halo axis ratio at $r = 0.52$ and $0.57$, respectively, well beyond what the conditioning field supplies.

We also show the orientations of the different fields with respect to the dark-matter-only conditioning (major-axis misalignment angle) in the bottom panels of Fig.~\ref{fig:shape_perhalo}. This is motivated by the literature studying the misalignment angle between the dark-matter-only halo and the baryonic components. Measured against the DMO axis, the generated baryons decorrelate from the dark-matter geometry by the same amount as the true baryons do (median $|\Delta\phi| = 13^\circ$ versus $15^\circ$ for gas and $11^\circ$ versus $12^\circ$ for stars). From this, we conclude that the model genuinely learns the orientation of the individual components and their decorrelation from the DMO conditioning. Moreover, this indicates that \textsc{BIND} can be used to investigate the misalignment angle between dark matter and hydrodynamical components, and potentially whether they vary as functions of astrophysics or cosmology.  

We close this subsection with three caveats when interpreting these results. First, we measure shapes from projected mass maps, whereas observational shape measurements are made in flux. A rigorous comparison with observational intrinsic alignment measurements would require modeling the halo SED and computing shapes from mock photometric images, or directly modeling the luminosity of stars in \textsc{BIND}ed halos \citep[e.g., as in][]{Lee-2026}. Second, our pixel scale of ${\sim}50\,h^{-1}\,{\rm kpc}$ is comparable to the effective radius of the brightest central galaxies at these halo masses, so the central stellar component is effectively unresolved and occupies roughly a single pixel. Third, we project along the full $50\,h^{-1}\,{\rm Mpc}$ line of sight, so foreground and background mass contribute to both the true and generated images, diluting the intrinsic halo ellipticity signal.

Intrinsic orientation alignments of galaxies are a key systematic in upcoming weak lensing analyses, and current models exclude dependencies on baryonic effects. The above caveats notwithstanding, the goal of this section is not to present \textsc{BIND} as a model for intrinsic alignments (yet). Instead, it shows that the generative model can capture the intrinsic shapes of these halos across the parameter space, with generated and true shapes in close agreement. This implies that \textsc{BIND} has learned the morphological mapping from dark-matter-only halos to their baryonic counterparts, a non-trivial result that potentially opens the door to large-scale intrinsic alignment emulation in the future.

\section{Fidelity across the Parameter Space}
\label{sec:param_response}

The previous section established that \textsc{BIND} reproduces the baryonic fields of individual halos when compared with the same halos in our three test sets. We now turn to the second question posed in the introduction. \textit{Has \textsc{BIND} learned how those fields change across the IllustrisTNG parameter space?} We test this at three levels of resolution (integrated masses, radial profiles, and the fields themselves) and close with the redshift dependence.

%-------------------------------------------------------------
\subsection{Parameter Response of Integrated Masses}
\label{subsec:mass_response}
%-------------------------------------------------------------
\begin{figure*}[!t]
    \centering
    \includegraphics[width=\linewidth]{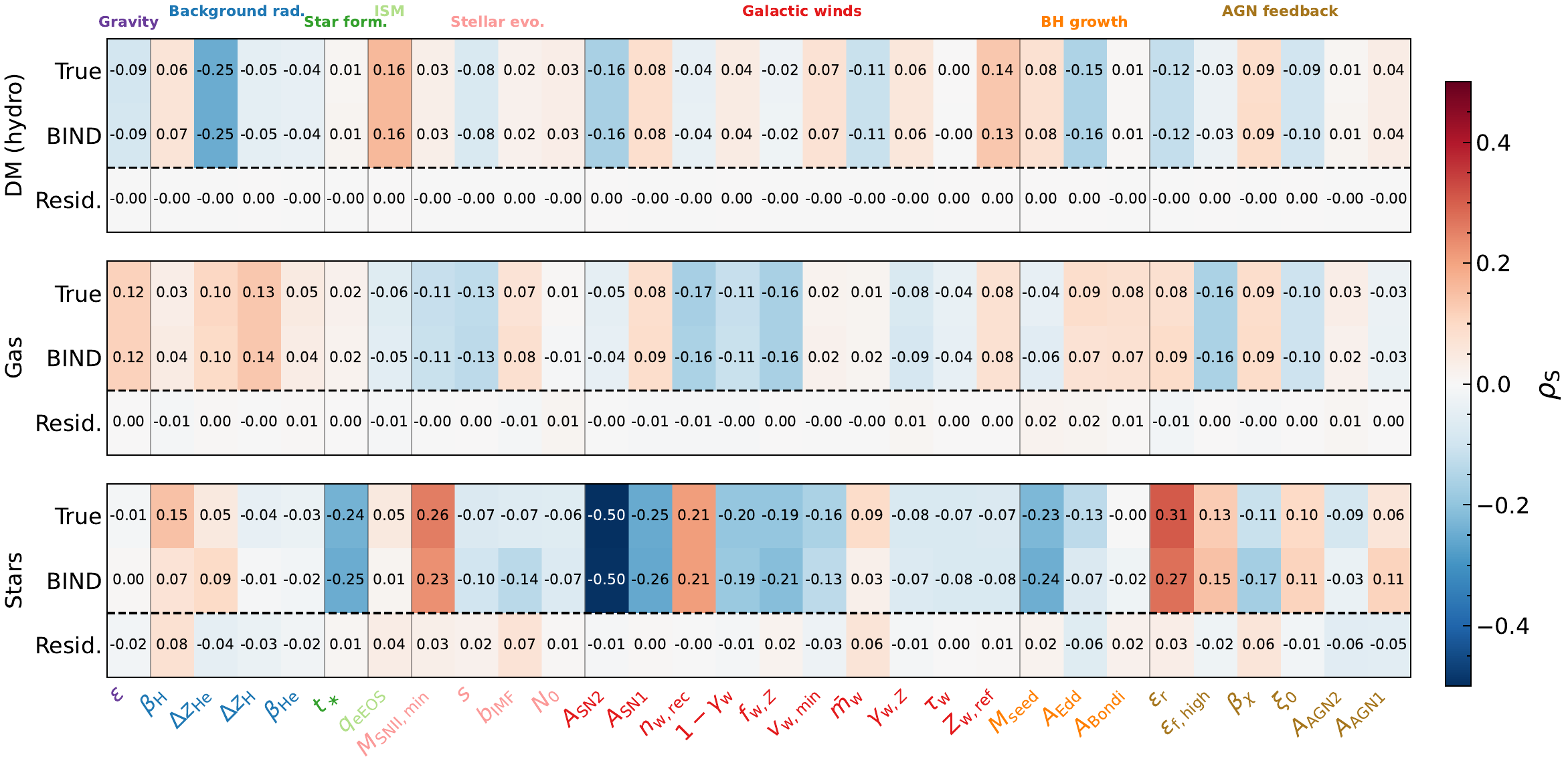}
    \caption{Spearman rank correlation between SB35 test set integrated masses per component and each of the astrophysical parameters. The top rows show the correlations measured in the true halos, while the middle shows those measured from the \textsc{BIND}ed halos, and the bottom rows show the residuals between the two. We group each parameter by type following the conventions of Table~1 of \citealt{Genel-2026}, and within each group, we sort parameters by their largest absolute correlation value. The strong agreement between truth and \textsc{BIND} demonstrates that the model has correctly learned the dependence of baryonic mass on the IllustrisTNG subgrid parameters. The stellar channel (stars) exhibits the strongest parameter sensitivity, followed by gas, while the dark-matter channel shows little response to astrophysical parameter variations.}
    \label{fig:mass_corr}
\end{figure*}

To measure how generated summaries change with parameters, we compute the Spearman rank correlation coefficient between the quantities explored in the previous section and the model parameters. In Fig.~\ref{fig:mass_corr} we show this correlation coefficient between the integrated mass in each component and each of the galaxy formation elements of $\boldsymbol{\theta}_{\rm TNG}$, for both the true and generated halos at identical parameter-space locations. To truly gauge the effect of feedback on these quantities, we exclude the $\Lambda$CDM parameters from the correlation analysis, though they are also captured with high fidelity. We use the parameter names defined in Table~1 of \citealt{Genel-2026}, which also provides the exact definitions and citations to the papers in which they are introduced. In Fig.~\ref{fig:mass_corr}, we group the parameters by parameter type, indicated by the color and the names at the top of the plot. In each group, we organize the columns by the maximum correlation observed in the stellar channel, which appears most sensitive to the parameters. The top rows show the true correlation, the central rows show \textsc{BIND}'s predicted correlation, and the third row shows the residual for each channel. 

Across the 102 SB35 test simulations, the true and generated correlations agree, with a root-mean-squared (RMS) residual of $0.014$ across all channels and parameters, dominated by deviations in the stellar channel. The largest deviation is $|\Delta\rho_S| = 0.08$, which appears in a few of the parameters. The strongest true responses are recovered almost perfectly. For example, stellar mass against the energy of stellar winds $A_{\rm SN2}$ gives $\rho_S = -0.50$ (true) versus $-0.50$ (\textsc{BIND}ed). The dark matter correlations are perfect for all parameters, and the gas correlations are within 0.02 for all parameters. 

Fig.~\ref{fig:mass_corr} shows that the stellar mass channel is the most sensitive to the parameters. This makes intuitive sense, as changing the parameters directly influences the star formation rate within the halos. For example, increasing $A_{\rm SN2}$ acts as preventative feedback, starving the central black hole of gas and disrupting star formation. As a result, we observe a negative correlation between stellar mass and the parameter. Stellar mass is then most dependent on the model parameters, whereas the gas channel shows intermediate sensitivity, as it correlates with the dark-matter structure yet remains responsive to variations in stellar and AGN feedback. The dark-matter channel shows little response to the astrophysical parameters. Gravitational collapse determines most of the dark-matter distribution and resulting halo masses, which are ultimately controlled by cosmological parameters such as the total matter density $\Omega_m$ and the variance of fluctuations $\sigma_8$. The feedback parameters affect the stellar and gas distributions, which in turn back-react on the dark matter distribution \citep{Gebhardt-2026}; however, this effect is much smaller than gravity.

%-------------------------------------------------------------
\subsection{Parameter Response of Radial Profiles}
\label{subsec:profile_response}
%-------------------------------------------------------------

\begin{figure*}
    \centering
    \includegraphics[width=\linewidth]{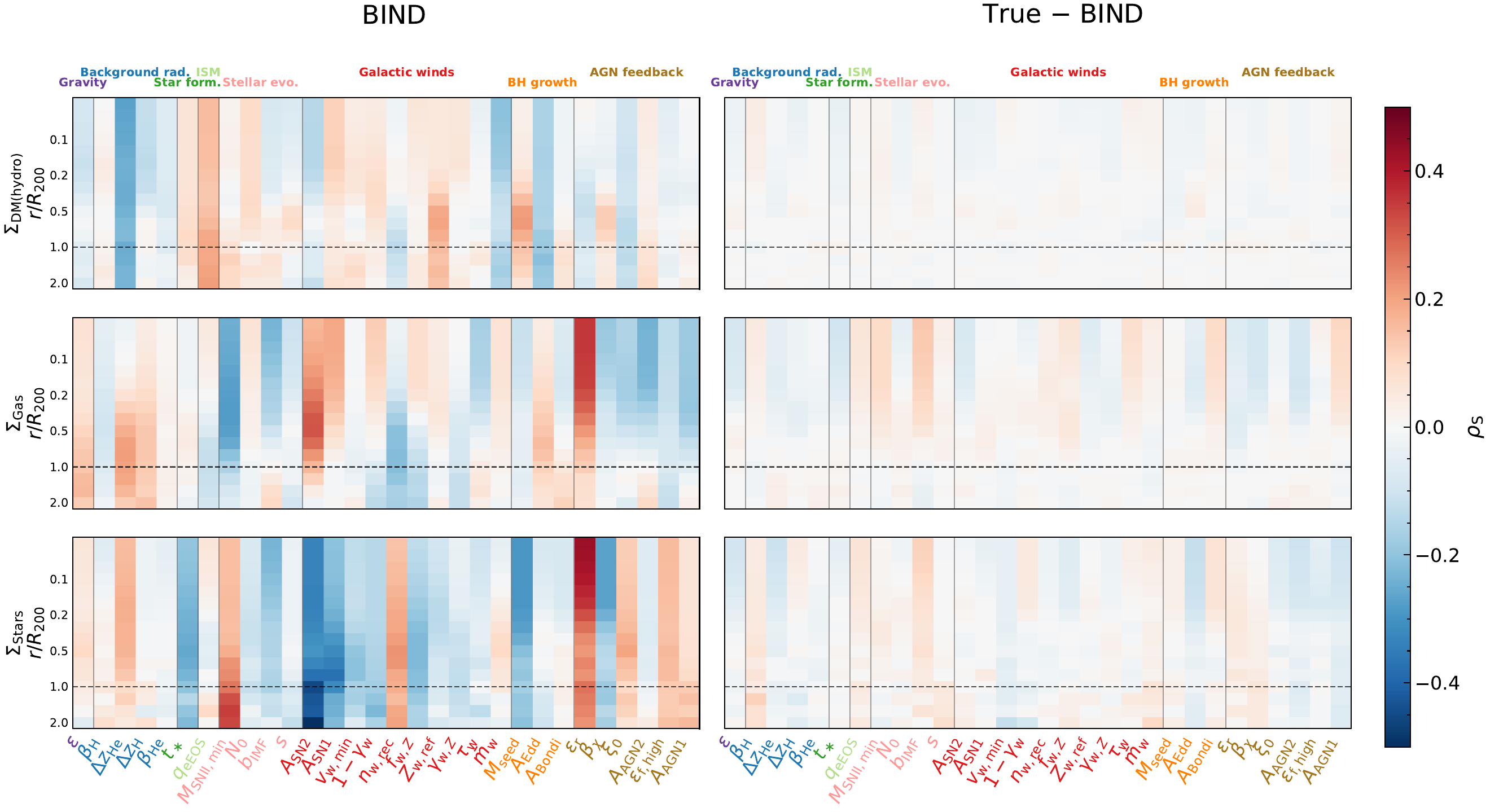}
    \caption{Spearman rank correlation between the azimuthally averaged surface density profile and the 30 astrophysical parameters as a function of radius $r/R_{200c}$ in 19 bins, computed over the 102 SB35 test simulations. For each channel, the left column shows the \textsc{BIND} correlations and the right column the residual (True $-$ \textsc{BIND}). Again, similar to Fig.~\ref{fig:mass_corr}, the stellar channel is most parameter-sensitive; however, we now see that this is localized mostly to the central regions of the halo. }

    \label{fig:dens_corr}
\end{figure*}

Fig.~\ref{fig:dens_corr} extends the Spearman correlation analysis of Section~\ref{subsec:mass_response} to the azimuthal radial profiles. For each annular bin and each matter component, we compute the Spearman correlation between the surface density and each element of $\boldsymbol{\theta}_{\rm TNG}$, for both the true and generated halos. Here, we show the correlation for \textsc{BIND} in the left panels and the residual in the right panels. Over the SB35 test set, the radially resolved grid is reproduced with an RMS residual of $0.029$, indicating that \textsc{BIND} has correctly learned the spatially resolved dependence of baryonic structure on the IllustrisTNG subgrid parameters beyond just the integrated masses.

As with the integrated mass comparison of Fig.~\ref{fig:mass_corr}, the stellar channel shows the strongest parameter sensitivity, particularly in the central regions of the halo, where the stellar surface density is highest, and AGN feedback most directly regulates the star formation rate. There is also a clear anticorrelation between the stellar and gas density responses for several feedback parameters (such as $A_{\rm SN1}$ and $A_{\rm SN2}$). Parameters that drive higher stellar densities tend to suppress the gas density at similar radii, and vice versa. This reflects the competition between gas cooling and star formation on the one hand and stellar- and AGN-feedback-driven outflows on the other, a balance that \textsc{BIND} appears to have learned from the training data. A more detailed investigation into the parameter dependencies and effects on galaxy group gas and stellar properties is saved for future work, as the goal here is to show that \textsc{BIND} is a tool that can be used to decipher this, given that it accurately captures the parameter dependencies of both the integrated masses and the radial profiles.

%-------------------------------------------------------------
\subsection{Field-Level Parameter Response}
\label{subsec:field_response}
%-------------------------------------------------------------
\begin{figure*}[!t]
    \centering
    \includegraphics[width=\linewidth]{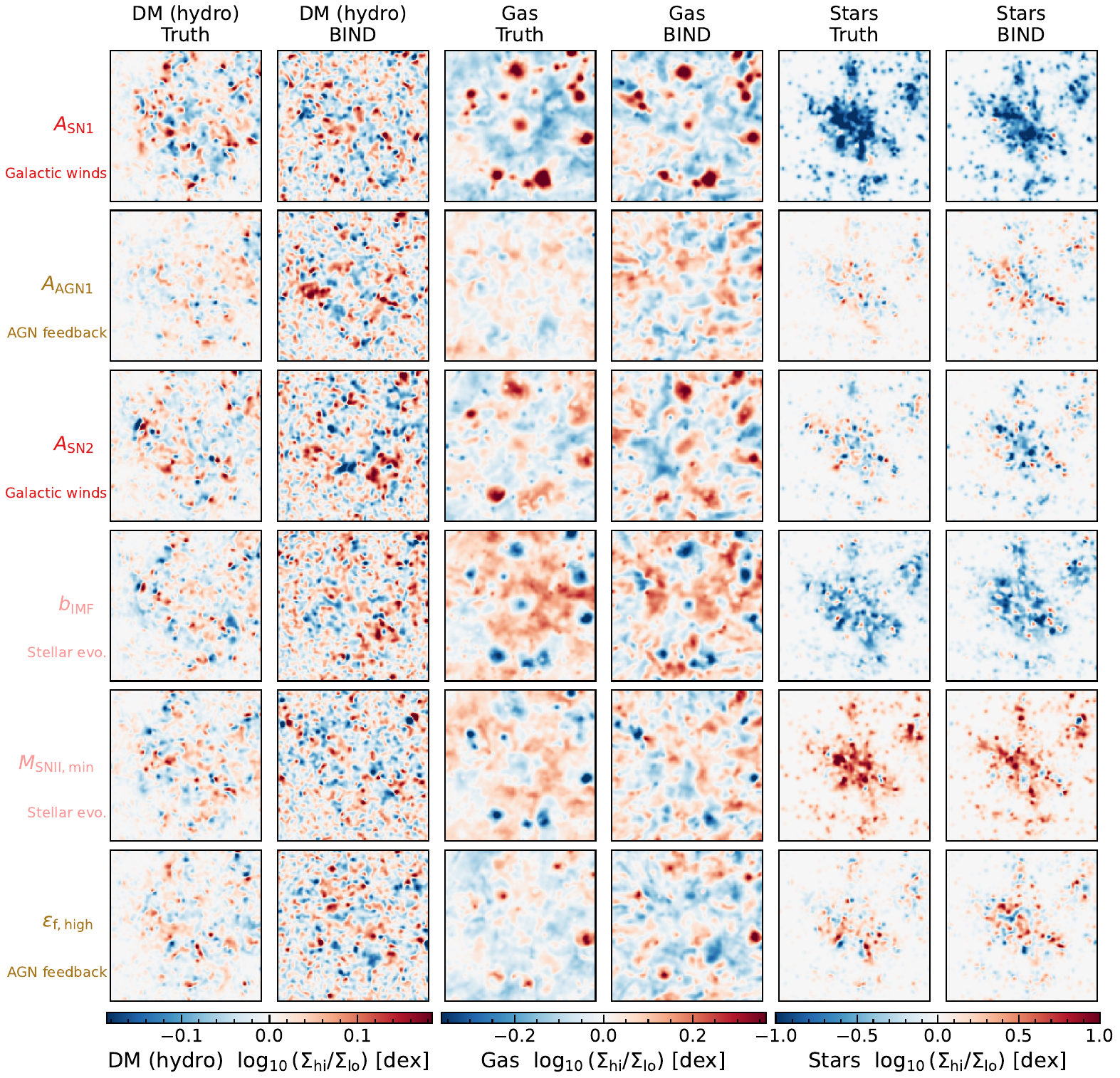}
        \caption{Field-level parameter response for the most massive 1P halo shown as the $\log_{10}$ difference between parameters at their upper and lower prior bounds. Each row is one IllustrisTNG parameter, and the six columns pair truth and \textsc{BIND} for the dark-matter, gas, and stellar channels. Each panel shows the response $\delta\bm{M}$ of Eq.~\ref{eq:field_residual} between the parameter's prior maximum and minimum on a symmetric per-channel color scale. The dark-matter responses are consistent with noise in both truth and \textsc{BIND}, as expected. For the wind and IMF parameters ($A_{\rm SN1}$, $A_{\rm SN2}$, \texttt{IMFslope}, \texttt{SNII\_MinMass\_Msun}) the generated gas response reproduces the sign and morphology of the truth (pixel-level $r = 0.3$--$0.6$), and for the AGN parameters, the true single-halo response is weaker than the stochastic floor of a single generated sample, and the corresponding \textsc{BIND} panels are noise-dominated.}

    \label{fig:butterfly}
\end{figure*}

One of the hopes for \textsc{BIND} is that, unlike radial baryon correction models, it captures the asymmetric response of fields to changes in the underlying galaxy-formation parameters. To test this, we use the 1P test set, in which we vary a single IllustrisTNG parameter at a time while holding all others at their fiducial values and keeping the initial conditions fixed. For each parameter, we select a halo at the two prior bounds (the minimum and maximum parameter values) and compute the signed field-level residual
\begin{equation}\label{eq:field_residual}
    \delta\bm{M}(\bm{x}) = \bm{M}_{\rm high}(\bm{x}) - \bm{M}_{\rm low}(\bm{x}),
\end{equation}
where $\bm{M}_{\rm high}$ and $\bm{M}_{\rm low}$ are the projected mass maps at the upper and lower parameter bounds, respectively. Because the initial conditions are identical across the 1P simulations, any structure in $\delta\bm{M}$ is attributable to baryonic processes alone rather than cosmic variance\footnote{However, there is a noise floor that is imposed by the butterfly effect in \citet{Genel-2019}. Comparing the butterfly-effect floor to these residuals would require understanding how the effect changes with model parameters, which we save for future work.}. We compute this difference for each of the three output channels (dark matter, gas, and stellar mass) for both the true hydrodynamical fields and the \textsc{BIND}ed counterparts and show the results in Fig.~\ref{fig:butterfly}. 

The left two columns of Fig.~\ref{fig:butterfly} show the dark matter residuals following Eq.~\ref{eq:field_residual} for both the true and generated fields. In each panel, we apply 1-pixel Gaussian smoothing to bring out the shapes of the effects. We find that the dark-matter residual maps are largely consistent with small pixel-level shifts between the parameter bounds, likely corresponding to subhalo positional shifts, and show no coherent large-scale structure. This is expected, as dark matter responds primarily to gravity, and, at fixed initial conditions, the dark-matter distribution is nearly insensitive to subgrid parameters over the mass and redshift range probed here. We note that this insensitivity is not exact. Baryonic processes do back-react on dark matter, and \citet{Gebhardt-2026} measure this effect across the CAMELS suites, finding a real but small redistribution compared to the gas response. Our result is consistent with that picture, in which any back-reaction present is at or below the noise level of a single generated sample. The back-reaction is also likely affecting the dark matter at pixel scales ($\sim50\,{\rm kpc}\,h^{-1}$), causing noisier patterns than the stronger gas and stellar redistribution driven by the parameters. We defer a detailed study to future work. 

The middle two panels of Fig.~\ref{fig:butterfly} show the gas residuals, which contain rich, spatially coherent structure varying significantly across the parameters. For example, in $A_{\rm SN1}$, the cores of both the central and exterior halos become more condensed as the parameter increases, while the halo exteriors are suppressed. This is in line with expectations from \citet{Lee-2024}, who have found that an increase in $A_{\rm SN1}$ in massive halos starves the black hole, preventing it from ejecting matter far outside the central galaxy.

In contrast, a shallower IMF slope allows more massive stars to form, leading to higher metallicity, cooling, and black hole growth. The core density of the halos is depleted, but feedback pushes gas further out of the halos, as shown by the red rings. Similarly, the stellar fields in the rightmost panels show that changes to the feedback parameters have net positive or negative impacts on total star formation across the fields, and that this response is consistent with changes in gas. Taken together, this figure demonstrates that \textsc{BIND} has encoded the directional dependence of baryonic processes at the field level.

\textsc{BIND} recovers the major trends across these parameters with good fidelity. The generated residual maps share the sign, spatial morphology, and approximate amplitude of the true residuals in both the gas and stellar channels, capturing features that span a dynamic range of order $10^{1}$--$10^{2}$ in surface density. The main discrepancy is that the \textsc{BIND} residuals are somewhat noisier than the ground truth. Small-scale features in the true maps that arise from the deterministic response of the same halo to a parameter change are partially washed out in the generated fields by the stochastic nature of the sampling. This is not unexpected, as the model generates independent samples from the learned posterior for each parameter value rather than computing a deterministic perturbation. The residual is therefore noisy at the single-sample level even if the mean response is correct. Still, the bulk spatial structure is preserved, and the qualitative parameter responses are clearly recognizable. This provides strong evidence that \textsc{BIND} has learned the mapping between the IllustrisTNG subgrid parameters and their field-level baryonic signatures, beyond what could be inferred from integrated-mass or azimuthally averaged statistics alone.

\subsection{Redshift Dependence}
\label{subsec:z_response}
Up until this point, we have presented results only at a single redshift, $z=0$. However, throughout this work and in training, we treat the redshift as a model parameter that \textsc{BIND} is trained on. Especially for our companion paper, \citet{Lee-2026c}, where an entire lightcone is \textsc{BIND}ed out to redshift $z=2.44$, we must validate that the redshift conditioning works as intended. We focus on whether \textsc{BIND} retains model fidelity at each redshift in isolation and whether it has learned the evolution of the baryonic fields with redshift. 

\begin{figure*}
    \centering
    \includegraphics[width=\linewidth]{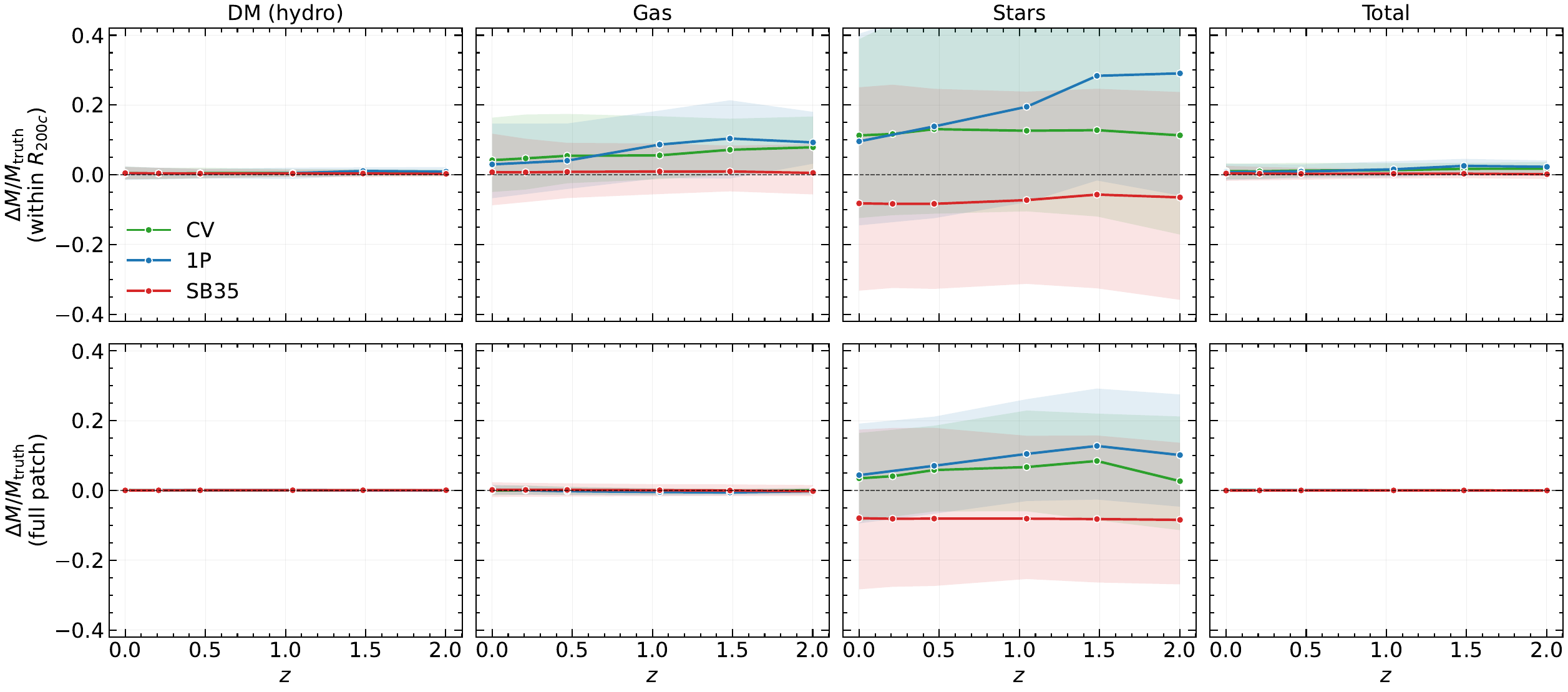}
    \caption{Relative error in integrated mass, $M_{\rm BIND}/M_{\rm hydro}-1$, as a function of redshift, for each matter component and for the total (columns), measured within $R_{200c}$ (top row) and over the full $128\times128$ patch (bottom row). Lines and points show the median over the halos of each suite at each snapshot, and shaded bands show the halo-to-halo 16th--84th percentile scatter. The suites are shown separately. Dark matter and total mass are flat and accurate everywhere, while the held-out SB35 suite is flat in every channel.}
    \label{fig:z_mass_error}
\end{figure*}

Fig.~\ref{fig:z_mass_error} repeats the integrated-mass test of Fig.~\ref{fig:mass_comp} at each of the eight redshifts, reported per set. The dark matter and total mass are accurate and flat at every epoch (medians within $+0.3$--$1.1\%$ and $+0.2$--$2.6\%$ over $0 \leq z \leq 2$), and the held-out SB35 suite shows the least evolution with redshift in every channel. One interesting trend is that the 1P set appears to degrade for the stellar channel as redshift increases. Given that the 1P set has all parameters fixed to the fiducial values, it makes sense that the CV and 1P lines track each other, but at higher redshifts, the effect of individual parameters on the stellar distribution is clearly degrading. The 1P set inherently probes the edges of the parameter prior where \textsc{BIND} has fewer training samples, which is the likely cause of this trend. 

We also verified directly that the conditioning is used and that it interpolates. Holding the dark-matter conditioning, the parameter vector, and the initial ODE noise fixed while varying only the scale-factor label produces a monotonic response, and evaluations at five redshifts that correspond to no training snapshot ($z = 0.10$--$1.75$) fall on the same smooth monotonic curve. This implies that the model has learned a continuous function of the scale factor rather than memorizing a discrete set of snapshots, which allows it to be used on lightcones and across redshift. Taken together, these tests validate the redshift conditioning over $0\leq z\leq 2.5$ spanned by the lightcones of the companion paper \citep{Lee-2026c}.

\section{Emergent Joint Structure: Component Correlations}
\label{sec:joint_structure}
%-------------------------------------------------------------

We now turn to the third question posed in the introduction. \textit{To what extent does \textsc{BIND} capture correlations between physically distinct baryonic components?} We begin by examining the halo baryon fraction across all three test sets to probe whether the model correctly partitions mass among gas, stars, and dark matter as a function of halo mass and cosmic baryon fraction. We then focus on the 1,111 CV halos generated at the fiducial IllustrisTNG parameters to investigate a set of inter-component scaling relations: the stellar-to-halo mass relation (SHMR), the baryonic-to-halo mass relation, and the gas-to-stellar mass relation. Finally, we decompose the scatter around these relations and show that \textsc{BIND} correctly reproduces the joint distribution of halos that deviate coherently from multiple relations simultaneously, including the most pathological cases.

%-------------------------------------------------------------
\subsection{Baryon Fractions}
\label{subsec:baryon_fraction}
%-------------------------------------------------------------

For each halo in the three test sets, we compute the projected baryon fraction within $R_{200c}$ as
\begin{equation}\label{eq:f_b}
    f_{b}(<r) = \frac{M_{\rm gas}(<r) + M_{\rm star}(<r)}{M_{\rm total}(<r)},
\end{equation}
and normalize by the cosmological baryon fraction $f_{b,0} \equiv \Omega_b/\Omega_m$ computed from the parameter tables of each simulation. We focus on $f_b/f_{b,0}$ as a function of halo mass because we expect groups to retain a smaller fraction of their baryons than clusters \citep[e.g.,][]{Lovisari-2021, Hadzhiyska-2025, Siegel-2025}, and because the mass dependence of $f_b$ is one of the quantities that varies strongly across subgrid models \citep{Medlock-2024b, Lau-2025}.

\begin{figure}
    \centering
    \includegraphics[width=\linewidth]{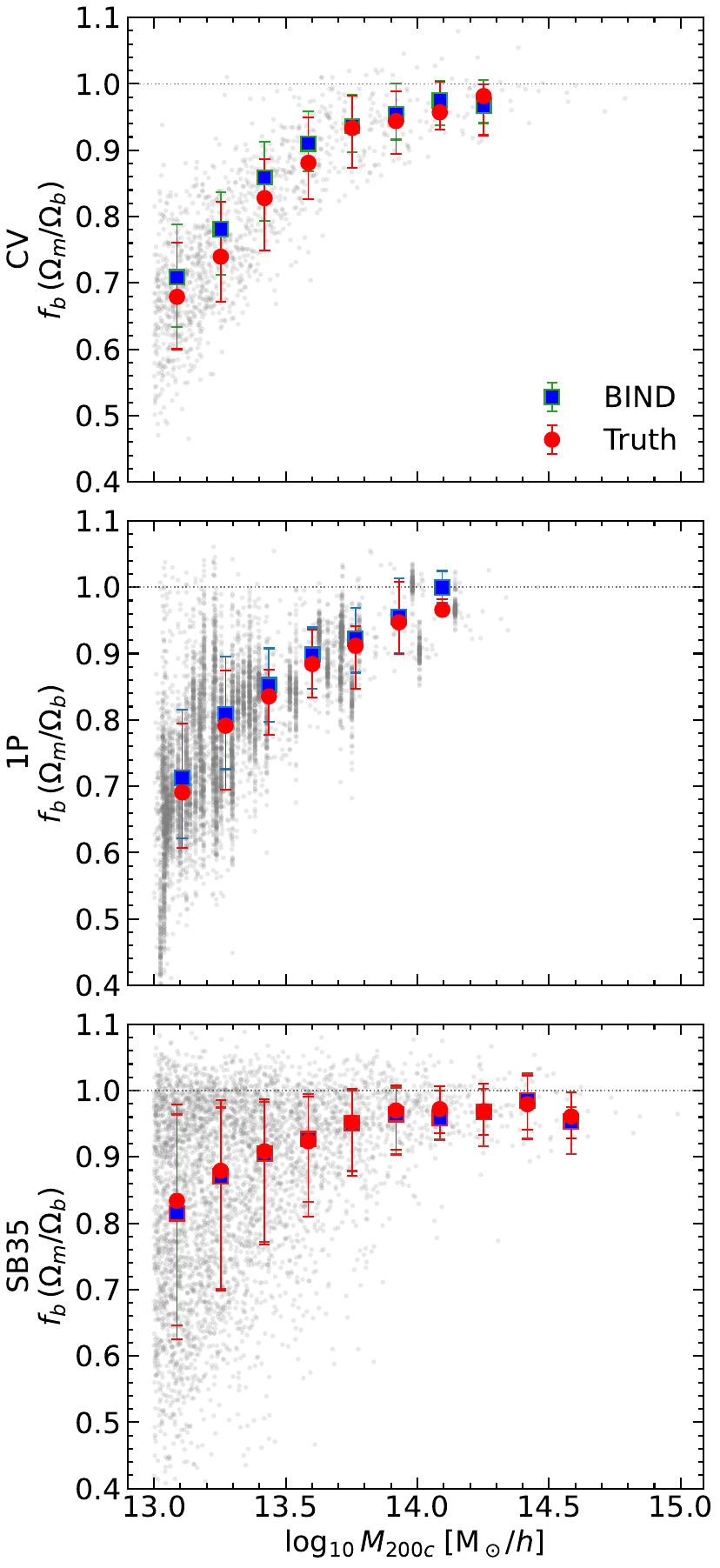}
    \caption{Normalized baryon fraction $f_b(<R_{200c})/f_{b,0}$ as a
    function of halo mass $M_{200c}$ for the three test sets. Points show individual halos, while lines and error bars show the median and 16th--84th percentile scatter computed in logarithmic mass bins. \emph{True} hydrodynamical halos are shown with black scatter points to help guide the eye. \textsc{BIND} reproduces both the median trend and scatter of the baryon fraction across the full range of halo masses and parameter values, despite halo mass never being used as a conditioning variable.} \label{fig:baryon_fraction}
\end{figure}

We bin the halos in logarithmic $M_{200c}$ intervals, computing the median normalized baryon fraction and the halo-to-halo scatter in each bin for both the true hydrodynamical halos and the \textsc{BIND}ed counterparts. Fig.~\ref{fig:baryon_fraction} shows the resulting $f_b/f_{b,0}$ vs.\ $M_{200c}$ relation for each test set. The \textsc{BIND}ed halos reproduce the median and scatter of the true relation with excellent fidelity across more than a decade in halo mass and across the diverse parameter coverage of the SB35 suite.

The baryon fraction involves the ratio of two separately generated fields (gas and stellar mass) to the total halo mass. We did not supply halo mass information as explicit conditioning to the network. Still, the model has learned the halo-mass dependence of the baryon fraction, the sensitivity of $f_b$ to the cosmic baryon fraction $\Omega_b/\Omega_m$, and the correct relative normalization between the gas, stellar, and dark-matter channels simultaneously. The close agreement between the true and generated relations across all three suites therefore reflects a non-trivial internal consistency in the generative model, similar to that shown in \S~\ref{subsec:integrated_mass}.

%-------------------------------------------------------------
\subsection{Inter-component Scaling Relations}
\label{subsec:scaling_relations}
%-------------------------------------------------------------

We next investigate whether \textsc{BIND} reproduces the joint structure of the baryonic mass components. For this analysis, we focus on the 1,111 CV halos generated from the 27 CV simulations. We measure three scaling relations within $R_{200c}$ of each halo: the gas-to-stellar mass relation, $M_{\rm star}$--$M_{\rm gas}$, the SHMR ($M_{\rm star}-M_{200c}$), and the baryonic mass $M_{\rm bar}-M_{200c}$ relation (unnormalized quantity shown in Fig.~\ref{fig:baryon_fraction}). These relations, and the SHMR in particular, are among the most heavily studied summaries of the galaxy-halo connection \citep[for a review, see][]{WechslerTinker-2018}. Each relation is roughly described by a power law over the mass range, so we fit a simple linear model in log-log space for each relation and overlay the best-fit regression on both the true and \textsc{BIND}ed halo populations.

\begin{figure*}
    \centering
    \includegraphics[width=\linewidth]{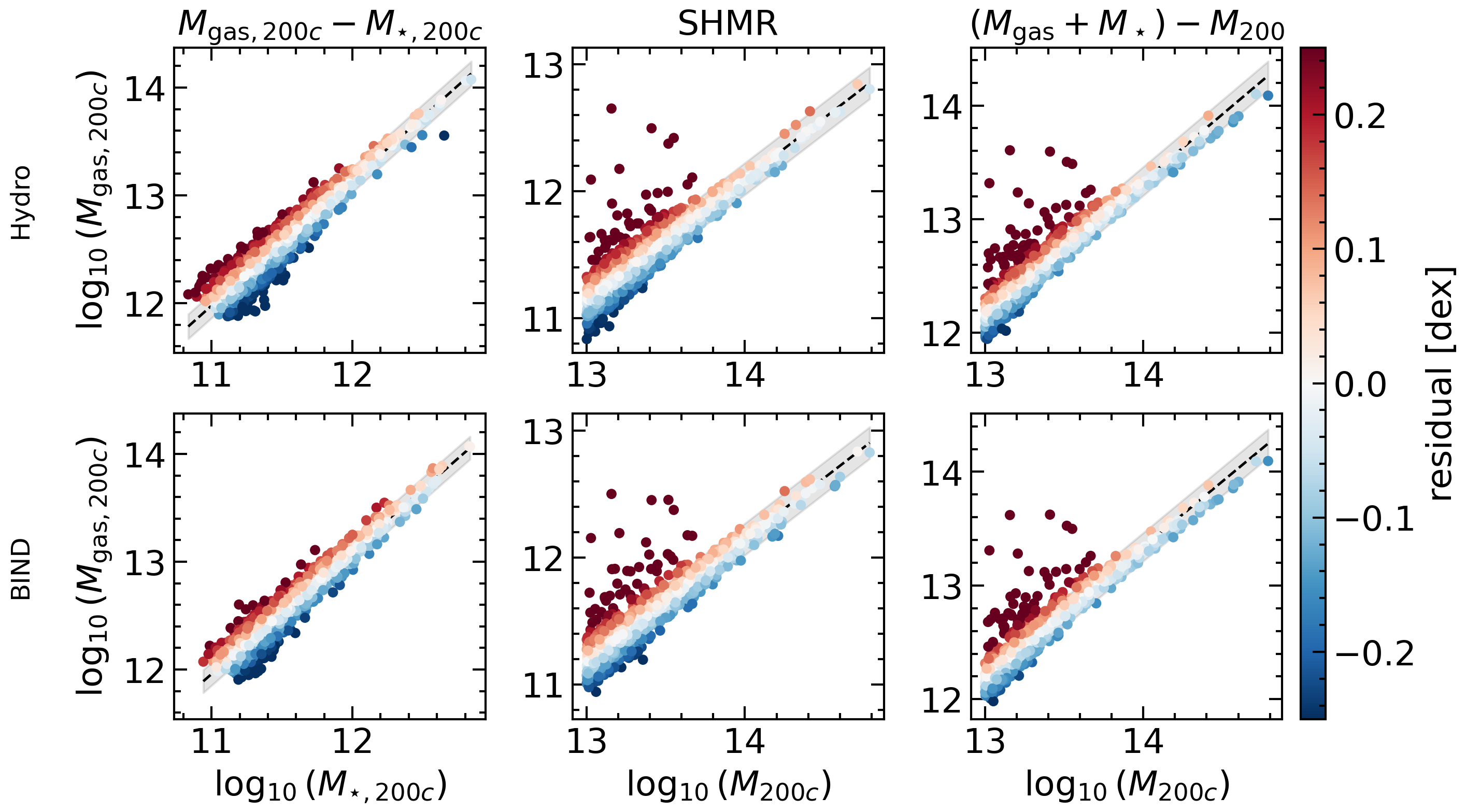}
    \caption{Inter-component mass scaling relations for the 1,111 CV halos at the fiducial IllustrisTNG parameters. Each column shows a different relation in log--log space: gas-to-stellar mass (\emph{left}), stellar-to-halo mass (SHMR; \emph{center}), and baryonic mass--$M_{200c}$ (\emph{right}). The \emph{top} panels show the relations measured from the hydro simulations in the CV set, while the bottom panels show the generated \textsc{BIND}ed realizations. Dashed lines show the best-fit linear regressions to the true and generated populations, with $\pm 1\sigma$ scatter indicated by shaded bands, and the colors represent the residuals of a given halo from the best-fit line. \textsc{BIND} accurately reproduces both the slopes and normalizations of all three relations as well as the amplitude and morphology of the scatter, indicating that the model has learned the joint covariance structure of the multi-component baryonic fields.}
    \label{fig:relations}
\end{figure*}

Fig.~\ref{fig:relations} shows these three panels for the true halos (top) and \textsc{BIND} halos (bottom). The generated population reproduces not only the slope and normalization of each scaling relation but also the scatter amplitude and asymmetry about the best-fit line. Halos that scatter significantly above or below the mean relation in the hydrodynamical simulations have corresponding analogs in the \textsc{BIND} output, indicating that the model has learned the physical correlations that drive halo-to-halo scatter in baryonic content at fixed mass. Scatter in these relations is driven by the assembly history of each halo, gas cooling, star formation, and feedback, none of which are directly encoded in the $N$-body input beyond the projected dark-matter distribution. That we can recover this almost perfectly suggests the model has, in some way, learned these non-linear effects.

%-------------------------------------------------------------
\subsection{Residual Correlations}
\label{subsec:corner}
%-------------------------------------------------------------

To examine the degree to which scatter in different scaling relations is correlated, we identify halos that lie more than $1\sigma$ above or below the best-fit linear relation in each of the three panels of Fig.~\ref{fig:relations}, classifying them as the ``high'' and ``low'' populations for each relation. For each pair of relations, we then visualize the joint distribution of the residuals $\Delta\log M_{\rm star}$, $\Delta\log M_{\rm gas}$, and $\Delta\log M_{\rm bar}$ for both the high and low subsamples as a corner plot. We smooth the distributions with a Gaussian KDE and overplot the $1\sigma$, $2\sigma$, and $3\sigma$ contours. We apply the same procedure to the \textsc{BIND}ed population and overlay it for direct comparison.

\begin{figure}[!t]
    \centering
    \includegraphics[width=\linewidth]{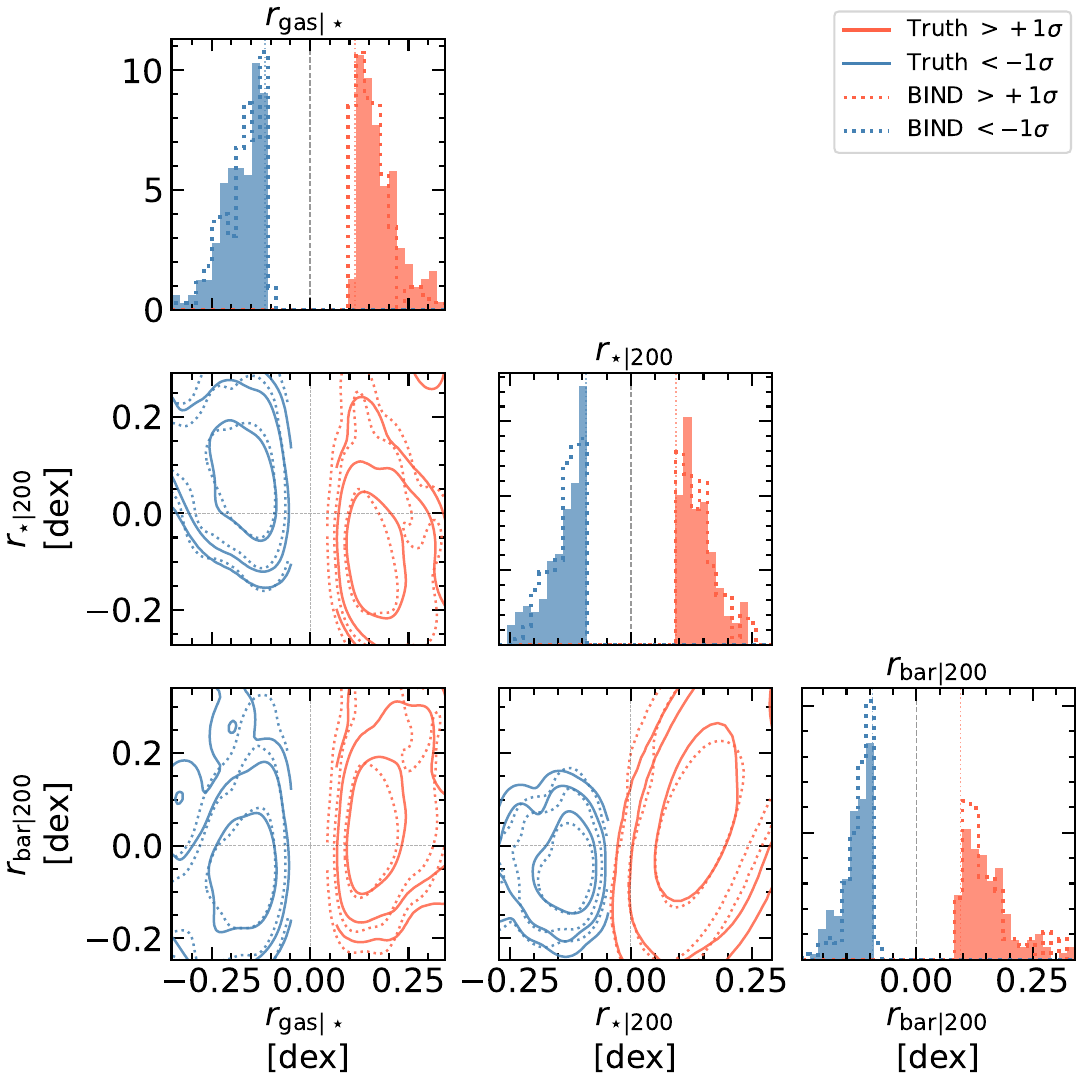}
    \caption{Corner plot of the residuals $\Delta\log M$ about the best-fit scaling relations of Fig.~\ref{fig:relations} for the CV halo population. Each off-diagonal panel shows the joint distribution of residuals for a pair of relations. The diagonal panels show the marginal distributions. Orange (blue) contours correspond to halos that scatter more than $1\sigma$ above (below) the mean scaling relation in any given panel. Filled contours show the true hydrodynamical halos, and open contours show the \textsc{BIND}-generated halos, smoothed with a Gaussian KDE. Contour levels correspond to the $1\sigma$, $2\sigma$, and $3\sigma$ enclosed probability. The agreement between the true and generated contours across all panels demonstrates that \textsc{BIND} has learned the full multivariate structure of the baryonic mass components.}
    \label{fig:corner}
\end{figure}

Fig.~\ref{fig:corner} presents this for the true and \textsc{BIND}ed CV halos. The generated contours reproduce the ground-truth distributions across all residual pairs and for both the high and low populations. The halos that scatter high in the SHMR also tend to scatter high in the baryonic mass relation, reflecting a correlated excess of both stellar and total baryonic content. The halos that scatter low in the gas-to-stellar mass relation tend to scatter high in the SHMR, consistent with AGN-driven gas expulsion redistributing baryons out of the halo while leaving the stellar component largely intact. These correlations span multiple physically distinct processes, making them non-trivial for the network to recover from projected dark-matter and IllustrisTNG parameter-set inputs alone.

The close agreement between the truth and \textsc{BIND} across all of these joint distributions is perhaps the most demanding consistency test. It demonstrates that the model has not merely learned marginal distributions in each channel independently but has captured the full multivariate structure of the baryonic fields as a function of halo properties. This follows directly from \textsc{BIND}'s joint, multi-channel generative architecture, which produces all three mass fields simultaneously from a single sample of the learned probability path and naturally preserves inter-channel correlations.

%-------------------------------------------------------------
\section{Using \textsc{BIND} in the Wild}
\label{sec:bind_in_the_wild}
%-------------------------------------------------------------
\begin{figure*}
    \centering
    \includegraphics[width=\linewidth]{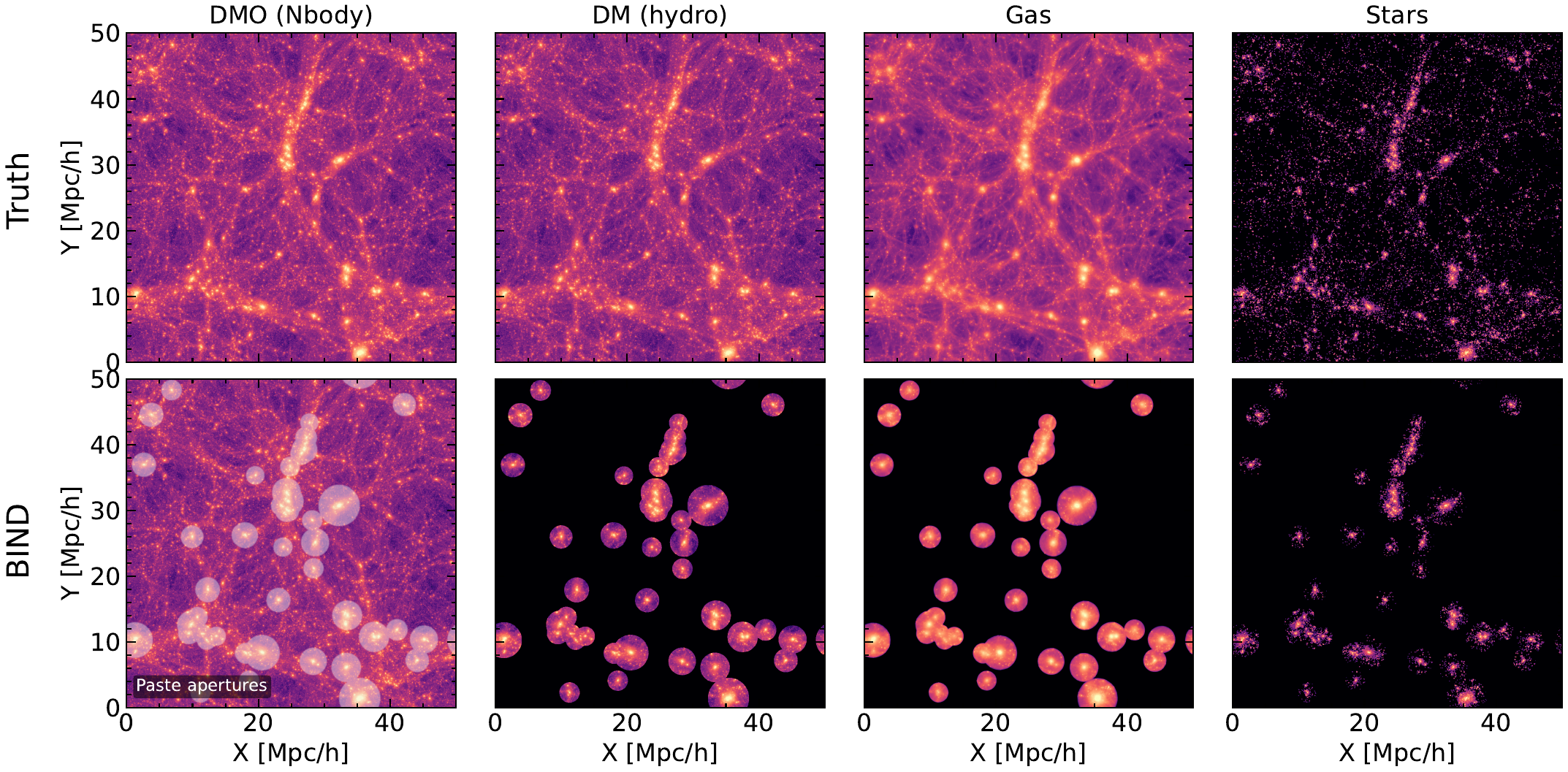}
    \caption{\textsc{BIND} applied to a full $N$-body volume. \textit{Top row:} The truth, showing the DMO projection of the $N$-body run followed by the dark matter, gas, and stellar surface densities of its paired hydrodynamical counterpart, each projected over the full $50\,h^{-1}\,\mathrm{Mpc}$ box onto a $1024^2$ grid. \textit{Bottom row:} the same DMO field, and the three channels generated by \textsc{BIND} from it and pasted back in, halo by halo, for every halo with $M_{200c}\geq10^{13}\,M_\odot\,h^{-1}$. We show the channels individually rather than cumulatively to make the procedure's scope explicit. \textsc{BIND} generates the baryonic content of the resolved halos and their immediate surroundings, and the field between them is inherited from the $N$-body run. Each bubble reproduces the morphology of its hydrodynamical counterpart, with the stellar component far more compact than the gas, and the largest halos tracing the same knots of the cosmic web that the truth does.}
    \label{fig:showcase}
\end{figure*}

One of the primary motivations for \textsc{BIND}, and the fourth question posed in the introduction, is to serve as a practical baryonification engine. \textit{Given an arbitrary $N$-body simulation and a choice of IllustrisTNG galaxy-formation parameters, can \textsc{BIND} produce a baryonified matter field whose statistical properties match those of a true hydrodynamical run at that parameter point?} The key downstream observable we target is the projected matter power spectrum suppression, 
\begin{equation}\label{eq:suppression}
    S(k) = \frac{P_{\rm hydro}(k)}{P_{\rm DMO}(k)},
\end{equation}
which encodes the net impact of baryonic feedback on the matter distribution as a function of scale. In the companion paper, we extend this to non-Gaussian statistics \citep{Lee-2026c}. Accurately reproducing $S(k)$ across the full IllustrisTNG parameter space is a prerequisite for using \textsc{BIND} as a baryonic correction tool in cosmological inference pipelines.

%-------------------------------------------------------------
\subsection{BINDing an $N$-body Simulation}
\label{subsec:binding}
%-------------------------------------------------------------

To construct a full baryonified field from an $N$-body simulation, we apply \textsc{BIND} halo by halo and paste the generated patches back into a global projected mass map. The procedure proceeds in four steps.

\begin{enumerate}
    \item \textbf{Halo extraction.} We read in the DMO particle catalog for each CV or SB35 test simulation and identify all halos with $M_{200c} \geq 10^{13}\,M_\odot\,h^{-1}$ using the same halo finder and centering procedure described in Section~\ref{subsec:data_processing}. For each halo, we extract the $128^2$-pixel central patch.

    \item \textbf{\textsc{BIND} generation.} We pass each extracted patch through \textsc{BIND}, conditioned on the cosmological and IllustrisTNG parameter vector $\boldsymbol{\theta}_{\rm TNG}$ of the parent simulation, and generate a single sample of the three-channel baryonic output (dark matter, gas, stellar mass).

    \item \textbf{Replacement and pasting.} The $N$-body DMO projection of the full $50\,h^{-1}\,\mathrm{Mpc}$ box is first projected onto a $1024^2$ grid. We then paste each generated halo patch back into the global map at the halo center, replacing the corresponding DMO pixels. Each generated field blends into the surrounding DMO field through a raised-cosine taper that begins at $3R_{200c}$ and fades the replacement to zero at the patch edge. In regions where multiple halo patches overlap, we apply a cosine taper function to blend the edges of adjacent patches and assign priority to the more massive halo so that the largest halos take precedence in contested regions. This strategy follows \citet{Lee-2026a}, adapted here for \textsc{BIND}ed fields rather than the true hydrodynamical patches used in that work. The total baryonic map is obtained by summing the gas, stellar, and modified dark-matter channels of the pasted field.

    \item \textbf{Power spectrum computation.} We compute the 2D projected matter power spectrum of the full baryonified map using \textsc{Pylians} \citep{Pylians} and compare it to the power spectra of (i) the DMO projection, (ii) the true hydrodynamical map, and (iii) a ``hydro-replace'' control map constructed by pasting the actual hydrodynamical halo patches in place of the \textsc{BIND} outputs using the same cosine-taper procedure. The hydro-replace map isolates the pasting algorithm from any inaccuracy in the \textsc{BIND} generations.
\end{enumerate}

Fig.~\ref{fig:showcase} shows the resulting maps from this procedure for each of the fields. Each bubble is centered on a halo extracted from the DMO simulation and generated with \textsc{BIND}. We then apply the pasting mechanism cumulatively, but here we show each channel individually to make clear what they look like after being placed back onto the map.

\begin{figure*}
    \centering
    \includegraphics[width=\linewidth]{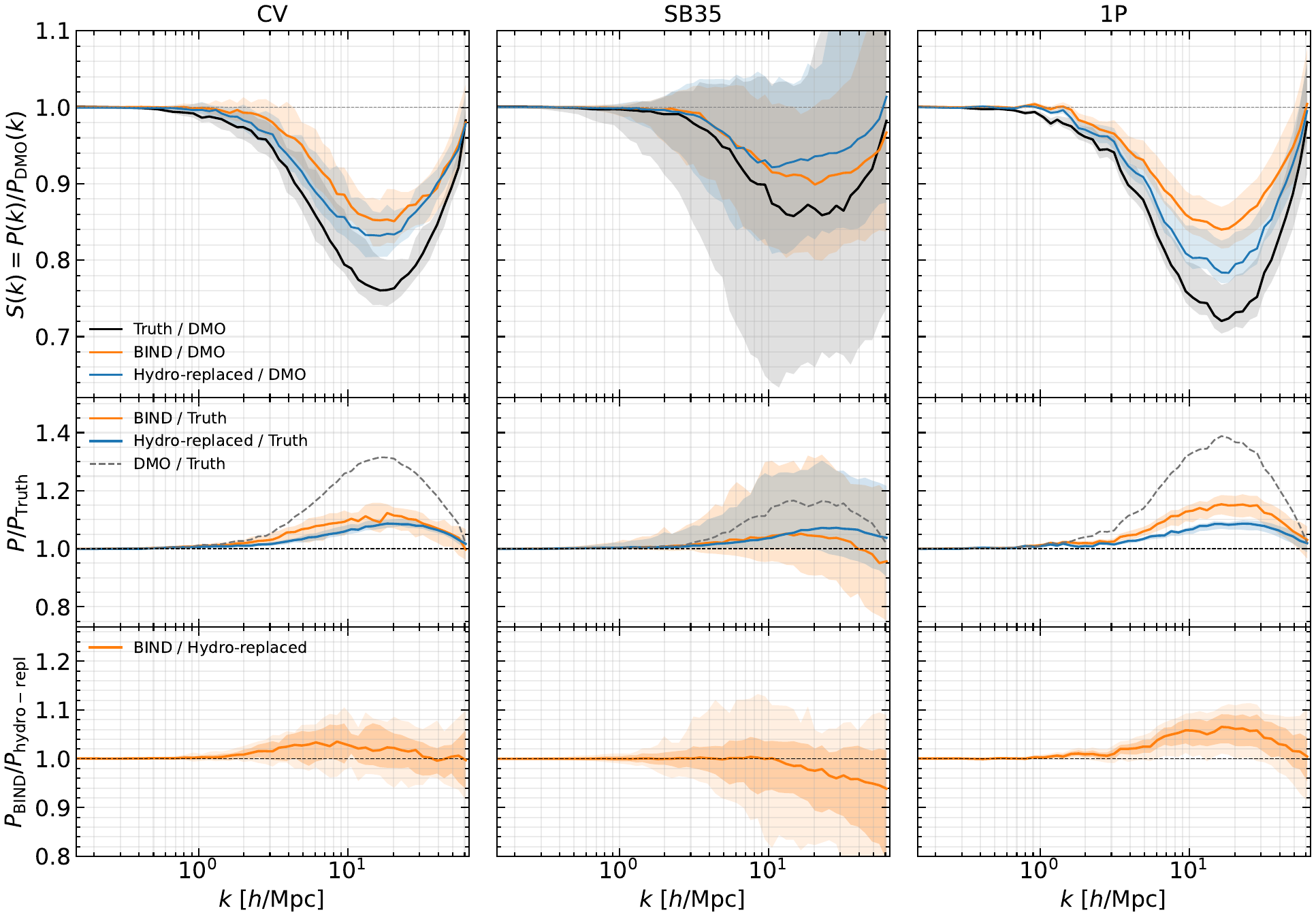}
    \caption{Projected matter power spectrum suppression for the CV, SB35 test, and 1P suites. All curves are medians over simulations of per-simulation ratios, truncated at the Nyquist scale $k = 64\,h\,{\rm Mpc}^{-1}$. \emph{Top row:} $S(k) = P(k)/P_{\rm DMO}(k)$ for the true hydrodynamical maps (black), the \textsc{BIND}-baryonified maps (orange), and the hydro-replace control where true hydrodynamical patches are pasted through the identical aperture and blending as \textsc{BIND} (blue), each with its 16--84 percentile band over simulations. \emph{Middle row:} the ratio of each to the true hydrodynamical spectrum, with 16--84 percentile bands over simulations and the DMO-to-truth ratio (gray dashed) for reference. \emph{Bottom row:} the ratio $P_{\rm BIND}/P_{\rm hydro\text{-}replace}$ with 16--84 (dark) and 5--95 (light) bands. Neither \textsc{BIND} nor the hydro-replace control reaches the full hydrodynamical suppression (black), because only halos with $M_{200c} \geq 10^{13}\,M_\odot\,h^{-1}$ are replaced, and only within a finite aperture; the hydro-replace control therefore sets the ceiling attainable with this halo sample, and the relevant comparison for \textsc{BIND} is orange versus blue rather than orange versus black.}

    \label{fig:suppression}
\end{figure*}

Fig.~\ref{fig:suppression} shows the resulting power spectrum suppression $S(k)$ for each of these maps, along with the ratio of each to the true hydrodynamical power spectrum, for the CV, SB35, and 1P test sets (left, middle, and right columns of Fig.~\ref{fig:suppression}). We stress up front that, because only halos above $10^{13}\,M_\odot\,h^{-1}$ are replaced, neither \textsc{BIND} nor the hydro-replace control is expected to recover the full suppression of the hydrodynamical box; the hydro-replace control sets the ceiling attainable with this halo sample and aperture, and it, rather than the true hydrodynamical curve, is the benchmark against which \textsc{BIND} should be judged. Several conclusions are immediately apparent.

First, the pasted maps recover only part of the suppression measured in the full hydrodynamical simulations, a consequence of the halo sample and paste aperture rather than of the \textsc{BIND} emulator. The hydro-replace control, which is derived from \emph{true} hydrodynamical patches pasted through the identical machinery, itself recovers only ${\sim}65\%$ (CV) of the true suppression $1-S(k)$ at $k = 10$--$20\,h\,{\rm Mpc}^{-1}$ at the $M_{200c} \geq 10^{13}\,M_\odot\,h^{-1}$ threshold, consistent with previous findings that the full suppression requires halos down to $M_{200c} \sim 10^{12}\,M_\odot$ and the baryonic field out to several $R_{200c}$ \citep{vanDaalen-2020, Miller-2026, Lee-2026a}. Extending \textsc{BIND} to lower halo masses is a straightforward avenue for future work.

Second, \textsc{BIND} approaches but does not fully saturate the ceiling set by halo replacement. Plotting the ratio $P_{\rm BIND}/P_{\rm hydro\text{-}replace}$ directly (bottom row of Fig.~\ref{fig:suppression}) shows \textsc{BIND} carrying $2.5\%$ more small-scale power than the control at $k = 10$--$20\,h\,{\rm Mpc}^{-1}$ in the CV suite ($1.025$, with a 16--84 simulation range of $[0.992, 1.067]$) and $6.0\%$ more in 1P ($1.060$ $[1.030, 1.084]$), while the SB35 test set is consistent with the control there ($0.986$ $[0.927, 1.034]$) and falls ${\sim}5\%$ below it at $k = 40$--$64\,h\,{\rm Mpc}^{-1}$. Because $P_{\rm BIND}/P_{\rm truth} > 1$ at these scales, the residual error is an \emph{under}-suppression, meaning that the generated halos remove slightly too little power relative to the truth, with the median $P_{\rm BIND}/P_{\rm truth}$ peaking at $+12\%$ (CV) and $+15\%$ (1P) near $k \approx 16$--$18\,h\,{\rm Mpc}^{-1}$, and its size tracks the set-dependent stellar and gas mass residuals of Section~\ref{subsec:integrated_mass}. Across simulations, the excess over the control correlates with the per-simulation stellar and gas mass residuals. 

Finally, for the SB35 test set (which spans the full 35-dimensional parameter space with parameter combinations and initial conditions entirely withheld from training), the median suppression is reproduced to better than ${\sim}5\%$ at all $k < 40\,h\,{\rm Mpc}^{-1}$ (the median $P_{\rm BIND}/P_{\rm truth}$ peaks at $1.05$ near $k \simeq 13\,h\,{\rm Mpc}^{-1}$), consistent with the parameter-sensitivity results in Section~\ref{sec:param_response}.

These results are encouraging. For a model to saturate the information ceiling exactly, it would have to reproduce the fields in and around halos perfectly. This is, of course, a tall order for any model, especially one expected to maintain fidelity across the full high-dimensional parameter space. The advantage of this approach is that, beyond achieving near-percent-level accuracy across scales relative to this ceiling, it simultaneously captures the profiles, integrated masses, and aspherical redistribution of mass. Future work will first improve the mass threshold of \textsc{BIND} to allow lower masses; this will allow the ceiling to approach that of the true hydrodynamical simulation. Future work will also improve the fidelity of the flow model.

%-------------------------------------------------------------
\subsection{Implications for the Future}
\label{subsec:implications}
%-------------------------------------------------------------

The results in Section~\ref{subsec:binding} establish that \textsc{BIND} can serve as a practical, fast baryonification engine for large-scale structure analyses. Generating baryonic fields for all halos in a single $50\,h^{-1}\,\mathrm{Mpc}$ simulation takes $\mathcal{O}(\mathrm{minutes})$ on a single GPU (see Table~\ref{tab:timing}), orders of magnitude faster than re-running the hydrodynamical simulation, which makes it feasible to baryonify thousands of $N$-body realizations in the time required to complete a single hydrodynamical run. 

The most immediate application is learning the relationship between baryonic effects in massive halos and weak lensing statistics. This application is already explored in the companion paper \citep{Lee-2026c}. However, that work uses a base DMO simulation from IllustrisTNG300-Dark, run at a single point in the cosmological parameter space.

By running a suite of $N$-body simulations at different cosmological parameters and baryonifying each with \textsc{BIND} at a range of $\boldsymbol{\theta}_{\rm TNG}$ values, we can construct a joint likelihood over cosmological and galaxy-formation parameters directly from field-level observables such as weak-lensing convergence maps, without relying on analytic profile prescriptions.

Alternatively, \textsc{BIND} can be used in a forward-modeling framework to constrain the IllustrisTNG subgrid parameters from observations. Multi-wavelength data such as X-ray gas fractions, the kinematic and thermal Sunyaev-Zel'dovich effect, and optical weak lensing jointly constrain different combinations of the feedback parameters \citep{Kovac-2025}, and fast radio burst dispersion measures are emerging as a further, largely independent handle on how far feedback pushes gas \citep{Medlock-2024, Medlock-2025}. Emulators built on CAMELS have already been used to turn such observables into constraints on the subgrid parameters directly \citep{Lau-2025}. Because \textsc{BIND} is conditioned on the full 35-dimensional parameter vector, it can, in principle, propagate observational constraints on these parameters into predictions for any field-level observable, which would provide a unified framework for multi-probe baryon modeling. This would require updates to the \textsc{BIND} methodology for the output fields and, ideally, the incorporation of 3D field generation.

A third direction is generating large training sets for summary-statistic emulators or simulation-based inference networks. Field-level neural network summaries trained on cosmological maps require thousands of simulation realizations across diverse parameter combinations, a task that \textsc{BIND} can perform at negligible cost once the $N$-body simulations are available. Ideally, we can use \textsc{BIND}ed simulations to generate accurate beyond-Gaussian statistics.

Finally, a joint training over multiple subgrid prescriptions is an enticing next step. When the CAMELS suite expands to include $50\,h^{-1}\,\mathrm{Mpc}$ boxes across a range of subgrid prescriptions, \textsc{BIND} can inherit all the different feedback models and, hopefully, become a unified halo-based generative model.

%-------------------------------------------------------------\
\section{Caveats with the \textsc{BIND} Approach}
\label{sec:caveats}
%-------------------------------------------------------------

Despite its encouraging performance, the \textsc{BIND} framework presented here has several limitations to keep in mind when applying it in practice.

The current model is trained and validated on halos with $M_{200c} \geq 10^{13}\,M_\odot\,h^{-1}$, which contribute the largest individual patches to the baryonified field but represent only a fraction of the halos responsible for the full power spectrum suppression \citep{vanDaalen-2011,Schneider-2015, Schneider-2019,Arico-2020,Lee-2026a}. Extending the training to group- and galaxy-scale halos ($10^{12}$--$10^{13}\,M_\odot$) would require handling a larger number of objects per simulation volume and a wider dynamic range of halo morphologies, but is a natural and necessary next step for achieving percent-level accuracy in $S(k)$ at all relevant scales.

\textsc{BIND} operates on projected 2D mass maps rather than 3D particle distributions, which introduces line-of-sight confusion from foreground and background matter unassociated with the target halo. While baryonification carried out directly on projected maps has been shown to reproduce two-point and several higher-order weak lensing statistics at the few-percent level \citep{Anbajagane-2024, Zhou-2025}, the approximation may be less accurate for higher-order statistics that are sensitive to the three-dimensional geometry of halos and filaments or quantities that care about the full three-dimensional density instead of projected surface density. Extending \textsc{BIND} to 3D inputs and outputs is conceptually straightforward but requires substantially more GPU memory and training data.

\textsc{BIND} is trained exclusively on the IllustrisTNG galaxy-formation model as implemented in the CAMELS SB35 suite. While this covers a broad, 35-dimensional subgrid parameter space, it is anchored to a single simulation code and a single set of physical prescriptions. Applying \textsc{BIND} to parameter combinations outside the SB35 training range, or to simulation codes with qualitatively different feedback implementations (e.g., SIMBA or EAGLE), would require either retraining on the appropriate simulation suite or demonstrating that the learned mapping generalizes across codes.

Similarly, \textsc{BIND} is trained only on a single map resolution. This limits its applicability to dark matter maps generated at the same resolution as those \textsc{BIND} was trained on. Increasing the training set to include multi-fidelity simulations will be an important next step to generalize \textsc{BIND}. It would also be interesting to see whether lower-resolution (and cheaper) simulations, gridded to the same map resolution as now, would work well for \textsc{BIND}.

%-------------------------------------------------------------
\section{Summary and Conclusions}
\label{sec:conclusions}
%-------------------------------------------------------------

We have introduced \textsc{BIND} (Baryonic INpainting with Deep learning), a conditional flow matching model that learns the field-level mapping from dark-matter-only halos to their hydrodynamical counterparts, conditioned on the full 35-dimensional cosmological and IllustrisTNG subgrid parameter space of the CAMELS SB35 suite. \textsc{BIND} takes as input a projected $N$-body patch centered on a halo, together with a parameter vector $\boldsymbol{\theta}_{\rm TNG}$, and outputs simultaneous, physically consistent realizations of the dark-matter, gas, and stellar mass fields without imposing any analytic functional form on any component. Applied patch-by-patch to the halos in an $N$-body simulation, \textsc{BIND} produces baryonified cosmological fields at $\mathcal{O}(\mathrm{minute})$ cost per simulation volume on a single GPU, orders of magnitude faster than a hydrodynamical re-simulation.

We validated \textsc{BIND} through a comprehensive suite of tests on three independent evaluation sets (the SB35 test set, the 1P single-parameter-variation set, and the CV cosmic-variance set), spanning a wide range of halo masses, cosmological parameters, and galaxy-formation physics. Our main findings are the following.

\begin{enumerate}

    \item \textbf{Accurate integrated masses and parameter sensitivity.}
    \textsc{BIND} recovers nearly unbiased total masses for all three baryonic components in hydrodynamical simulations across the full evaluation parameter space, with median relative errors below $1\%$ for dark matter and total mass, $\sim 1\%$ for gas, and $\sim 10\%$ for stars. Spearman rank correlations between the generated integrated masses and the IllustrisTNG subgrid parameters closely match the ground-truth correlations across all channels, showing that the model has learned the dependence of baryonic content on galaxy-formation physics.

    \item \textbf{Radially resolved profile accuracy.}
    Azimuthally averaged surface density profiles are reproduced to better than $10\%$ accuracy at all projected radii out to $3.125\,h^{-1}\,\mathrm{Mpc}$ across all test sets. The radially resolved Spearman correlations with the IllustrisTNG parameters match the ground truth at virtually all radii, capturing the physically motivated competition between gas cooling and feedback-driven expulsion as a function of projected radius.

    \item \textbf{Projected halo morphology.}
    Distributions of the projected axis ratio $q$ for all three mass channels are reproduced faithfully across all test sets, correctly recovering the ordering of shapes among components. Dark matter peaks near $q \approx 0.87$, the gas is more circular, and stars exhibit the broadest distribution and largest ellipticities. This demonstrates that \textsc{BIND} has learned the anisotropic morphological mapping from DMO to baryonic fields, a non-trivial capability with direct implications for weak-lensing intrinsic alignment modeling, including the direction of the ellipse for stellar fields.

    \item \textbf{Field-level parameter response.}
    Using the 1P suite, we showed that \textsc{BIND} correctly reproduces the signed, spatially resolved response of the gas and stellar fields to variations in individual subgrid parameters, including the central gas deficit and peripheral excess driven by AGN feedback (\texttt{AAGN1}, \texttt{AAGN2}) and the enhanced infall gas density and stellar suppression driven by stronger stellar winds (\texttt{ASN1}). The dark-matter channel correctly shows little coherent parameter response, consistent with its gravitational origin.

    \item \textbf{Inter-component correlations and scaling relations.}
    Across the 1,111 CV halos at the fiducial parameters, \textsc{BIND} reproduces the slope, normalization, and scatter of all three inter-component scaling relations ($M_{\rm gas}$--$M_{\rm star}$, SHMR, $M_{\rm bar}$--$M_{200c}$). Corner plots of the residuals confirm that the joint multivariate distribution of halos that deviate coherently from multiple relations simultaneously is also recovered, including physically motivated correlations such as the coherent SHMR-baryon excess and the gas-stellar anticorrelation driven by AGN feedback. Halo mass was never used as an explicit conditioning variable, yet the model emergently learned the correct halo-mass dependence of the baryon fraction across more than a decade in $M_{200c}$.

    \item \textbf{Power spectrum and baryonification.}
    Baryonifying full $50\,h^{-1}\,\mathrm{Mpc}$ $N$-body projections with \textsc{BIND} reproduces the projected matter power spectrum suppression $S(k) = P_{\rm hydro}(k)/P_{\rm DMO}(k)$ to within a few percent across the full range of $k$ probed, for both the CV and SB35 test sets. The \textsc{BIND} suppression curve closely tracks the hydro-replace control, indicating that the model approaches the halo-pasting procedure's performance ceiling.
\end{enumerate}

Together, these results answer the four questions posed in the introduction. \textsc{BIND} can learn the field-level DMO-to-hydro mapping at the halo level, correctly encode the dependence of baryonic structure on the IllustrisTNG galaxy-formation parameter space, capture the inter-component correlations and their parameter dependence, and can be deployed as a practical baryonification engine with percent-level accuracy for two-point statistics.

The present work has several limitations that define a natural roadmap for future development. The most important is the halo mass threshold of $M_{200c} \geq 10^{13}\,M_\odot\,h^{-1}$: extending \textsc{BIND} to group- and galaxy-scale halos ($10^{12}$--$10^{13}\, M_\odot$) would recover the full power spectrum suppression and open the model to a wider range of observational comparisons. \textsc{BIND} currently operates in 2D projection, a choice that reduced computational costs and simplified the halo-pasting step, but extending to 3D fields is a natural next iteration. The model is also tied to the IllustrisTNG subgrid framework, and extending it to other galaxy-formation models such as SIMBA or EAGLE would require retraining on the appropriate CAMELS suite, though the architecture and training procedure are directly transferable. 

Looking forward, \textsc{BIND} is well positioned to address some of the most pressing systematic challenges for upcoming weak-lensing surveys. The model's halo-centric, field-level, parameter-conditioned, and generative design allows it to serve as a forward-modeling engine for joint constraints on cosmological and galaxy-formation parameters using non-Gaussian statistics such as peak counts, Minkowski functionals, and field-level neural summaries, for which analytic baryonic correction tools have been shown to be insufficient \citep{Lee-2023,Lee-2026a}. When combined with a fast $N$-body simulator and a flexible posterior sampler, \textsc{BIND} could enable full field-level inference pipelines that marginalize over the IllustrisTNG parameter space at speeds that were previously infeasible, unlocking the full statistical power of surveys such as \textit{Euclid} \citep{Laureijs-2011}, the Vera C.\ Rubin Observatory LSST \citep{Ivezic-2019}, and the Nancy Grace Roman Space Telescope \citep{Spergel-2015}. The companion paper \citep{Lee-2026c} takes the first step.

%% Please use the acknowledgment and contribution environments. This will 
%% be anonomyized when the "anonymous" style option is used. 
\begin{acknowledgments}
We thank Colin Hill, Carolina Cuesta-Lazaro, Daisuke Nagai, Erwin Lau, and Amanda Lue for useful discussions during this project. MEL is supported by NSF grant DGE-2036197. ZH acknowledges financial support from NASA ATP grant 80NSSC24K1093. The Flatiron Institute is supported by the Simons Foundation. GLB acknowledges support from the NSF (AST-2307419) and NASA (80NSSC21K1053), as well as support from the Simons Foundation through the Learning the Universe Collaboration. C. Lovell was supported by the research environment and infrastructure of the Handley Lab at the University of Cambridge. MEL also thanks Erin Walter for helpful comments and editing. The authors used Claude Opus and sonnet to refine sections of code and text, and Grammarly was used to refine portions of the draft with grammatical edits. The authors take full responsibility for the final content.

\end{acknowledgments}

\begin{contribution}
MEL performed the analysis, code generation, testing, and text writing under the mentorship and support of SG, ZH, GLB, CL and BH.
%%This section gives authors the space to recognize author contributions. The text inside this environment is NOT counted towards the total word quanta. At a minimum, manuscripts are expected to include this text:

%% But authors are expected to provide more specific details, e.g. 
%%
%%SC was responsible for writing and submitting the manuscript.
%%WWM came up with the initial research concept and edited the manuscript.
%%OTS obtained the funding and edited the manuscript.
%%EBF provided the formal analysis and validation. He also edited the manuscript.
%%GEH Supervised the undergraduates, wrote the software and administers the project github and Zenodo repositories.
%%
%% Authors can use the Contributor Role Taxonomy (CRediT) at
%% for ideas on how write a good statement tailored to their needs.

\end{contribution}

% \facilities{} --- simulation-based work; no observing facilities.

\software{astropy \citep{2013A&A...558A..33A,2018AJ....156..123A,2022ApJ...935..167A},
          Pylians \citep{Pylians},
          PyTorch \citep{paszke2019pytorch}
          }


\appendix

\section{Architecture}\label{appendix:architecture}
\subsection{Conditioning Embeddings}

The flow time $t$ is encoded with sinusoidal positional embeddings \citep{dhariwal2021diffusion} of dimension 128, then projected to a 512-dimensional embedding $\bm{e}_t$ via a two-layer MLP with SiLU activations. The scale factor $a$ is encoded in the same way (128-dimensional sinusoidal features followed by a two-layer MLP with SiLU activations) to yield $\bm{e}_a$. The parameter vector $\boldsymbol{\theta}_{\rm TNG}$ is independently processed by a third two-layer MLP ($35 \to 512 \to 512$, SiLU activations) to yield $\bm{e}_\theta$. The three embeddings are summed to form a single conditioning vector
\begin{equation}
    \bm{e} = \bm{e}_t + \bm{e}_a + \bm{e}_\theta,
    \label{eq:embedding}
\end{equation}
which is broadcast into every residual block of the U-Net.

\subsection{Encoder--Decoder with Adaptive Group Normalization}

The encoder consists of four resolution stages with channel multipliers $(1,2,4,8)$ relative to a base width of 128, giving feature-map widths of 128, 256, 512, and 1024. Each stage contains two residual blocks, and multi-head self-attention layers \citep[4 heads;][]{vaswani2017attention} are inserted at resolutions $32^2$ and $16^2$. Spatial downsampling uses strided convolutions; upsampling uses nearest-neighbor interpolation followed by convolution. Encoder activations are passed to the decoder via skip connections by channel-wise concatenation. Each residual block conditions on $\bm{e}$ through Adaptive Group Normalization \citep{dhariwal2021diffusion},
\begin{equation}
    {\rm AdaGN}(\bm{h},\bm{e})
    =
    \bigl(1 + \boldsymbol{\gamma}(\bm{e})\bigr)
    \odot {\rm GN}(\bm{h})
    + \boldsymbol{\beta}(\bm{e}),
    \label{eq:adagn}
\end{equation}
where ${\rm GN}$ denotes Group Normalization with 32 groups \citep{wu2018groupnorm}, and $\boldsymbol{\gamma}$, $\boldsymbol{\beta}$ are learned linear projections of $\bm{e}$. This allows the network to modulate its internal representations as a joint function of flow time, scale factor, and model parameters. The output projection layer is zero-initialized to stabilize early training.

The full architecture is summarized in Table~\ref{tab:architecture}.

\begin{table}
\centering
\caption{U-Net architecture hyperparameters.}
\label{tab:architecture}
\begin{tabular}{lc}
\hline
Parameter & Value \\
\hline
Base channels              & 128 \\
Channel multipliers        & $(1, 2, 4, 8)$ \\
Residual blocks per stage  & 2 \\
Attention heads            & 4 \\
Attention resolutions      & $32^2,\;16^2$ \\
Embedding dimension        & 512 \\
Dropout                    & 0.1 \\
Output initialization      & Zero \\
\hline
\end{tabular}
\end{table}

\section{Optimization details}\label{appendix:training}

The learning rate follows a two-phase schedule: a linear warm-up from zero to $\eta_0 = 10^{-4}$ over the first $N_{\rm warm} = 1,000$ gradient steps, followed by cosine annealing to zero over the remainder of training,
\begin{equation}
    \eta(k) =
    \begin{cases}
        \eta_0\,\dfrac{k}{N_{\rm warm}},
            & k < N_{\rm warm}, \\[8pt]
        \dfrac{\eta_0}{2}
        \left[1 + \cos\left(
            \pi\,\dfrac{k - N_{\rm warm}}{K_{\rm tot} - N_{\rm warm}}
        \right)\right],
            & k \geq N_{\rm warm},
    \end{cases}
    \label{eq:lrsched}
\end{equation}
where $k$ is the current gradient step and $K_{\rm tot}$ is the total number of steps \citep{loshchilov2017sgdr}. Weight decay is set to $\lambda = 10^{-4}$, and gradients are clipped to a maximum $\ell_2$ norm of 1.0.

An exponential moving average (EMA) of the network weights is maintained throughout training with decay $\rho = 0.9999$,
\begin{equation}
    \bar{\phi}_k = \rho\,\bar{\phi}_{k-1} + (1-\rho)\,\phi_k,
    \label{eq:ema}
\end{equation}
and the EMA weights $\bar{\phi}$ are used exclusively at inference \citep{karras2022edm}. During training, the conditioning vector is dropped with probability $0.1$, following the classifier-free guidance construction of \citet{ho2022cfg}, so that the network also learns an unconditional velocity field. We do not apply guidance at inference in this work, and all results shown use the conditional model. Table~\ref{tab:training} collects the full set of training hyperparameters.

\begin{table}
\centering
\caption{Training hyperparameters.}
\label{tab:training}
\begin{tabular}{lc}
\hline
Hyperparameter & Value \\
\hline
Learning rate $\eta_0$      & $10^{-4}$ \\
Weight decay                & $10^{-4}$ \\
Gradient clip norm          & 1.0 \\
Warm-up steps               & 1,000 \\
Max epochs                  & 200 \\
Batch size (per GPU)        & 64 \\
EMA decay $\rho$            & 0.9999 \\
CFG dropout probability     & 0.1 \\
Euler steps at inference    & 20 \\
\hline
\end{tabular}
\end{table}

\section{Characterizing the Line-of-Sight Projection}\label{cube_comp}

\begin{figure*}
\centering
\includegraphics[width=\textwidth]{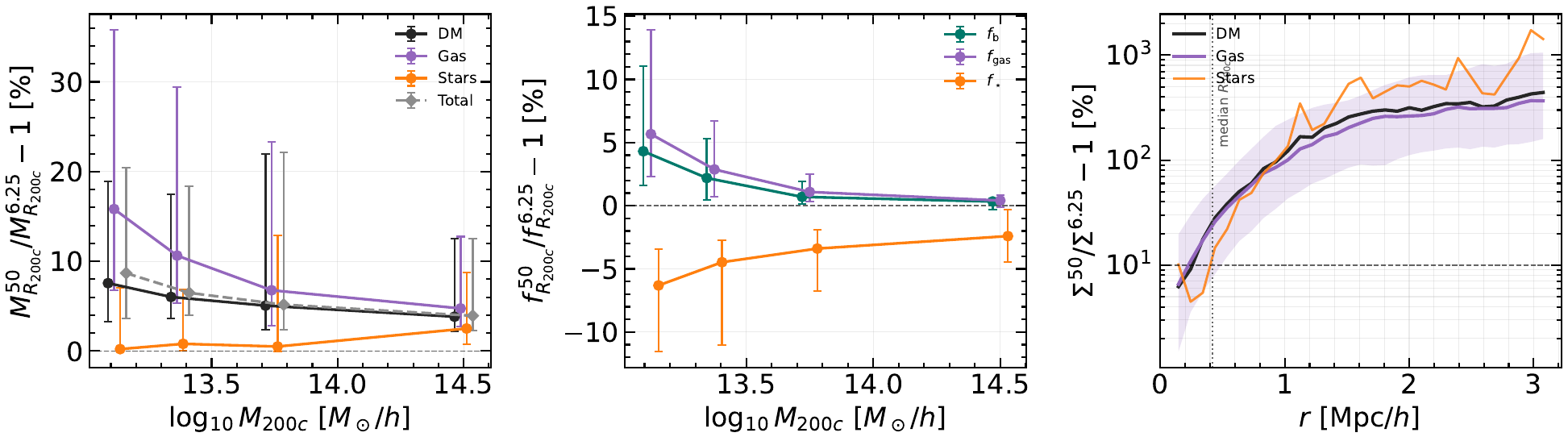}
\caption{Projection-depth characterization on $8,273$ matched halos: the full $50\,h^{-1}\,{\rm Mpc}$-depth truth measured against the halo-local $6.25\,h^{-1}\,{\rm Mpc}$ cube truth of the same halos. \textit{Left:} median $R_{200c}$-aperture line-of-sight excess $M^{50}_{R_{200c}}/M^{6.25}_{R_{200c}} - 1$ per component as a function of halo mass; bars span the 16th--84th percentiles. \textit{Center:} the induced bias of the projected aperture fractions, $f^{50}_{R_{200c}}/f^{6.25}_{R_{200c}} - 1$. \textit{Right:} median surface-density excess $\Sigma^{50}/\Sigma^{6.25} - 1$ versus radius (32 linear annuli; shaded: gas 16th--84th range; the horizontal line marks $10\%$); the innermost annulus is omitted because it is dominated by sub-pixel registration differences between the two projections.}
\label{fig:cube_comp}
\end{figure*}

The fiducial training and evaluation maps integrate the density fields over the full $50\,h^{-1}\,{\rm Mpc}$ line of sight (Section~\ref{subsec:data_processing}), so every halo aperture contains some foreground and background material unassociated with the halo. Here we quantify the projection contribution directly. For every evaluation halo we construct a halo-local reference by re-projecting only the $(6.25\,h^{-1}\,{\rm Mpc})^3$ cube centered on the halo, and match halos one-to-one between the two projections (identical friends-of-friends masses and centers), giving $8,273$ matched halos with $M_{200c} \geq 10^{13}\,M_\odot\,h^{-1}$ ($1,154$ CV, $6,823$ 1P, $296$ SB35-test). The line-of-sight excess within $R_{200c}$, $X = M^{50}/M^{6.25} - 1$, has a pooled median (16th--84th percentile range) of $+6.1\%$ $[+3.1, +18.5]$ for dark matter, $+12.0\%$ $[+4.6, +30.8]$ for gas, $+0.4\%$ $[-0.0, +8.1]$ for stars, and $+6.6\%$ $[+3.3, +20.1]$ for the total mass, essentially identical across the three suites (Fig.~\ref{fig:cube_comp}). The excess falls with halo mass as the halo increasingly dominates its own line of sight: the median gas excess drops from $+15.8\%$ at $M_{200c} = 10^{13}$--$10^{13.25}\,M_\odot\,h^{-1}$ to $+4.7\%$ above $10^{14}\,M_\odot\,h^{-1}$ (total: $+8.7\%$ to $+3.9\%$). A small tail is heavily contaminated --- $X_{\rm tot} > 0.5$ for $0.8\%$ of halos and $X_{\rm tot} > 1$ for $0.25\%$ --- and it is these systems that appear as outliers above the stellar-to-halo and baryonic scaling relations of Section~\ref{subsec:scaling_relations}. The stellar apertures are essentially projection-free, so projection biases the \emph{composition} of the aperture: relative to the halo-local values, the projected baryon fraction is high by a median $+2.4\%$ $[+0.4, +7.9]$ (rising to $+4.3\%$ in the lowest mass bin), the gas fraction by $+3.2\%$, and the stellar fraction is diluted by $-5.2\%$ $[-10.3, -2.7]$. Radially, the contamination lives in the aperture outskirts: the median surface-density excess $\Sigma^{50}/\Sigma^{6.25} - 1$ is below $10\%$ inside $0.25\,h^{-1}\,{\rm Mpc}$, crosses $+50\%$ near $0.7\,h^{-1}\,{\rm Mpc}$, and reaches $300$--$400\%$ at the patch edge, while the aperture-integrated excess stays modest because the aperture mass is core-dominated.

Because the true and generated maps share the identical projection (the same rotation matrices and line of sight, Section~\ref{subsec:data_processing}), every truth-versus-\textsc{BIND} comparison in this paper is internally consistent. The line-of-sight term appears on both sides and cancels out in the relative statements. The absolute aperture quantities, however, include it. Aperture masses overestimate the halo-associated mass by the $X$ values above, the baryon fractions of Section~\ref{subsec:baryon_fraction} carry the $+2$--$4\%$ relative offset, and the projected shapes of Section~\ref{subsec:shapes} are diluted, as already noted there. Comparisons with three-dimensional, halo-intrinsic quantities from other simulations or from observations should therefore either apply these corrections or forward-model the same projection depth. Finally, the projection depth is not forced on the model by the learning problem. An otherwise identical flow-matching model trained on the $6.25\,h^{-1}\,{\rm Mpc}$ cube projections (without the large-scale context channels) performs nearly as well against its own halo-local truth, with median total-mass residuals within $R_{200c}$ of $+1.9\%$ versus $+0.5\%$ for the fiducial model (CV suite; gas $+8.8\%$ versus $+3.4\%$), so the full-depth choice is driven by the conditioning and pasting workflow of Section~\ref{sec:bind_in_the_wild} rather than by model accuracy alone.

%% For this sample we use BibTeX plus aasjournalv7.bst to generate the

\bibliography{biblio}{}

@ARTICLE{Hadzhiyska-2025,
       author = {{Hadzhiyska}, Boryana and {Ferraro}, Simone and {Farren}, Gerrit S. and {Sailer}, Noah and {Zhou}, Rongpu},
        title = "{Missing baryons recovered: a measurement of the gas fraction in galaxies and groups with the kinematic Sunyaev-Zel'dovich effect and CMB lensing}",
      journal = {arXiv e-prints},
         year = 2025,
        month = jul,
          eid = {arXiv:2507.14136},
        pages = {arXiv:2507.14136},
          doi = {10.48550/arXiv.2507.14136},
archivePrefix = {arXiv},
       eprint = {2507.14136},
 primaryClass = {astro-ph.CO},
       adsurl = {https://ui.adsabs.harvard.edu/abs/2025arXiv250714136H}
}

@ARTICLE{Leung-2025,
       author = {{Leung}, Calvin and {Borrow}, Josh and {Masui}, Kiyoshi W. and {Andrew}, Shion and {Chen}, Kai-Feng and {Schaye}, Joop and {Schaller}, Matthieu},
        title = "{Nulling baryonic feedback in weak lensing surveys using cross-correlations with fast radio bursts}",
      journal = {arXiv e-prints},
         year = 2025,
        month = sep,
          eid = {arXiv:2509.19514},
        pages = {arXiv:2509.19514},
          doi = {10.48550/arXiv.2509.19514},
archivePrefix = {arXiv},
       eprint = {2509.19514},
 primaryClass = {astro-ph.CO}
}

@ARTICLE{Wayland-2026,
       author = {{Wayland}, Amy and {Alonso}, David and {Reischke}, Robert},
        title = "{Probing baryonic feedback with fast radio bursts: joint analyses with cosmic shear and galaxy clustering}",
      journal = {\mnras},
         year = 2026,
        month = feb,
       volume = {547},
       number = {4},
          eid = {stag557},
        pages = {stag557},
          doi = {10.1093/mnras/stag557},
archivePrefix = {arXiv},
       eprint = {2602.12174},
 primaryClass = {astro-ph.CO}
}

@ARTICLE{Lovell-2022,
       author = {{Lovell}, Christopher C. and {Wilkins}, Stephen M. and {Thomas}, Peter A. and {Schaller}, Matthieu and {Baugh}, Carlton M. and {Fabbian}, Giulio and {Bah{\'e}}, Yannick},
        title = "{A machine learning approach to mapping baryons on to dark matter haloes using the EAGLE and C-EAGLE simulations}",
      journal = {\mnras},
         year = 2022,
        month = feb,
       volume = {509},
       number = {4},
        pages = {5046-5061},
          doi = {10.1093/mnras/stab3221},
archivePrefix = {arXiv},
       eprint = {2106.04980},
 primaryClass = {astro-ph.GA},
       adsurl = {https://ui.adsabs.harvard.edu/abs/2022MNRAS.509.5046L}
}

@ARTICLE{Miller-2026,
       author = {{Miller}, Kyle and {More}, Surhud and {Jain}, Bhuvnesh},
        title = "{Baryonic feedback across halo mass: impact on the matter power spectrum}",
      journal = {\jcap},
         year = 2026,
        month = may,
       volume = {2026},
       number = {5},
          eid = {092},
        pages = {092},
          doi = {10.1088/1475-7516/2026/05/092},
archivePrefix = {arXiv},
       eprint = {2511.10634},
 primaryClass = {astro-ph.CO},
       adsurl = {https://ui.adsabs.harvard.edu/abs/2026JCAP...05..092M}
}

@ARTICLE{Villaescusa-Navarro-2023,
       author = {{Villaescusa-Navarro}, Francisco and {Genel}, Shy and {Angl{\'e}s-Alc{\'a}zar}, Daniel and {Perez}, Lucia A. and {Villanueva-Domingo}, Pablo and {Wadekar}, Digvijay and {Shao}, Helen and {Mohammad}, Faizan G. and {Hassan}, Sultan and {Moser}, Emily and {Lau}, Erwin T. and {Machado Poletti Valle}, Luis Fernando and {Nicola}, Andrina and {Thiele}, Leander and {Jo}, Yongseok and {Philcox}, Oliver H.~E. and {Oppenheimer}, Benjamin D. and {Tillman}, Megan and {Hahn}, ChangHoon and {Kaushal}, Neerav and {Pisani}, Alice and {Gebhardt}, Matthew and {Delgado}, Ana Maria and {Caliendo}, Joyce and {Kreisch}, Christina and {Wong}, Kaze W.~K. and {Coulton}, William R. and {Eickenberg}, Michael and {Parimbelli}, Gabriele and {Ni}, Yueying and {Steinwandel}, Ulrich P. and {La Torre}, Valentina and {Dave}, Romeel and {Battaglia}, Nicholas and {Nagai}, Daisuke and {Spergel}, David N. and {Hernquist}, Lars and {Burkhart}, Blakesley and {Narayanan}, Desika and {Wandelt}, Benjamin and {Somerville}, Rachel S. and {Bryan}, Greg L. and {Viel}, Matteo and {Li}, Yin and {Irsic}, Vid and {Kraljic}, Katarina and {Marinacci}, Federico and {Vogelsberger}, Mark},
        title = "{The CAMELS Project: Public Data Release}",
      journal = {\apjs},
         year = 2023,
        month = apr,
       volume = {265},
       number = {2},
          eid = {54},
        pages = {54},
          doi = {10.3847/1538-4365/acbf47},
archivePrefix = {arXiv},
       eprint = {2201.01300},
 primaryClass = {astro-ph.CO},
       adsurl = {https://ui.adsabs.harvard.edu/abs/2023ApJS..265...54V}
}

@ARTICLE{Ni-2023,
       author = {{Ni}, Yueying and {Genel}, Shy and {Angl{\'e}s-Alc{\'a}zar}, Daniel and {Villaescusa-Navarro}, Francisco and {Jo}, Yongseok and {Bird}, Simeon and {Di Matteo}, Tiziana and {Croft}, Rupert and {Chen}, Nianyi and {de Santi}, Natal{\'\i} S.~M. and {Gebhardt}, Matthew and {Shao}, Helen and {Pandey}, Shivam and {Hernquist}, Lars and {Dave}, Romeel},
        title = "{The CAMELS Project: Expanding the Galaxy Formation Model Space with New ASTRID and 28-parameter TNG and SIMBA Suites}",
      journal = {\apj},
         year = 2023,
        month = dec,
       volume = {959},
       number = {2},
          eid = {136},
        pages = {136},
          doi = {10.3847/1538-4357/ad022a},
archivePrefix = {arXiv},
       eprint = {2304.02096},
 primaryClass = {astro-ph.CO},
       adsurl = {https://ui.adsabs.harvard.edu/abs/2023ApJ...959..136N}
}

@ARTICLE{Colibre,
       author = {{Schaye}, Joop and {Chaikin}, Evgenii and {Schaller}, Matthieu and {Ploeckinger}, Sylvia and {Hu{\v{s}}ko}, Filip and {McGibbon}, Robert J. and {Trayford}, James W. and {Ben{\'\i}tez-Llambay}, Alejandro and {Correa}, Camila and {Frenk}, Carlos S. and {Richings}, Alexander J. and {Forouhar Moreno}, Victor J. and {Bah{\'e}}, Yannick M. and {Borrow}, Josh and {Durrant}, Anna and {Gebek}, Andrea and {Helly}, John C. and {Jenkins}, Adrian and {Lacey}, Cedric G. and {Ludlow}, Aaron and {Nobels}, Folkert S.~J.},
        title = "{The COLIBRE project: cosmological hydrodynamical simulations of galaxy formation and evolution}",
      journal = {\mnras},
         year = 2026,
        month = may,
       volume = {548},
       number = {1},
          eid = {stag375},
        pages = {stag375},
          doi = {10.1093/mnras/stag375},
archivePrefix = {arXiv},
       eprint = {2508.21126},
 primaryClass = {astro-ph.GA},
       adsurl = {https://ui.adsabs.harvard.edu/abs/2026MNRAS.548ag375S}
}

@ARTICLE{Fang-2007,
       author = {{Fang}, Wenjuan and {Haiman}, Zolt{\'a}n},
        title = "{Constraining dark energy by combining cluster counts and shear-shear correlations in a weak lensing survey}",
      journal = {\prd},
         year = 2007,
        month = feb,
       volume = {75},
       number = {4},
          eid = {043010},
        pages = {043010},
          doi = {10.1103/PhysRevD.75.043010},
archivePrefix = {arXiv},
       eprint = {astro-ph/0612187},
 primaryClass = {astro-ph},
       adsurl = {https://ui.adsabs.harvard.edu/abs/2007PhRvD..75d3010F}
}

@ARTICLE{Lee-2023,
       author = {{Lee}, Max E. and {Lu}, Tianhuan and {Haiman}, Zolt{\'a}n and {Liu}, Jia and {Osato}, Ken},
        title = "{Comparing weak lensing peak counts in baryonic correction models to hydrodynamical simulations}",
      journal = {\mnras},
         year = 2023,
        month = feb,
       volume = {519},
       number = {1},
        pages = {573-584},
          doi = {10.1093/mnras/stac3592},
archivePrefix = {arXiv},
       eprint = {2201.08320},
 primaryClass = {astro-ph.CO},
       adsurl = {https://ui.adsabs.harvard.edu/abs/2023MNRAS.519..573L}
}

@ARTICLE{Lee-2024,
       author = {{Lee}, Max E. and {Genel}, Shy and {Wandelt}, Benjamin D. and {Zhang}, Benjamin and {Delgado}, Ana Maria and {Pandey}, Shivam and {Lau}, Erwin T. and {Carr}, Christopher and {Cook}, Harrison and {Nagai}, Daisuke and {Angl{\'e}s-Alc{\'a}zar}, Daniel and {Villaescusa-Navarro}, Francisco and {Bryan}, Greg L.},
        title = "{Zooming by in the CARPoolGP Lane: New CAMELS-TNG Simulations of Zoomed-in Massive Halos}",
      journal = {\apj},
         year = 2024,
        month = jun,
       volume = {968},
       number = {1},
          eid = {11},
        pages = {11},
          doi = {10.3847/1538-4357/ad3d4a},
archivePrefix = {arXiv},
       eprint = {2403.10609},
 primaryClass = {astro-ph.CO},
       adsurl = {https://ui.adsabs.harvard.edu/abs/2024ApJ...968...11L}
}

@ARTICLE{Lee-2026a,
       author = {{Lee}, Max E. and {Haiman}, Zoltan and {Genel}, Shy},
        title = "{The impact of baryons on weak lensing statistics as a function of halo mass and radius}",
      journal = {arXiv e-prints},
         year = 2026,
        month = mar,
          eid = {arXiv:2603.11815},
        pages = {arXiv:2603.11815},
          doi = {10.48550/arXiv.2603.11815},
archivePrefix = {arXiv},
       eprint = {2603.11815},
 primaryClass = {astro-ph.CO},
       adsurl = {https://ui.adsabs.harvard.edu/abs/2026arXiv260311815L}
}

@ARTICLE{Lee-2026c,
       author = {{Lee}, Max E. and {Genel}, Shy and {Haiman}, Zolt{\'a}n and {Bryan}, Greg L.},
        title = "{BINDing the lightcone: A Feedback Atlas for Stage-IV Weak Lensing and the tSZ Cross-Correlation}",
      journal = {arXiv e-prints},
         year = 2026,
         note = {in preparation}
}

@ARTICLE{Genel-2019,
       author = {{Genel}, Shy and {Bryan}, Greg L. and {Springel}, Volker and {Hernquist}, Lars and {Nelson}, Dylan and {Pillepich}, Annalisa and {Weinberger}, Rainer and {Pakmor}, Ruediger and {Marinacci}, Federico and {Vogelsberger}, Mark},
        title = "{A Quantification of the Butterfly Effect in Cosmological Simulations and Implications for Galaxy Scaling Relations}",
      journal = {\apj},
         year = 2019,
        month = jan,
       volume = {871},
       number = {1},
          eid = {21},
        pages = {21},
          doi = {10.3847/1538-4357/aaf4bb},
archivePrefix = {arXiv},
       eprint = {1807.07084},
 primaryClass = {astro-ph.GA},
       adsurl = {https://ui.adsabs.harvard.edu/abs/2019ApJ...871...21G}
}

@ARTICLE{Genel-2026,
       author = {{Genel}, Shy and {Jo}, Yongseok and {Oh}, Boon Kiat and {Tillman}, Megan Taylor and {Lee}, Max E. and {Lee}, Jun-Young and {Hern{\'a}ndez-Mart{\'\i}nez}, Elena and {Lovell}, Christopher C. and {Sims}, Xavier and {Burkhart}, Blakesley and {Nagamine}, Kentaro and {Angl{\'e}s-Alc{\'a}zar}, Daniel and {Villaescusa-Navarro}, Francisco},
        title = "{Learning the Universe with the 2nd Generation of CAMELS: Varying 35 parameters of the IllustrisTNG model in (50Mpc/h)\^3 boxes}",
      journal = {arXiv e-prints},
         year = 2026,
        month = jun,
          eid = {arXiv:2606.10038},
        pages = {arXiv:2606.10038},
          doi = {10.48550/arXiv.2606.10038},
archivePrefix = {arXiv},
       eprint = {2606.10038},
 primaryClass = {astro-ph.CO},
       adsurl = {https://ui.adsabs.harvard.edu/abs/2026arXiv260610038G}
}

@ARTICLE{VillaescusaNavarro-2021,
       author = {{Villaescusa-Navarro}, Francisco and {Angl{\'e}s-Alc{\'a}zar}, Daniel and {Genel}, Shy and others},
        title = "{The CAMELS Project: Cosmology and Astrophysics with Machine-learning Simulations}",
      journal = {\apj},
         year = 2021,
        month = jul,
       volume = {915},
       number = {1},
          eid = {71},
        pages = {71},
          doi = {10.3847/1538-4357/abf7ba},
archivePrefix = {arXiv},
       eprint = {2010.00619},
 primaryClass = {astro-ph.CO},
       adsurl = {https://ui.adsabs.harvard.edu/abs/2021ApJ...915...71V}
}

@ARTICLE{Vogelsberger-2014,
       author = {{Vogelsberger}, Mark and {Genel}, Shy and {Springel}, Volker and others},
        title = "{Introducing the Illustris Project: simulating the coevolution of dark and visible matter in the Universe}",
      journal = {\mnras},
         year = 2014,
        month = oct,
       volume = {444},
       number = {2},
        pages = {1518-1547},
          doi = {10.1093/mnras/stu1536},
archivePrefix = {arXiv},
       eprint = {1405.2921},
 primaryClass = {astro-ph.CO},
       adsurl = {https://ui.adsabs.harvard.edu/abs/2014MNRAS.444.1518V}
}

@ARTICLE{Schaye-2015,
       author = {{Schaye}, Joop and {Crain}, Robert A. and {Bower}, Richard G. and others},
        title = "{The EAGLE project: simulating the evolution and assembly of galaxies and their environments}",
      journal = {\mnras},
         year = 2015,
        month = jan,
       volume = {446},
       number = {1},
        pages = {521-554},
          doi = {10.1093/mnras/stu2058},
archivePrefix = {arXiv},
       eprint = {1407.7040},
 primaryClass = {astro-ph.GA},
       adsurl = {https://ui.adsabs.harvard.edu/abs/2015MNRAS.446..521S}
}

@ARTICLE{Pillepich-2018,
       author = {{Pillepich}, Annalisa and {Nelson}, Dylan and {Hernquist}, Lars and others},
        title = "{First results from the IllustrisTNG simulations: the stellar mass content of groups and clusters of galaxies}",
      journal = {\mnras},
         year = 2018,
        month = mar,
       volume = {475},
       number = {1},
        pages = {648-675},
          doi = {10.1093/mnras/stx3112},
archivePrefix = {arXiv},
       eprint = {1707.03406},
 primaryClass = {astro-ph.GA},
       adsurl = {https://ui.adsabs.harvard.edu/abs/2018MNRAS.475..648P}
}

@ARTICLE{Dave-2019,
       author = {{Dav{\'e}}, Romeel and {Angl{\'e}s-Alc{\'a}zar}, Daniel and {Narayanan}, Desika and {Li}, Qi and {Rafieferantsoa}, Mika H. and {Appleby}, Sarah},
        title = "{SIMBA: Cosmological simulations with black hole growth and feedback}",
      journal = {\mnras},
         year = 2019,
        month = jun,
       volume = {486},
       number = {2},
        pages = {2827-2849},
          doi = {10.1093/mnras/stz937},
archivePrefix = {arXiv},
       eprint = {1901.10203},
 primaryClass = {astro-ph.GA},
       adsurl = {https://ui.adsabs.harvard.edu/abs/2019MNRAS.486.2827D}
}

@ARTICLE{Schaye-2023,
       author = {{Schaye}, Joop and {Kugel}, Roi and {Schaller}, Matthieu and others},
        title = "{The FLAMINGO project: cosmological hydrodynamical simulations for large-scale structure and galaxy cluster surveys}",
      journal = {\mnras},
         year = 2023,
        month = dec,
       volume = {526},
       number = {4},
        pages = {4978-5020},
          doi = {10.1093/mnras/stad2419},
archivePrefix = {arXiv},
       eprint = {2306.04024},
 primaryClass = {astro-ph.CO},
       adsurl = {https://ui.adsabs.harvard.edu/abs/2023MNRAS.526.4978S}
}

@ARTICLE{Vogelsberger-2020,
       author = {{Vogelsberger}, Mark and {Marinacci}, Federico and {Torrey}, Paul and {Puchwein}, Ewald},
        title = "{Cosmological simulations of galaxy formation}",
      journal = {Nature Reviews Physics},
         year = 2020,
        month = jan,
       volume = {2},
       number = {1},
        pages = {42-66},
          doi = {10.1038/s42254-019-0127-2},
archivePrefix = {arXiv},
       eprint = {1909.07976},
 primaryClass = {astro-ph.GA},
       adsurl = {https://ui.adsabs.harvard.edu/abs/2020NatRP...2...42V}
}

@ARTICLE{springel2010arepo,
       author = {{Springel}, Volker},
        title = "{E pur si muove: Galilean-invariant cosmological hydrodynamical simulations on a moving mesh}",
      journal = {\mnras},
         year = 2010,
        month = jan,
       volume = {401},
       number = {2},
        pages = {791-851},
          doi = {10.1111/j.1365-2966.2009.15715.x},
archivePrefix = {arXiv},
       eprint = {0901.4107},
 primaryClass = {astro-ph.CO},
       adsurl = {https://ui.adsabs.harvard.edu/abs/2010MNRAS.401..791S}
}

@ARTICLE{weinberger2017supermassive,
       author = {{Weinberger}, Rainer and {Springel}, Volker and {Hernquist}, Lars and others},
        title = "{Simulating galaxy formation with black hole driven thermal and kinetic feedback}",
      journal = {\mnras},
         year = 2017,
        month = feb,
       volume = {465},
       number = {3},
        pages = {3291-3308},
          doi = {10.1093/mnras/stw2944},
archivePrefix = {arXiv},
       eprint = {1607.03486},
 primaryClass = {astro-ph.GA},
       adsurl = {https://ui.adsabs.harvard.edu/abs/2017MNRAS.465.3291W}
}

@ARTICLE{pillepich2018simulating,
       author = {{Pillepich}, Annalisa and {Springel}, Volker and {Nelson}, Dylan and others},
        title = "{Simulating galaxy formation with the IllustrisTNG model}",
      journal = {\mnras},
         year = 2018,
        month = jan,
       volume = {473},
       number = {3},
        pages = {4077-4106},
          doi = {10.1093/mnras/stx2656},
archivePrefix = {arXiv},
       eprint = {1703.02970},
 primaryClass = {astro-ph.GA},
       adsurl = {https://ui.adsabs.harvard.edu/abs/2018MNRAS.473.4077P}
}

@ARTICLE{Sobol-1967,
       author = {{Sobol'}, Ilya M.},
        title = "{On the distribution of points in a cube and the approximate evaluation of integrals}",
      journal = {USSR Computational Mathematics and Mathematical Physics},
         year = 1967,
       volume = {7},
       number = {4},
        pages = {86-112},
          doi = {10.1016/0041-5553(67)90144-9}
}

@ARTICLE{Jing-2006,
       author = {{Jing}, Y.~P. and {Zhang}, Pengjie and {Lin}, W.~P. and {Gao}, Liang and {Springel}, Volker},
        title = "{The Influence of Baryons on the Clustering of Matter and Weak-Lensing Surveys}",
      journal = {\apjl},
         year = 2006,
        month = apr,
       volume = {640},
       number = {2},
        pages = {L119-L122},
          doi = {10.1086/503547},
archivePrefix = {arXiv},
       eprint = {astro-ph/0512426},
       adsurl = {https://ui.adsabs.harvard.edu/abs/2006ApJ...640L.119J}
}

@ARTICLE{vanDaalen-2011,
       author = {{van Daalen}, Marcel P. and {Schaye}, Joop and {Booth}, C.~M. and {Dalla Vecchia}, Claudio},
        title = "{The effects of galaxy formation on the matter power spectrum: a challenge for precision cosmology}",
      journal = {\mnras},
         year = 2011,
        month = aug,
       volume = {415},
       number = {4},
        pages = {3649-3665},
          doi = {10.1111/j.1365-2966.2011.18981.x},
archivePrefix = {arXiv},
       eprint = {1104.1174},
 primaryClass = {astro-ph.CO},
       adsurl = {https://ui.adsabs.harvard.edu/abs/2011MNRAS.415.3649V}
}

@ARTICLE{vanDaalen-2020,
       author = {{van Daalen}, Marcel P. and {McCarthy}, Ian G. and {Schaye}, Joop},
        title = "{Exploring the effects of galaxy formation on matter clustering through a library of simulation power spectra}",
      journal = {\mnras},
         year = 2020,
        month = jan,
       volume = {491},
       number = {2},
        pages = {2424-2446},
          doi = {10.1093/mnras/stz3199},
archivePrefix = {arXiv},
       eprint = {1906.00968},
 primaryClass = {astro-ph.CO},
       adsurl = {https://ui.adsabs.harvard.edu/abs/2020MNRAS.491.2424V}
}

@ARTICLE{Chisari-2019,
       author = {{Chisari}, Nora Elisa and {Mead}, Alexander J. and {Joudaki}, Shahab and others},
        title = "{Modelling baryonic feedback for survey cosmology}",
      journal = {The Open Journal of Astrophysics},
         year = 2019,
        month = jun,
       volume = {2},
       number = {1},
          eid = {4},
        pages = {4},
          doi = {10.21105/astro.1905.06082},
archivePrefix = {arXiv},
       eprint = {1905.06082},
 primaryClass = {astro-ph.CO},
       adsurl = {https://ui.adsabs.harvard.edu/abs/2019OJAp....2E...4C}
}

@ARTICLE{Schneider-2015,
       author = {{Schneider}, Aurel and {Teyssier}, Romain},
        title = "{A new method to quantify the effects of baryons on the matter power spectrum}",
      journal = {\jcap},
         year = 2015,
        month = dec,
       volume = {2015},
       number = {12},
          eid = {049},
        pages = {049},
          doi = {10.1088/1475-7516/2015/12/049},
archivePrefix = {arXiv},
       eprint = {1510.06034},
 primaryClass = {astro-ph.CO},
       adsurl = {https://ui.adsabs.harvard.edu/abs/2015JCAP...12..049S}
}

@ARTICLE{Schneider-2019,
       author = {{Schneider}, Aurel and {Teyssier}, Romain and {Stadel}, Joachim and {Chisari}, Nora Elisa and {Le Brun}, Amandine M.~C. and {Amara}, Adam and {Refregier}, Alexandre},
        title = "{Quantifying baryon effects on the matter power spectrum and the weak lensing shear correlation}",
      journal = {\jcap},
         year = 2019,
        month = mar,
       volume = {2019},
       number = {3},
          eid = {020},
        pages = {020},
          doi = {10.1088/1475-7516/2019/03/020},
archivePrefix = {arXiv},
       eprint = {1810.08629},
 primaryClass = {astro-ph.CO},
       adsurl = {https://ui.adsabs.harvard.edu/abs/2019JCAP...03..020S}
}

@ARTICLE{Arico-2020,
       author = {{Aric{\`o}}, Giovanni and {Angulo}, Raul E. and {Hern{\'a}ndez-Monteagudo}, Carlos and {Contreras}, Sergio and {Zennaro}, Matteo and {Pellejero-Iba{\~n}ez}, Marcos and {Rosas-Guevara}, Yetli},
        title = "{Modelling the large-scale mass density field of the universe as a function of cosmology and baryonic physics}",
      journal = {\mnras},
         year = 2020,
        month = jul,
       volume = {495},
       number = {5},
        pages = {4800-4819},
          doi = {10.1093/mnras/staa1478},
archivePrefix = {arXiv},
       eprint = {1911.08471},
 primaryClass = {astro-ph.CO},
       adsurl = {https://ui.adsabs.harvard.edu/abs/2020MNRAS.495.4800A}
}

@ARTICLE{Arico-2021a,
       author = {{Aric{\`o}}, Giovanni and {Angulo}, Raul E. and {Contreras}, Sergio and {Ondaro-Mallea}, Lurdes and {Pellejero-Iba{\~n}ez}, Marcos and {Zennaro}, Matteo},
        title = "{Simultaneous modelling of matter power spectrum and bispectrum in the presence of baryons}",
      journal = {\mnras},
         year = 2021,
        month = may,
       volume = {503},
       number = {3},
        pages = {3596-3609},
          doi = {10.1093/mnras/stab699},
archivePrefix = {arXiv},
       eprint = {2009.14225},
 primaryClass = {astro-ph.CO},
       adsurl = {https://ui.adsabs.harvard.edu/abs/2021MNRAS.503.3596A}
}

@ARTICLE{Arico-2021,
       author = {{Aric{\`o}}, Giovanni and {Angulo}, Raul E. and {Contreras}, Sergio and {Ondaro-Mallea}, Lurdes and {Pellejero-Iba{\~n}ez}, Marcos and {Zennaro}, Matteo},
        title = "{The BACCO simulation project: a baryonification emulator with neural networks}",
      journal = {\mnras},
         year = 2021,
        month = sep,
       volume = {506},
       number = {3},
        pages = {4070-4082},
          doi = {10.1093/mnras/stab1911},
archivePrefix = {arXiv},
       eprint = {2011.15018},
 primaryClass = {astro-ph.CO},
       adsurl = {https://ui.adsabs.harvard.edu/abs/2021MNRAS.506.4070A}
}

@ARTICLE{Mead-2021,
       author = {{Mead}, Alexander J. and {Brieden}, Samuel and {Tr{\"o}ster}, Tilman and {Heymans}, Catherine},
        title = "{HMCODE-2020: improved modelling of non-linear cosmological power spectra with baryonic feedback}",
      journal = {\mnras},
         year = 2021,
        month = mar,
       volume = {502},
       number = {1},
        pages = {1401-1422},
          doi = {10.1093/mnras/stab082},
archivePrefix = {arXiv},
       eprint = {2009.01858},
 primaryClass = {astro-ph.CO},
       adsurl = {https://ui.adsabs.harvard.edu/abs/2021MNRAS.502.1401M}
}

@ARTICLE{Schneider-2025,
       author = {{Schneider}, Aurel and {Kova{\v{c}}}, Michael and {Bucko}, Jozef and {Nicola}, Andrina and {Reischke}, Robert and {Giri}, Sambit K. and {Teyssier}, Romain and {Tr{\"o}ster}, Tilman and {Refregier}, Alexandre and {Schaller}, Matthieu and {Schaye}, Joop},
        title = "{Baryonification: an alternative to hydrodynamical simulations for cosmological studies}",
      journal = {\jcap},
         year = 2025,
        month = dec,
       volume = {2025},
       number = {12},
          eid = {043},
        pages = {043},
          doi = {10.1088/1475-7516/2025/12/043},
archivePrefix = {arXiv},
       eprint = {2507.07892},
 primaryClass = {astro-ph.CO},
       adsurl = {https://ui.adsabs.harvard.edu/abs/2025JCAP...12..043S}
}

@ARTICLE{Kovac-2025,
       author = {{Kova{\v{c}}}, Michael and {Schneider}, Aurel and {Giri}, Sambit K. and others},
        title = "{Baryonification II: Constraining feedback with X-ray and kinematic Sunyaev-Zel'dovich observations}",
      journal = {\jcap},
         year = 2025,
        month = nov,
       volume = {2025},
       number = {11},
          eid = {046},
        pages = {046},
          doi = {10.1088/1475-7516/2025/11/046},
archivePrefix = {arXiv},
       eprint = {2507.07991},
 primaryClass = {astro-ph.CO},
       adsurl = {https://ui.adsabs.harvard.edu/abs/2025JCAP...11..046K}
}

@ARTICLE{Hadzhiyska-2024,
       author = {{Hadzhiyska}, Boryana and {Ferraro}, Simone and {Ried Guachalla}, Bernardita and others},
        title = "{Evidence for large baryonic feedback at low and intermediate redshifts from kinematic Sunyaev-Zel'dovich observations with ACT and DESI photometric galaxies}",
      journal = {arXiv e-prints},
         year = 2024,
        month = jul,
          eid = {arXiv:2407.07152},
        pages = {arXiv:2407.07152},
          doi = {10.48550/arXiv.2407.07152},
archivePrefix = {arXiv},
       eprint = {2407.07152},
 primaryClass = {astro-ph.CO},
       adsurl = {https://ui.adsabs.harvard.edu/abs/2024arXiv240707152H}
}

@ARTICLE{Bigwood-2024,
       author = {{Bigwood}, L. and {Amon}, A. and {Schneider}, A. and others},
        title = "{Weak lensing combined with the kinetic Sunyaev-Zel'dovich effect: a study of baryonic feedback}",
      journal = {\mnras},
         year = 2024,
        month = oct,
       volume = {534},
       number = {1},
        pages = {655-682},
          doi = {10.1093/mnras/stae2100},
archivePrefix = {arXiv},
       eprint = {2404.06098},
 primaryClass = {astro-ph.CO},
       adsurl = {https://ui.adsabs.harvard.edu/abs/2024MNRAS.534..655B}
}

@ARTICLE{Siegel-2025,
       author = {{Siegel}, Jared and {Amon}, Alexandra and {McCarthy}, Ian G. and {Bigwood}, Leah and {Yamamoto}, Masaya and others},
        title = "{Joint X-ray, kinetic Sunyaev-Zeldovich, and weak lensing measurements: toward a consensus picture of efficient gas expulsion from groups and clusters}",
      journal = {arXiv e-prints},
         year = 2025,
        month = sep,
          eid = {arXiv:2509.10455},
        pages = {arXiv:2509.10455},
          doi = {10.48550/arXiv.2509.10455},
archivePrefix = {arXiv},
       eprint = {2509.10455},
 primaryClass = {astro-ph.CO},
       adsurl = {https://ui.adsabs.harvard.edu/abs/2025arXiv250910455S}
}

@ARTICLE{Secco-2022,
       author = {{Secco}, L.~F. and {Samuroff}, S. and {Krause}, E. and others},
        title = "{Dark Energy Survey Year 3 results: Cosmology from cosmic shear and robustness to modeling uncertainty}",
      journal = {\prd},
         year = 2022,
        month = jan,
       volume = {105},
       number = {2},
          eid = {023515},
        pages = {023515},
          doi = {10.1103/PhysRevD.105.023515},
archivePrefix = {arXiv},
       eprint = {2105.13544},
 primaryClass = {astro-ph.CO},
       adsurl = {https://ui.adsabs.harvard.edu/abs/2022PhRvD.105b3515S}
}

@ARTICLE{Li-2023,
       author = {{Li}, Xiangchong and {Zhang}, Tianqing and {Sugiyama}, Sunao and others},
        title = "{Hyper Suprime-Cam Year 3 results: Cosmology from cosmic shear two-point correlation functions}",
      journal = {\prd},
         year = 2023,
        month = dec,
       volume = {108},
       number = {12},
          eid = {123518},
        pages = {123518},
          doi = {10.1103/PhysRevD.108.123518},
archivePrefix = {arXiv},
       eprint = {2304.00702},
 primaryClass = {astro-ph.CO},
       adsurl = {https://ui.adsabs.harvard.edu/abs/2023PhRvD.108l3518L}
}

@ARTICLE{Dalal-2023,
       author = {{Dalal}, Roohi and {Li}, Xiangchong and {Nicola}, Andrina and others},
        title = "{Hyper Suprime-Cam Year 3 results: Cosmology from cosmic shear power spectra}",
      journal = {\prd},
         year = 2023,
        month = dec,
       volume = {108},
       number = {12},
          eid = {123519},
        pages = {123519},
          doi = {10.1103/PhysRevD.108.123519},
archivePrefix = {arXiv},
       eprint = {2304.00701},
 primaryClass = {astro-ph.CO},
       adsurl = {https://ui.adsabs.harvard.edu/abs/2023PhRvD.108l3519D}
}

@ARTICLE{Sharma-2024,
       author = {{Sharma}, Divij and {Dai}, Biwei and {Villaescusa-Navarro}, Francisco and {Seljak}, Uro{\v{s}}},
        title = "{A field-level emulator for modelling baryonic effects across hydrodynamic simulations}",
      journal = {\mnras},
         year = 2025,
        month = apr,
       volume = {538},
       number = {3},
        pages = {1415-1426},
          doi = {10.1093/mnras/staf294},
archivePrefix = {arXiv},
       eprint = {2401.15891},
 primaryClass = {astro-ph.CO},
       adsurl = {https://ui.adsabs.harvard.edu/abs/2025MNRAS.538.1415S}
}

@ARTICLE{Chadayammuri-2023,
       author = {{Chadayammuri}, Urmila and {Ntampaka}, Michelle and {ZuHone}, John and {Bogd{\'a}n}, {\'A}kos and {Kraft}, Ralph P.},
        title = "{Painting baryons onto N-body simulations of galaxy clusters with image-to-image deep learning}",
      journal = {\mnras},
         year = 2023,
        month = dec,
       volume = {526},
       number = {2},
        pages = {2812-2829},
          doi = {10.1093/mnras/stad2596},
archivePrefix = {arXiv},
       eprint = {2307.16733},
 primaryClass = {astro-ph.CO},
       adsurl = {https://ui.adsabs.harvard.edu/abs/2023MNRAS.526.2812C}
}

@ARTICLE{Moews-2021,
       author = {{Moews}, Ben and {Dav{\'e}}, Romeel and {Mitchell}, Sourav and others},
        title = "{Hybrid analytic and machine-learned baryonic property insertion into galactic dark matter haloes}",
      journal = {\mnras},
         year = 2021,
        month = jul,
       volume = {504},
       number = {3},
        pages = {4024-4038},
          doi = {10.1093/mnras/stab1120},
archivePrefix = {arXiv},
       eprint = {2012.05820},
 primaryClass = {astro-ph.GA},
       adsurl = {https://ui.adsabs.harvard.edu/abs/2021MNRAS.504.4024M}
}

@ARTICLE{Mishra-2026,
       author = {{Mishra}, A. and others},
        title = "{Cosmo-FOLD: Fast generation and upscaling of field-level cosmological maps with overlap latent diffusion}",
      journal = {arXiv e-prints},
         year = 2026,
        month = jan,
          eid = {arXiv:2601.14377},
        pages = {arXiv:2601.14377},
          doi = {10.48550/arXiv.2601.14377},
archivePrefix = {arXiv},
       eprint = {2601.14377},
 primaryClass = {astro-ph.CO},
       adsurl = {https://ui.adsabs.harvard.edu/abs/2026arXiv260114377M}
}

@ARTICLE{troxel2015intrinsic,
       author = {{Troxel}, Michael A. and {Ishak}, Mustapha},
        title = "{The intrinsic alignment of galaxies and its impact on weak gravitational lensing in an era of precision cosmology}",
      journal = {\physrep},
         year = 2015,
        month = feb,
       volume = {558},
        pages = {1-59},
          doi = {10.1016/j.physrep.2014.11.001},
archivePrefix = {arXiv},
       eprint = {1407.6990},
 primaryClass = {astro-ph.CO},
       adsurl = {https://ui.adsabs.harvard.edu/abs/2015PhR...558....1T}
}

@ARTICLE{joachimi2015intrinsic,
       author = {{Joachimi}, Benjamin and {Cacciato}, Marcello and {Kitching}, Thomas D. and others},
        title = "{Galaxy Alignments: An Overview}",
      journal = {\ssr},
         year = 2015,
        month = nov,
       volume = {193},
       number = {1-4},
        pages = {1-65},
          doi = {10.1007/s11214-015-0177-4},
archivePrefix = {arXiv},
       eprint = {1504.05456},
 primaryClass = {astro-ph.GA},
       adsurl = {https://ui.adsabs.harvard.edu/abs/2015SSRv..193....1J}
}

@INPROCEEDINGS{lipman2023flowmatching,
       author = {{Lipman}, Yaron and {Chen}, Ricky T.~Q. and {Ben-Hamu}, Heli and {Nickel}, Maximilian and {Le}, Matt},
        title = "{Flow Matching for Generative Modeling}",
    booktitle = {The Eleventh International Conference on Learning Representations (ICLR)},
         year = 2023,
          doi = {10.48550/arXiv.2210.02747},
archivePrefix = {arXiv},
       eprint = {2210.02747},
 primaryClass = {cs.LG}
}

@INPROCEEDINGS{ronneberger2015unet,
       author = {{Ronneberger}, Olaf and {Fischer}, Philipp and {Brox}, Thomas},
        title = "{U-Net: Convolutional Networks for Biomedical Image Segmentation}",
    booktitle = {Medical Image Computing and Computer-Assisted Intervention (MICCAI)},
         year = 2015,
        pages = {234-241},
          doi = {10.1007/978-3-319-24574-4_28},
archivePrefix = {arXiv},
       eprint = {1505.04597},
 primaryClass = {cs.CV}
}

@INPROCEEDINGS{dhariwal2021diffusion,
       author = {{Dhariwal}, Prafulla and {Nichol}, Alexander},
        title = "{Diffusion Models Beat GANs on Image Synthesis}",
    booktitle = {Advances in Neural Information Processing Systems 34 (NeurIPS)},
         year = 2021,
          doi = {10.48550/arXiv.2105.05233},
archivePrefix = {arXiv},
       eprint = {2105.05233},
 primaryClass = {cs.LG}
}

@INPROCEEDINGS{karras2022edm,
       author = {{Karras}, Tero and {Aittala}, Miika and {Aila}, Timo and {Laine}, Samuli},
        title = "{Elucidating the Design Space of Diffusion-Based Generative Models}",
    booktitle = {Advances in Neural Information Processing Systems 35 (NeurIPS)},
         year = 2022,
          doi = {10.48550/arXiv.2206.00364},
archivePrefix = {arXiv},
       eprint = {2206.00364},
 primaryClass = {cs.CV}
}

@INPROCEEDINGS{loshchilov2019decoupled,
       author = {{Loshchilov}, Ilya and {Hutter}, Frank},
        title = "{Decoupled Weight Decay Regularization}",
    booktitle = {The Seventh International Conference on Learning Representations (ICLR)},
         year = 2019,
          doi = {10.48550/arXiv.1711.05101},
archivePrefix = {arXiv},
       eprint = {1711.05101},
 primaryClass = {cs.LG}
}

@INPROCEEDINGS{loshchilov2017sgdr,
       author = {{Loshchilov}, Ilya and {Hutter}, Frank},
        title = "{SGDR: Stochastic Gradient Descent with Warm Restarts}",
    booktitle = {The Fifth International Conference on Learning Representations (ICLR)},
         year = 2017,
          doi = {10.48550/arXiv.1608.03983},
archivePrefix = {arXiv},
       eprint = {1608.03983},
 primaryClass = {cs.LG}
}

@INPROCEEDINGS{wu2018groupnorm,
       author = {{Wu}, Yuxin and {He}, Kaiming},
        title = "{Group Normalization}",
    booktitle = {Proceedings of the European Conference on Computer Vision (ECCV)},
         year = 2018,
          doi = {10.48550/arXiv.1803.08494},
archivePrefix = {arXiv},
       eprint = {1803.08494},
 primaryClass = {cs.CV}
}

@BOOK{hockney,
       author = {{Hockney}, R.~W. and {Eastwood}, J.~W.},
        title = "{Computer Simulation Using Particles}",
    publisher = {Bristol: Hilger},
         year = 1988,
       adsurl = {https://ui.adsabs.harvard.edu/abs/1988csup.book.....H}
}

@MISC{Pylians,
       author = {{Villaescusa-Navarro}, Francisco},
        title = "{Pylians: Python libraries for the analysis of numerical simulations}",
 howpublished = {Astrophysics Source Code Library, record ascl:1811.008},
         year = 2018,
        month = nov,
          eid = {ascl:1811.008},
       adsurl = {https://ui.adsabs.harvard.edu/abs/2018ascl.soft11008V}
}

@ARTICLE{2022ApJ...935..167A,
       author = {{Astropy Collaboration} and {Price-Whelan}, Adrian M. and {Lim}, Pey Lian and others},
        title = "{The Astropy Project: Sustaining and Growing a Community-oriented Open-source Project and the Latest Major Release (v5.0) of the Core Package}",
      journal = {\apj},
         year = 2022,
        month = aug,
       volume = {935},
       number = {2},
          eid = {167},
        pages = {167},
          doi = {10.3847/1538-4357/ac7c74},
archivePrefix = {arXiv},
       eprint = {2206.14220},
 primaryClass = {astro-ph.IM},
       adsurl = {https://ui.adsabs.harvard.edu/abs/2022ApJ...935..167A}
}

@ARTICLE{2018AJ....156..123A,
       author = {{Astropy Collaboration} and {Price-Whelan}, A.~M. and {Sip{\H{o}}cz}, B.~M. and others},
        title = "{The Astropy Project: Building an Open-science Project and Status of the v2.0 Core Package}",
      journal = {\aj},
         year = 2018,
        month = sep,
       volume = {156},
       number = {3},
          eid = {123},
        pages = {123},
          doi = {10.3847/1538-3881/aabc4f},
archivePrefix = {arXiv},
       eprint = {1801.02634},
 primaryClass = {astro-ph.IM},
       adsurl = {https://ui.adsabs.harvard.edu/abs/2018AJ....156..123A}
}

@ARTICLE{2013A&A...558A..33A,
       author = {{Astropy Collaboration} and {Robitaille}, Thomas P. and {Tollerud}, Erik J. and others},
        title = "{Astropy: A community Python package for astronomy}",
      journal = {\aap},
         year = 2013,
        month = oct,
       volume = {558},
          eid = {A33},
        pages = {A33},
          doi = {10.1051/0004-6361/201322068},
archivePrefix = {arXiv},
       eprint = {1307.6212},
 primaryClass = {astro-ph.IM},
       adsurl = {https://ui.adsabs.harvard.edu/abs/2013A&A...558A..33A}
}

@INPROCEEDINGS{perez2018film,
       author = {{Perez}, Ethan and {Strub}, Florian and {de Vries}, Harm and {Dumoulin}, Vincent and {Courville}, Aaron C.},
        title = "{FiLM: Visual Reasoning with a General Conditioning Layer}",
    booktitle = {Proceedings of the AAAI Conference on Artificial Intelligence},
         year = 2018,
          doi = {10.48550/arXiv.1709.07871},
archivePrefix = {arXiv},
       eprint = {1709.07871},
 primaryClass = {cs.CV}
}

@INPROCEEDINGS{vaswani2017attention,
       author = {{Vaswani}, Ashish and {Shazeer}, Noam and {Parmar}, Niki and {Uszkoreit}, Jakob and {Jones}, Llion and {Gomez}, Aidan N. and {Kaiser}, Lukasz and {Polosukhin}, Illia},
        title = "{Attention Is All You Need}",
    booktitle = {Advances in Neural Information Processing Systems 30 (NeurIPS)},
         year = 2017,
        pages = {5998-6008},
          doi = {10.48550/arXiv.1706.03762},
archivePrefix = {arXiv},
       eprint = {1706.03762},
 primaryClass = {cs.CL}
}

@ARTICLE{ho2022cfg,
       author = {{Ho}, Jonathan and {Salimans}, Tim},
        title = "{Classifier-Free Diffusion Guidance}",
      journal = {arXiv e-prints},
         year = 2022,
        month = jul,
          eid = {arXiv:2207.12598},
        pages = {arXiv:2207.12598},
          doi = {10.48550/arXiv.2207.12598},
archivePrefix = {arXiv},
       eprint = {2207.12598},
 primaryClass = {cs.LG}
}

@INPROCEEDINGS{albergo2023interpolants,
       author = {{Albergo}, Michael S. and {Vanden-Eijnden}, Eric},
        title = "{Building Normalizing Flows with Stochastic Interpolants}",
    booktitle = {The Eleventh International Conference on Learning Representations (ICLR)},
         year = 2023,
          doi = {10.48550/arXiv.2209.15571},
archivePrefix = {arXiv},
       eprint = {2209.15571},
 primaryClass = {cs.LG}
}

@INPROCEEDINGS{liu2022rectified,
       author = {{Liu}, Xingchao and {Gong}, Chengyue and {Liu}, Qiang},
        title = "{Flow Straight and Fast: Learning to Generate and Transfer Data with Rectified Flow}",
    booktitle = {The Eleventh International Conference on Learning Representations (ICLR)},
         year = 2023,
          doi = {10.48550/arXiv.2209.03003},
archivePrefix = {arXiv},
       eprint = {2209.03003},
 primaryClass = {cs.LG}
}

@INPROCEEDINGS{paszke2019pytorch,
       author = {{Paszke}, Adam and {Gross}, Sam and {Massa}, Francisco and {Lerer}, Adam and {Bradbury}, James and {Chanan}, Gregory and {Killeen}, Trevor and {Lin}, Zeming and {Gimelshein}, Natalia and {Antiga}, Luca and others},
        title = "{PyTorch: An Imperative Style, High-Performance Deep Learning Library}",
    booktitle = {Advances in Neural Information Processing Systems 32 (NeurIPS)},
         year = 2019,
        pages = {8024-8035},
          doi = {10.48550/arXiv.1912.01703},
archivePrefix = {arXiv},
       eprint = {1912.01703},
 primaryClass = {cs.LG}
}

@ARTICLE{Davis-1985,
       author = {{Davis}, M. and {Efstathiou}, G. and {Frenk}, C.~S. and {White}, S.~D.~M.},
        title = "{The evolution of large-scale structure in a universe dominated by cold dark matter}",
      journal = {\apj},
         year = 1985,
        month = may,
       volume = {292},
        pages = {371-394},
          doi = {10.1086/163168},
       adsurl = {https://ui.adsabs.harvard.edu/abs/1985ApJ...292..371D}
}

@ARTICLE{Springel-2001,
       author = {{Springel}, Volker and {White}, Simon D.~M. and {Tormen}, Giuseppe and {Kauffmann}, Guinevere},
        title = "{Populating a cluster of galaxies - I. Results at z=0}",
      journal = {\mnras},
         year = 2001,
        month = dec,
       volume = {328},
       number = {3},
        pages = {726-750},
          doi = {10.1046/j.1365-8711.2001.04912.x},
archivePrefix = {arXiv},
       eprint = {astro-ph/0012055},
       adsurl = {https://ui.adsabs.harvard.edu/abs/2001MNRAS.328..726S}
}

@ARTICLE{Nelson-2019,
       author = {{Nelson}, Dylan and {Springel}, Volker and {Pillepich}, Annalisa and others},
        title = "{The IllustrisTNG simulations: public data release}",
      journal = {Computational Astrophysics and Cosmology},
         year = 2019,
        month = may,
       volume = {6},
       number = {1},
          eid = {2},
        pages = {2},
          doi = {10.1186/s40668-019-0028-x},
archivePrefix = {arXiv},
       eprint = {1812.05609},
 primaryClass = {astro-ph.GA},
       adsurl = {https://ui.adsabs.harvard.edu/abs/2019ComAC...6....2N}
}

@ARTICLE{Ivezic-2019,
       author = {{Ivezi{\'c}}, {\v{Z}}eljko and {Kahn}, Steven M. and {Tyson}, J. Anthony and others},
        title = "{LSST: From Science Drivers to Reference Design and Anticipated Data Products}",
      journal = {\apj},
         year = 2019,
        month = mar,
       volume = {873},
       number = {2},
          eid = {111},
        pages = {111},
          doi = {10.3847/1538-4357/ab042c},
archivePrefix = {arXiv},
       eprint = {0805.2366},
 primaryClass = {astro-ph.IM},
       adsurl = {https://ui.adsabs.harvard.edu/abs/2019ApJ...873..111I}
}

@ARTICLE{Laureijs-2011,
       author = {{Laureijs}, R. and {Amiaux}, J. and {Arduini}, S. and others},
        title = "{Euclid Definition Study Report}",
      journal = {arXiv e-prints},
         year = 2011,
        month = oct,
          eid = {arXiv:1110.3193},
        pages = {arXiv:1110.3193},
          doi = {10.48550/arXiv.1110.3193},
archivePrefix = {arXiv},
       eprint = {1110.3193},
 primaryClass = {astro-ph.CO},
       adsurl = {https://ui.adsabs.harvard.edu/abs/2011arXiv1110.3193L}
}

@ARTICLE{Spergel-2015,
       author = {{Spergel}, D. and {Gehrels}, N. and {Baltay}, C. and others},
        title = "{Wide-Field InfraRed Survey Telescope-Astrophysics Focused Telescope Assets WFIRST-AFTA 2015 Report}",
      journal = {arXiv e-prints},
         year = 2015,
        month = mar,
          eid = {arXiv:1503.03757},
        pages = {arXiv:1503.03757},
          doi = {10.48550/arXiv.1503.03757},
archivePrefix = {arXiv},
       eprint = {1503.03757},
 primaryClass = {astro-ph.IM},
       adsurl = {https://ui.adsabs.harvard.edu/abs/2015arXiv150303757S}
}

@ARTICLE{WechslerTinker-2018,
       author = {{Wechsler}, Risa H. and {Tinker}, Jeremy L.},
        title = "{The Connection Between Galaxies and Their Dark Matter Halos}",
      journal = {\araa},
         year = 2018,
        month = sep,
       volume = {56},
        pages = {435-487},
          doi = {10.1146/annurev-astro-081817-051756},
archivePrefix = {arXiv},
       eprint = {1804.03097},
 primaryClass = {astro-ph.GA},
       adsurl = {https://ui.adsabs.harvard.edu/abs/2018ARA&A..56..435W}
}

@ARTICLE{Velliscig-2015,
       author = {{Velliscig}, Marco and {Cacciato}, Marcello and {Schaye}, Joop and {Crain}, Robert A. and {Bower}, Richard G. and {van Daalen}, Marcel P. and {Dalla Vecchia}, Claudio and {Frenk}, Carlos S. and {Furlong}, Michelle and {McCarthy}, I.~G. and {Schaller}, Matthieu and {Theuns}, Tom},
        title = "{The alignment and shape of dark matter, stellar, and hot gas distributions in the EAGLE and cosmo-OWLS simulations}",
      journal = {\mnras},
         year = 2015,
        month = oct,
       volume = {453},
       number = {1},
        pages = {721-738},
          doi = {10.1093/mnras/stv1690},
archivePrefix = {arXiv},
       eprint = {1504.04025},
 primaryClass = {astro-ph.CO},
       adsurl = {https://ui.adsabs.harvard.edu/abs/2015MNRAS.453..721V}
}

@ARTICLE{Lovisari-2021,
       author = {{Lovisari}, Lorenzo and {Ettori}, Stefano and {Gaspari}, Massimo and {Giles}, Paul A.},
        title = "{Scaling Properties of Galaxy Groups}",
      journal = {Universe},
         year = 2021,
        month = may,
       volume = {7},
       number = {5},
          eid = {139},
        pages = {139},
          doi = {10.3390/universe7050139},
archivePrefix = {arXiv},
       eprint = {2106.13256},
 primaryClass = {astro-ph.CO},
       adsurl = {https://ui.adsabs.harvard.edu/abs/2021Univ....7..139L}
}

@ARTICLE{Anbajagane-2024,
       author = {{Anbajagane}, Dhayaa and {Pandey}, Shivam and {Chang}, Chihway},
        title = "{Map-level baryonification: Efficient modelling of higher-order correlations in the weak lensing and thermal Sunyaev-Zeldovich fields}",
      journal = {The Open Journal of Astrophysics},
         year = 2024,
        month = dec,
       volume = {7},
          eid = {108},
        pages = {108},
          doi = {10.33232/001c.126788},
archivePrefix = {arXiv},
       eprint = {2409.03822},
 primaryClass = {astro-ph.CO},
       adsurl = {https://ui.adsabs.harvard.edu/abs/2024OJAp....7E.108A}
}

@ARTICLE{Zhou-2025,
       author = {{Zhou}, Alan Junzhe and {Gatti}, Marco and {Anbajagane}, Dhayaa and {Dodelson}, Scott and {Schaller}, Matthieu and {Schaye}, Joop},
        title = "{Map-level baryonification: unified treatment of weak lensing two-point and higher-order statistics}",
      journal = {\jcap},
         year = 2025,
        month = sep,
       volume = {2025},
       number = {9},
          eid = {073},
        pages = {073},
          doi = {10.1088/1475-7516/2025/09/073},
archivePrefix = {arXiv},
       eprint = {2505.07949},
 primaryClass = {astro-ph.CO},
       adsurl = {https://ui.adsabs.harvard.edu/abs/2025arXiv250507949Z}
}

@ARTICLE{Lee-2026,
       author = {{Lee}, Max E. and {Haiman}, Zolt{\'a}n and {Pandey}, Shivam and {Genel}, Shy},
        title = "{The Effect of Intrinsic Alignments on Weak-lensing Statistics in Hydrodynamical Simulations}",
      journal = {\apj},
         year = 2026,
        month = jan,
       volume = {996},
       number = {1},
          eid = {36},
        pages = {36},
          doi = {10.3847/1538-4357/ae1ca7},
archivePrefix = {arXiv},
       eprint = {2504.12460},
 primaryClass = {astro-ph.CO},
       adsurl = {https://ui.adsabs.harvard.edu/abs/2026ApJ...996...36L}
}

@ARTICLE{Ono-2024,
       author = {{Ono}, Victoria and {Park}, Core Francisco and {Mudur}, Nayantara and {Ni}, Yueying and {Cuesta-Lazaro}, Carolina and {Villaescusa-Navarro}, Francisco},
        title = "{Debiasing with Diffusion: Probabilistic Reconstruction of Dark Matter Fields from Galaxies with CAMELS}",
      journal = {arXiv e-prints},
         year = 2024,
        month = mar,
          eid = {arXiv:2403.10648},
        pages = {arXiv:2403.10648},
          doi = {10.48550/arXiv.2403.10648},
archivePrefix = {arXiv},
       eprint = {2403.10648},
 primaryClass = {astro-ph.CO},
       adsurl = {https://ui.adsabs.harvard.edu/abs/2024arXiv240310648O}
}

@ARTICLE{Mudur-2024,
       author = {{Mudur}, Nayantara and {Cuesta-Lazaro}, Carolina and {Finkbeiner}, Douglas P.},
        title = "{Diffusion-HMC: Parameter Inference with Diffusion-model-driven Hamiltonian Monte Carlo}",
      journal = {arXiv e-prints},
         year = 2024,
        month = may,
          eid = {arXiv:2405.05255},
        pages = {arXiv:2405.05255},
          doi = {10.48550/arXiv.2405.05255},
archivePrefix = {arXiv},
       eprint = {2405.05255},
 primaryClass = {astro-ph.CO},
       adsurl = {https://ui.adsabs.harvard.edu/abs/2024arXiv240505255M}
}

@ARTICLE{Pandey-2025,
       author = {{Pandey}, Shivam and {Lovell}, Christopher C. and {Modi}, Chirag and {Wandelt}, Benjamin D.},
        title = "{Galactification: painting galaxies onto dark matter only simulations using a transformer-based model}",
      journal = {arXiv e-prints},
         year = 2025,
        month = nov,
          eid = {arXiv:2511.08438},
        pages = {arXiv:2511.08438},
          doi = {10.48550/arXiv.2511.08438},
archivePrefix = {arXiv},
       eprint = {2511.08438},
 primaryClass = {astro-ph.CO},
       adsurl = {https://ui.adsabs.harvard.edu/abs/2025arXiv251108438P}
}

@ARTICLE{Hsu-2024,
       author = {{Hsu}, Alan and {Ho}, Matthew and {Lin}, Joyce and {Markey}, Carleen and {Ntampaka}, Michelle and {Trac}, Hy and {P{\'o}czos}, Barnab{\'a}s},
        title = "{Reconstructing Galaxy Cluster Mass Maps using Score-based Generative Modeling}",
      journal = {arXiv e-prints},
         year = 2024,
        month = oct,
          eid = {arXiv:2410.02857},
        pages = {arXiv:2410.02857},
          doi = {10.48550/arXiv.2410.02857},
archivePrefix = {arXiv},
       eprint = {2410.02857},
 primaryClass = {astro-ph.CO},
       adsurl = {https://ui.adsabs.harvard.edu/abs/2024arXiv241002857H}
}

@ARTICLE{Adam-2022,
       author = {{Adam}, Alexandre and {Coogan}, Adam and {Malkin}, Nikolay and {Legin}, Ronan and {Perreault-Levasseur}, Laurence and {Hezaveh}, Yashar and {Bengio}, Yoshua},
        title = "{Posterior samples of source galaxies in strong gravitational lenses with score-based priors}",
      journal = {arXiv e-prints},
         year = 2022,
        month = nov,
          eid = {arXiv:2211.03812},
        pages = {arXiv:2211.03812},
          doi = {10.48550/arXiv.2211.03812},
archivePrefix = {arXiv},
       eprint = {2211.03812},
 primaryClass = {astro-ph.IM},
       adsurl = {https://ui.adsabs.harvard.edu/abs/2022arXiv221103812A}
}

@ARTICLE{Medlock-2024,
       author = {{Medlock}, Isabel and {Nagai}, Daisuke and {Singh}, Priyanka and {Oppenheimer}, Benjamin and {Angl{\'e}s-Alc{\'a}zar}, Daniel and {Villaescusa-Navarro}, Francisco},
        title = "{Probing the Circumgalactic Medium with Fast Radio Bursts: Insights from CAMELS}",
      journal = {arXiv e-prints},
         year = 2024,
        month = mar,
          eid = {arXiv:2403.02313},
        pages = {arXiv:2403.02313},
          doi = {10.48550/arXiv.2403.02313},
archivePrefix = {arXiv},
       eprint = {2403.02313},
 primaryClass = {astro-ph.CO},
       adsurl = {https://ui.adsabs.harvard.edu/abs/2024arXiv240302313M}
}

@ARTICLE{Medlock-2024b,
       author = {{Medlock}, Isabel and {Neufeld}, Chloe and {Nagai}, Daisuke and {Angl{\'e}s-Alc{\'a}zar}, Daniel and {Genel}, Shy and {Oppenheimer}, Benjamin D. and {Sims}, Xavier and {Singh}, Priyanka and {Villaescusa-Navarro}, Francisco},
        title = "{Quantifying Baryonic Feedback on the Warm-Hot Circumgalactic Medium in CAMELS Simulations}",
      journal = {arXiv e-prints},
         year = 2024,
        month = oct,
          eid = {arXiv:2410.16361},
        pages = {arXiv:2410.16361},
          doi = {10.48550/arXiv.2410.16361},
archivePrefix = {arXiv},
       eprint = {2410.16361},
 primaryClass = {astro-ph.GA},
       adsurl = {https://ui.adsabs.harvard.edu/abs/2024arXiv241016361M}
}

@ARTICLE{Medlock-2025,
       author = {{Medlock}, Isabel and {Nagai}, Daisuke and {Angl{\'e}s-Alc{\'a}zar}, Daniel and {Gebhardt}, Matthew},
        title = "{Constraining Baryonic Feedback Effects on the Matter Power Spectrum with Fast Radio Bursts}",
      journal = {arXiv e-prints},
         year = 2025,
        month = jan,
          eid = {arXiv:2501.17922},
        pages = {arXiv:2501.17922},
          doi = {10.48550/arXiv.2501.17922},
archivePrefix = {arXiv},
       eprint = {2501.17922},
 primaryClass = {astro-ph.CO},
       adsurl = {https://ui.adsabs.harvard.edu/abs/2025arXiv250117922M}
}

@ARTICLE{Lau-2025,
       author = {{Lau}, Erwin T. and {Nagai}, Daisuke and {Bogd{\'a}n}, {\'A}kos and {Medlock}, Isabel and {Oppenheimer}, Benjamin D. and {Battaglia}, Nicholas and {Angl{\'e}s-Alc{\'a}zar}, Daniel and {Genel}, Shy and {Ni}, Yueying and {Villaescusa-Navarro}, Francisco},
        title = "{X-raying CAMELS: Constraining Baryonic Feedback in the Circum-Galactic Medium with the CAMELS Simulations and eRASS X-ray Observations}",
      journal = {arXiv e-prints},
         year = 2024,
        month = dec,
          eid = {arXiv:2412.04559},
        pages = {arXiv:2412.04559},
          doi = {10.48550/arXiv.2412.04559},
archivePrefix = {arXiv},
       eprint = {2412.04559},
 primaryClass = {astro-ph.GA},
       adsurl = {https://ui.adsabs.harvard.edu/abs/2024arXiv241204559L}
}

@ARTICLE{Gebhardt-2026,
       author = {{Gebhardt}, Matthew and {Angl{\'e}s-Alc{\'a}zar}, Daniel and {Genel}, Shy and {Nagai}, Daisuke and {Oh}, Boon Kiat and {Medlock}, Isabel and {Mercedes-Feliz}, Jonathan and {Sutherland}, Sagan and {Lee}, Max E. and {Sims}, Xavier and {Lovell}, Christopher C. and {Spergel}, David N. and {Dav{\'e}}, Romeel and {Schaller}, Matthieu and {Schaye}, Joop and {Villaescusa-Navarro}, Francisco},
        title = "{Cosmological back-reaction of baryons on dark matter in the CAMELS simulations}",
      journal = {arXiv e-prints},
         year = 2026,
        month = jan,
          eid = {arXiv:2601.06258},
        pages = {arXiv:2601.06258},
          doi = {10.48550/arXiv.2601.06258},
archivePrefix = {arXiv},
       eprint = {2601.06258},
 primaryClass = {astro-ph.CO},
       adsurl = {https://ui.adsabs.harvard.edu/abs/2026arXiv260106258G}
}

@INPROCEEDINGS{ho2020ddpm,
       author = {{Ho}, Jonathan and {Jain}, Ajay and {Abbeel}, Pieter},
        title = "{Denoising Diffusion Probabilistic Models}",
    booktitle = {Advances in Neural Information Processing Systems},
       volume = {33},
         year = 2020,
        pages = {6840--6851},
          doi = {10.48550/arXiv.2006.11239},
archivePrefix = {arXiv},
       eprint = {2006.11239},
 primaryClass = {cs.LG},
       adsurl = {https://ui.adsabs.harvard.edu/abs/2020arXiv200611239H}
}

@INPROCEEDINGS{song2021sde,
       author = {{Song}, Yang and {Sohl-Dickstein}, Jascha and {Kingma}, Diederik P. and {Kumar}, Abhishek and {Ermon}, Stefano and {Poole}, Ben},
        title = "{Score-Based Generative Modeling through Stochastic Differential Equations}",
    booktitle = {International Conference on Learning Representations (ICLR)},
         year = 2021,
          doi = {10.48550/arXiv.2011.13456},
archivePrefix = {arXiv},
       eprint = {2011.13456},
 primaryClass = {cs.LG},
       adsurl = {https://ui.adsabs.harvard.edu/abs/2020arXiv201113456S}
}

@ARTICLE{CuestaLazaro-2024,
       author = {{Cuesta-Lazaro}, Carolina and {Mishra-Sharma}, Siddharth},
        title = "{Point cloud approach to generative modeling for galaxy surveys at the field level}",
      journal = {\prd},
         year = 2024,
        month = jun,
       volume = {109},
       number = {12},
          eid = {123531},
        pages = {123531},
          doi = {10.1103/PhysRevD.109.123531},
archivePrefix = {arXiv},
       eprint = {2311.17141},
 primaryClass = {astro-ph.CO},
       adsurl = {https://ui.adsabs.harvard.edu/abs/2024PhRvD.109l3531C}
}

@ARTICLE{Legin-2023,
       author = {{Legin}, Ronan and {Ho}, Matthew and {Lemos}, Pablo and {Perreault-Levasseur}, Laurence and {Ho}, Shirley and {Hezaveh}, Yashar and {Wandelt}, Benjamin},
        title = "{Posterior sampling of the initial conditions of the universe from non-linear large scale structures using score-based generative models}",
      journal = {\mnras},
         year = 2023,
        month = nov,
       volume = {527},
       number = {1},
        pages = {L173--L178},
          doi = {10.1093/mnrasl/slad152},
archivePrefix = {arXiv},
       eprint = {2304.03788},
 primaryClass = {astro-ph.CO},
       adsurl = {https://ui.adsabs.harvard.edu/abs/2024MNRAS.527L.173L}
}

@ARTICLE{Remy-2023,
       author = {{Remy}, Beno{\^\i}t and {Lanusse}, Fran{\c{c}}ois and {Jeffrey}, Niall and {Liu}, Jia and {Starck}, Jean-Luc and {Osato}, Ken and {Schrabback}, Tim},
        title = "{Probabilistic mass-mapping with neural score estimation}",
      journal = {\aap},
         year = 2023,
        month = apr,
       volume = {672},
          eid = {A51},
        pages = {A51},
          doi = {10.1051/0004-6361/202243054},
archivePrefix = {arXiv},
       eprint = {2201.05561},
 primaryClass = {astro-ph.CO},
       adsurl = {https://ui.adsabs.harvard.edu/abs/2023A&A...672A..51R}
}
\bibliographystyle{aasjournalv7}



\end{document}